\documentclass{cernrep}
\usepackage{graphicx,here,rotate,epsf,epsfig}
\usepackage{axodraw}
 
\begin{document}

\title{LIMITS OF THE STANDARD MODEL}
 
\author{John Ellis}
 
\institute{CERN, Geneva, Switzerland}
%
%
%
 
\newcommand{\mycomm}[1]{\hfill\break{ \tt===$>$ \bf #1}\hfill\break}

\def\ga{\mathrel{\raise.3ex\hbox{$>$\kern-.75em\lower1ex\hbox{$\sim$}}}}
\def\la{\mathrel{\raise.3ex\hbox{$<$\kern-.75em\lower1ex\hbox{$\sim$}}}}
\def\gev{{\rm \, Ge\kern-0.125em V}}
\def\tev{{\rm \, Te\kern-0.125em V}}
\def\beq{\begin{equation}}
\def\eeq{\end{equation}}
\def\st{\scriptstyle}
\def\ss{\scriptscriptstyle}
\def\mb{m_{\widetilde B}}
\def\msf{m_{\tilde f}}
\def\mst{m_{\tilde t}}
\def\mf{m_{\ss{f}}}
\def\mpar{m_{\ss\|}^2}
\def\mpl{M_{\rm Pl}}
\def\mchi{m_{\chi}}
\def\ohsq{\Omega_{\chi} h^2}
\def\msn{m_{\tilde\nu}}
\def\m12{m_{1\!/2}}
\def\mstpl{m_{\tilde t_{\ss 1}}^2}
\def\mstpr{m_{\tilde t_{\ss 2}}^2}

\def\ga{\mathrel{\raise.3ex\hbox{$>$\kern-.75em\lower1ex\hbox{$\sim$}}}}
\def\la{\mathrel{\raise.3ex\hbox{$<$\kern-.75em\lower1ex\hbox{$\sim$}}}}
\def\gyr{{\rm \, G\kern-0.125em yr}}
\def\gev{{\rm \, Ge\kern-0.125em V}}
\def\tev{{\rm \, Te\kern-0.125em V}}
\def\beq{\begin{equation}}
\def\eeq{\end{equation}}
\def\ss{\scriptscriptstyle}
\def\scs{\scriptstyle}
\def\mb{m_{\widetilde B}}
\def\mst{m_{\tilde\tau_R}}
\def\mstop{m_{\tilde t_1}}
\def\msl{m_{\tilde{\ell}_1}}
\def\stau{\tilde \tau}
\def\stop{\tilde t}
\def\sbot{\tilde b}
\def\mchi{m_{\tilde \chi}}
\def\mxi{m_{\tilde{\chi}_i^0}}
\def\mxj{m_{\tilde{\chi}_j^0}}
\def\mchari{m_{\tilde{\chi}_i^+}}
\def\mcharj{m_{\tilde{\chi}_j^+}}
\def\mgluino{m_{\tilde g}}
\def\msf{m_{\tilde f}}
\def\m12{m_{1\!/2}}
\def\mtb{\overline{m}_{\ss t}}
\def\mbb{\overline{m}_{\ss b}}
\def\mfb{\overline{m}_{\ss f}}
\def\mf{m_{\ss{f}}}
\def\gt{\gamma_t}
\def\gb{\gamma_b}
\def\gf{\gamma_f}
\def\thm{\theta_\mu}
\def\tha{\theta_A}
\def\thb{\theta_B}
\def\mgl{m_{\ss \tilde g}}
\def\cp{C\!P}
\def\ch{{\widetilde \chi}} 
\def\st{{\widetilde \tau}_{\scriptscriptstyle\rm 1}}
\def\sm{{\widetilde \mu}_{\scriptscriptstyle\rm R}}
\def\sel{{\widetilde e}_{\scriptscriptstyle\rm R}}
\def\sl{{\widetilde \ell}_{\scriptscriptstyle\rm R}}
\def\msn{m_{\ch}}
\def\tsq{|{\cal T}|^2}
\def\tcm{\theta_{\rm\scriptscriptstyle CM}}
\def\half{{\textstyle{1\over2}}}
\def\neq{n_{\rm eq}}
\def\qeq{q_{\rm eq}}
\def\slash#1{\rlap{\hbox{$\mskip 1 mu /$}}#1}%
\def\mw{m_W}
\def\mz{m_Z}
\def\mhb{m_{H}}
\def\mhl{m_{h}}
\newcommand\f[1]{f_#1}
\def\nl{\hfill\nonumber\\&&}

\def\gappeq{\mathrel{\rlap {\raise.5ex\hbox{$>$}}
{\lower.5ex\hbox{$\sim$}}}}

\def\lappeq{\mathrel{\rlap{\raise.5ex\hbox{$<$}}
{\lower.5ex\hbox{$\sim$}}}}

\def\Toprel#1\over#2{\mathrel{\mathop{#2}\limits^{#1}}}
\def\FF{\Toprel{\hbox{$\scriptscriptstyle(-)$}}\over{$\nu$}}

\newcommand{\Zee}{$Z^0$}


\def\Yi{\eta^{\ast}_{11} \left( \frac{y_{i}}{2} g' Z_{\chi 1} + 
        g T_{3i} Z_{\chi 2} \right) + \eta^{\ast}_{12} 
        \frac{g m_{q_{i}} Z_{\chi 5-i}}{2 m_{W} B_{i}}}

\def\Xii{\eta^{\ast}_{11} 
        \frac{g m_{q_{i}}Z_{\chi 5-i}^{\ast}}{2 m_{W} B_{i}} - 
        \eta_{12}^{\ast} e_{i} g' Z_{\chi 1}^{\ast}}

\def\Wi{\eta_{21}^{\ast}
        \frac{g m_{q_{i}}Z_{\chi 5-i}^{\ast}}{2 m_{W} B_{i}} -
        \eta_{22}^{\ast} e_{i} g' Z_{\chi 1}^{\ast}}

\def\Vi{\eta_{22}^{\ast} \frac{g m_{q_{i}} Z_{\chi 5-i}}{2 m_{W} B_{i}}
        + \eta_{21}^{\ast}\left( \frac{y_{i}}{2} g' Z_{\chi 1}
        + g T_{3i} Z_{\chi 2} \right)}

\def\zthree{\delta_{1i} [g Z_{\chi 2} - g' Z_{\chi 1}]}

\def\zfour{\delta_{2i} [g Z_{\chi 2} - g' Z_{\chi 1}]}

 
\maketitle 

\begin{flushright}
CERN-TH/2002-320 \\
hep-ph/0211168
\end{flushright}
 
\begin{abstract}

Supersymmetry is one of the most plausible extensions of the Standard
Model, since it is well motivated by the hierarchy problem, supported by
measurements of the gauge coupling strengths, consistent with the
suggestion from precision electroweak data that the Higgs boson may be
relatively light, and provides a ready-made candidate for astrophysical
cold dark matter. In the first lecture, constraints on supersymmetric
models are reviewed, the problems of fine-tuning the electroweak scale and
the dark matter density are discussed, and a number of benchmark scenarios
are proposed. Then the prospects for discovering and measuring
supersymmetry at the LHC, linear colliders and in non-accelerator
experiments are presented. In the second lecture, the evidence for   
neutrino oscillations is recalled, and the parameter space of the seesaw
model is explained. It is shown how these parameters may be explored in a
supersymmetric model via the flavour-changing decays and electric dipole
moments of charged leptons. It is shown that leptogenesis does not relate
the baryon asymmetry of the Universe directly to CP violation in neutrino
oscillations. Finally, possible CERN projects beyond the LHC are
mentioned.

\end{abstract}

\begin{center}
{\it Lectures given at the PSI Summer School, Zuoz, August 2002}
\end{center}

\section{Supersymmetry}

\subsection{Parameters and Problems of the Standard Model}

The Standard Model agrees with all confirmed experimental data from
accelerators, but is theoretically very unsatisfactory~\cite{StAnd}. It
does not explain the particle quantum numbers, such as the electric charge
$Q$, weak isospin $I$, hypercharge $Y$ and colour, and contains at least
19 arbitrary parameters. These include three independent gauge couplings
and a possible CP-violating strong-interaction parameter, six quark and
three charged-lepton masses, three generalized Cabibbo weak mixing angles
and the CP-violating Kobayashi-Maskawa phase, as well as two independent
masses for weak bosons.

As if 19 parameters were insufficient to appall you, at least nine more
parameters must be introduced to accommodate neutrino oscillations: three
neutrino masses, three real mixing angles, and three CP-violating phases,
of which one is in principle observable in neutrino-oscillation
experiments and the other two in neutrinoless double-beta decay
experiments. Even more parameters would be needed to generate masses for 
all the neutrinos~\cite{EHLR}, as discussed in Lecture 2.

The Big Issues in physics beyond the Standard Model are conveniently
grouped into three categories~\cite{StAnd}. These include the problem of
{\bf Mass}: what is the origin of particle masses, are they due to a Higgs
boson, and, if so, why are the masses so small, {\bf Unification}: is
there a simple group framework for unifying all the particle interactions,
a so-called Grand Unified Theory (GUT), and {\bf Flavour}: why are there
so many different types of quarks and leptons and why do their weak
interactions mix in the peculiar way observed? Solutions to all these
problems should eventually be incorporated in a Theory of Everything (TOE)
that also includes gravity, reconciles it with quantum mechanics, explains
the origin of space-time and why it has four dimensions, etc. String
theory, perhaps in its current incarnation of M theory, is the best
(only?) candidate we have for such a TOE~\cite{TOE}, but we do not yet
understand it well enough to make clear experimental predictions.

Supersymmetry is thought to play a r\^ole in solving many of these
problems beyond the Standard Model. The hierarchy of mass scales in
physics, and particularly the fact that $m_W \ll m_P$, appears to require
relatively light supersymmetric particles: $M \lappeq 1$~TeV for its
stabilization~\cite{hierarchy}. As discussed later, GUT predictions for
the unification of gauge couplings work best if the effects of relatively
light supersymmetric particles are included~\cite{GUTs}.  Finally,
supersymmetry seems to be essential for the consistency of string
theory~\cite{GSW}, although this argument does not really restrict the
mass scale at which supersymmetric particles should appear.

Thus there are plenty of good reasons to study
supersymmetry~\cite{CHschool}, so this is the subject of Lecture~1, and it
reappears in Lecture~2 in connection with the observability of
charged-lepton flavour violation.

\subsection{Why Supersymmetry?}

The main theoretical reason to expect supersymmetry at an accessible
energy scale is provided by the {\it hierarchy problem}~\cite{hierarchy}:  
why is $m_W \ll m_P$, or equivalently why is $G_F \sim 1 / m_W^2 \gg G_N =
1 / m_P^2$? Another equivalent question is why the Coulomb potential in an
atom is so much greater than the Newton potential: $e^2 \gg G_N m^2 = m^2
/ m_P^2$, where $m$ is a typical particle mass?

Your first thought might simply be to set $m_P \gg m_W$ by hand, and 
forget about the problem. Life is not so simple, because quantum 
corrections to $m_H$ and hence $m_W$ are quadratically divergent in the 
Standard Model:
\begin{equation}
\delta m_{H,W}^2 \; \simeq \; {\cal O}({\alpha \over \pi}) \Lambda^2,
\label{Qdgt}
\end{equation}
which is $\gg m_W^2$ if the cutoff $\Lambda$, which represents the scale 
where new physics beyond the Standard Model appears, is comparable to the 
GUT or Planck scale. For example, if the 
Standard Model were to hold unscathed all the way up the Planck mass $m_P 
\sim 10^{19}$~GeV, the radiative correction (\ref{Qdgt}) would be 36 
orders of magnitude greater than the physical values of $m_{H,W}^2$! 

In principle, this is not a problem from the mathematical point of view of
renormalization theory. All one has to do is postulate a tree-level value
of $m_H^2$ that is (very nearly) equal and opposite to the `correction'
(\ref{Qdgt}), and the correct physical value may be obtained. However,
this fine tuning strikes many physicists as rather unnatural: they would
prefer a mechanism that keeps the `correction' (\ref{Qdgt}) comparable at
most to the physical value~\cite{hierarchy}.

This is possible in a supersymmetric theory, in which there are equal numbers 
of bosons and fermions with identical couplings. Since bosonic and fermionic 
loops have opposite signs, the residual one-loop correction is of the form
\begin{equation}
\delta m_{H,W}^2 \; \simeq \; {\cal O}({\alpha \over \pi}) (m_B^2 - 
m_F^2),
\label{susy}
\end{equation}
which is $\lappeq m_{H,W}^2$ and hence naturally small if the 
supersymmetric partner bosons $B$ and fermions $F$ have similar masses:
\begin{equation}
|m_B^2 - m_F^2| \; \lappeq \; 1~{\rm TeV}^2.
\label{natural}
\end{equation}
This is the best motivation we have for finding supersymmetry at 
relatively low energies~\cite{hierarchy}.
In addition to this first supersymmetric miracle of removing (\ref{susy})
the quadratic divergence (\ref{Qdgt}), many logarithmic divergences are
also absent in a supersymmetric theory~\cite{noren}, a property that also 
plays a r\^ole in the construction of supersymmetric GUTs~\cite{StAnd}. 

Could any of the known particles in the Standard Model be paired up
in supermultiplets? Unfortunately, none of the known fermions $q, \ell$
can be paired with any of the `known' bosons $\gamma, W^\pm Z^0, g, H$,
because their internal quantum numbers do not match~\cite{Fayet}. For
example, quarks $q$ sit in triplet representations of colour, whereas the
known bosons are either singlets or octets of colour. Then again, leptons
$\ell$ have non-zero lepton number $L = 1$, whereas the known bosons have
$L = 0$. Thus, the only possibility seems to be to introduce new
supersymmetric partners (spartners) for all the known particles: quark
$\to$ squark, lepton $\to$ slepton, photon $\to$ photino, Z $\to$ Zino, W
$\to$ Wino, gluon $\to$ gluino, Higgs $\to$ Higgsino. The best that one
can say for supersymmetry is that it economizes on principle, not on
particles!

\subsection{Hints of Supersymmetry}

There are some phenomenological hints that supersymmetry may, indeed,
appear at the Tev scale. 
One is provided by the strengths of the different gauge
interactions, as measured at LEP~\cite{GUTs}. These may be run up to high
energy scales using the renormalization-group equations, to see whether
they unify as predicted in a GUT. The answer is no, if supersymmetry is
not included in the calculations. In that case, GUTs would require
\begin{equation} 
\sin^2 \theta_W \; = \; 0.214 \pm 0.004, 
\label{GUT}
\end{equation} 
whereas the experimental value of the effective neutral
weak mixing parameter at the $Z^0$ peak is $\sin^2 \theta = 0.23149 \pm
0.00017$~\cite{LEPEWWG}. On the other hand, minimal supersymmetric GUTs
predict 
\begin{equation} 
\sin^2 \theta_W \; \simeq \; 0.232,
\label{susyGUT} 
\end{equation} 
where the error depends on the assumed
sparticle masses, the preferred value being around 1~TeV~\cite{GUTs}, as 
suggested completely independently by the naturalness of the electroweak mass
hierarchy. 

A second hint is the fact that precision electroweak data prefer a
relatively light Higgs boson weighing less than about
200~GeV~\cite{LEPEWWG}. This is perfectly consistent with calculations in
the minimal supersymmetric extension of the Standard Model (MSSM), in
which the lightest Higgs boson weighs less than about
130~GeV~\cite{susyHiggs}.

A third hint is provided by the astrophysical necessity of cold dark
matter. This could be provided by a neutral, weakly-interacting particle
weighing less than about 1~TeV, such as the lightest supersymmetric
particle (LSP) $\chi$~\cite{EHNOS}.

\subsection{Building Supersymmetric Models}

Any supersymmetric model is based on a Lagrangian that contains a
supersymmetric part and a supersym- metry-breaking 
part~\cite{FF,CHschool}:
\begin{equation}
{\cal L} \; = \; {\cal L}_{susy} \; + \; {\cal L}_{susy \times}.
\label{twobits}
\end{equation}
We concentrate here on the supersymmetric part ${\cal
L}_{susy}$. The minimal supersymmetric extension of the Standard Model
(MSSM) has the same gauge interactions as the Standard Model, and Yukawa
interactions that are closely related. They are based on a superpotential
$W$ that is a cubic function of complex superfields corresponding to
left-handed fermion fields. Conventional left-handed lepton and quark
doublets are denoted $L, Q$, and right-handed fermions are introduced via
their conjugate fields, which are left-handed, $e_R \to E^c, u_R \to U^c,
d_R \to D^c$. In terms of these,
\begin{equation}
W \; = \; \Sigma_{L,E^c} \lambda_L L E^c H_1 \; + \; \Sigma_{Q, U^c} 
\lambda_U Q U^c H_2 \; + \; \Sigma_{Q, D^c} \lambda_D Q D^c H_1 \; 
+ \mu H_1 H_2.
\label{SMW}
\end{equation}
A few words of explanation are warranted. The first three terms in
(\ref{SMW}) yield masses for the charged leptons, charge-$(+2/3)$ quarks
and charge-$(-1/3)$ quarks respectively. All of the Yukawa couplings
$\lambda_{L,U,D}$ are $3 \times 3$ matrices in flavour space, whose
diagonalizations yield the mass eigenstates and Cabibbo-Kobayashi-Maskawa
mixing angles for quarks.

Note that two distinct Higgs doublets $H_{1,2}$ have been introduced, for
two important reasons. One reason is that the superpotential must be an
analytic polynomial: it cannot contain both $H$ and $H^*$, whereas the
Standard Model uses both of these to give masses to all the quarks and
leptons with just a single Higgs doublet. The other reason for introducing
two Higgs doublets $H_{1,2}$ is to cancel the triangle anomalies that
destroy the renormalizability of a gauge theory. Ordinary Higgs boson
doublets do not contribute to these anomalies, but the fermions in Higgs
supermultiplets do, and pairs of doublets are required to cancel each
others' contributions. Once two Higgs supermultiplets have been
introduced, there must in general be a bilinear term $\mu H_1 H_2$
coupling them together.

In general, the supersymmetric partners of the $W^\pm$ and charged Higgs
bosons $H^\pm$ (the `charginos' $\chi^\pm$) mix, as do those of the
$\gamma, Z^0$ and $H^0_{1,2}$ (the `neutralinos' $\chi^0_i$):
see~\cite{StAnd}. The lightest neutralino $\chi$ is a likely candidate 
to be the Lightest Supersymmetric Particle (LSP), and hence constitute the 
astrophysical cold dark matter~\cite{EHNOS}.

Once the MSSM superpotential (\ref{SMW}) has been specified, the effective
potential is also fixed:
\begin{equation}
V \; = \; \Sigma_i |F^i|^2 \; + \; 
{1 \over 2} \Sigma_a (D^a)^2: \; \; 
F^*_i \equiv {\partial W \over \partial \phi^i}, \; 
D^a \equiv g_a \phi^*_i (T^a)^i_j \phi^j,
\label{SMV}
\end{equation}
where the sums
run over the different chiral fields $i$ and the $SU(3), SU(2)$ and $U(1)$
gauge-group factors $a$. Thus, the quartic terms in the effective Higgs 
potential are completely fixed, which leads to the prediction that the 
lightest Higgs boson should weigh $\lappeq 130$~GeV~\cite{susyHiggs}.

In addition to the supersymmetric part ${\cal L}_{susy}$ of the lagrangian
(\ref{twobits}) above, there is also the superym- metry-breaking piece
${\cal L}_{susy \times}$. The origin of this piece is unclear, and in
these lectures we shall just assume a suitable phenomenological
parameterization. In order not to undo the supersymmetric miracles
mentioned above, the breaking of supersymmetry should be `soft', in the
sense that it does not reintroduce any unwanted quadratic or logarithmic
divergences. The candidates for such soft superymmetry breaking are
gaugino masses $M_{a}$ for each of the gauge group factors $a$ in
the Standard Model, scalar masses-squared $m_{0}^2$ that should be 
regarded as matrices in the flavour index $i$ of the
matter supermultiplets, and trilinear scalar couplings $A_{ijk}$
corresponding to each of the Yukawa couplings $\lambda_{ijk}$ in the 
Standard Model.

There are very many such soft superymmetry-breaking terms. Upper limits on 
flavour-changing neutral interactions suggest~\cite{FCNI} that the scalar 
masses-squared $m_{0}^2$ are (approximately) independent of generation 
for particles with the same quantum numbers, e.g., sleptons, and that the 
$A_{ijk}$ are related to the $\lambda_{ijk}$ by a universal constant of 
proportionality $A$. In these lectures, for definiteness, we assume 
universality at the input GUT scale for all the gaugino masses:
\begin{equation}
M_{a} \; = \; m_{1/2},
\end{equation}
and likewise for the scalar masses-squared and trilinear parameters:
\begin{equation}
m_{0}^2 \; = m_{0}^2 \delta^i_j, \; A_{ijk} \; = \; A \lambda_{ijk}.
\end{equation}
This is known as the constrained MSSM (CMSSM). The values of the soft
supersymmetry-breaking parameters at observable energies $\sim 1$~TeV are
renormalized by calculable factors~\cite{softren}, in a similar manner to
the gauge couplings and fermion masses. These renormalization factors are
included in the subsequent discussions, and play a key r\^ole in 
Lecture~2.
The physical value of $\mu$ is fixed up to a sign in the CMSSM, as is the
pseudoscalar Higgs mass $m_A$, by the electroweak vacuum conditions.

\subsection{Constraints on the MSSM}

\begin{figure}
\vskip 0.5in
\vspace*{-0.75in}
\begin{minipage}{8in}
\epsfig{file=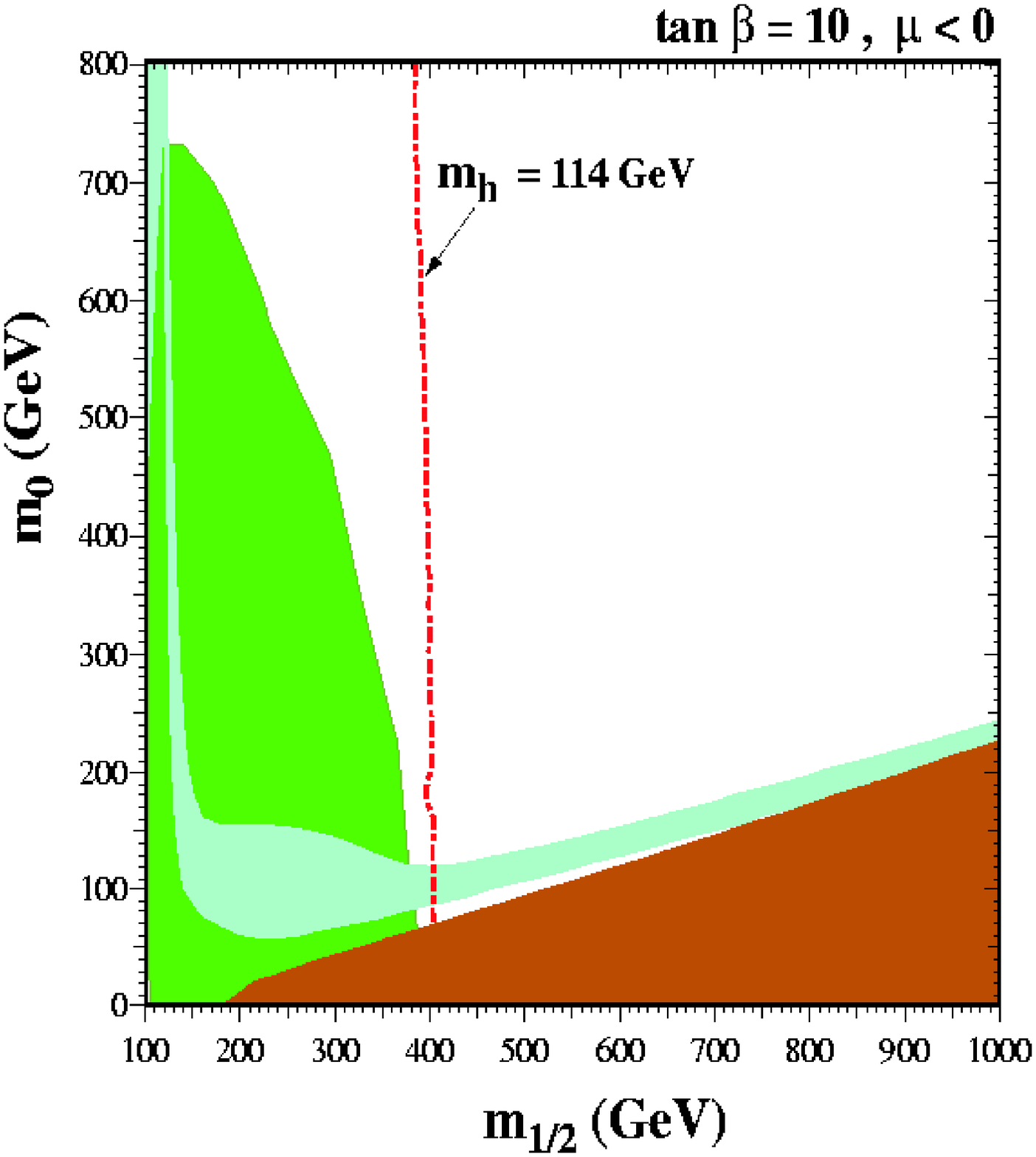,height=3.3in}
\epsfig{file=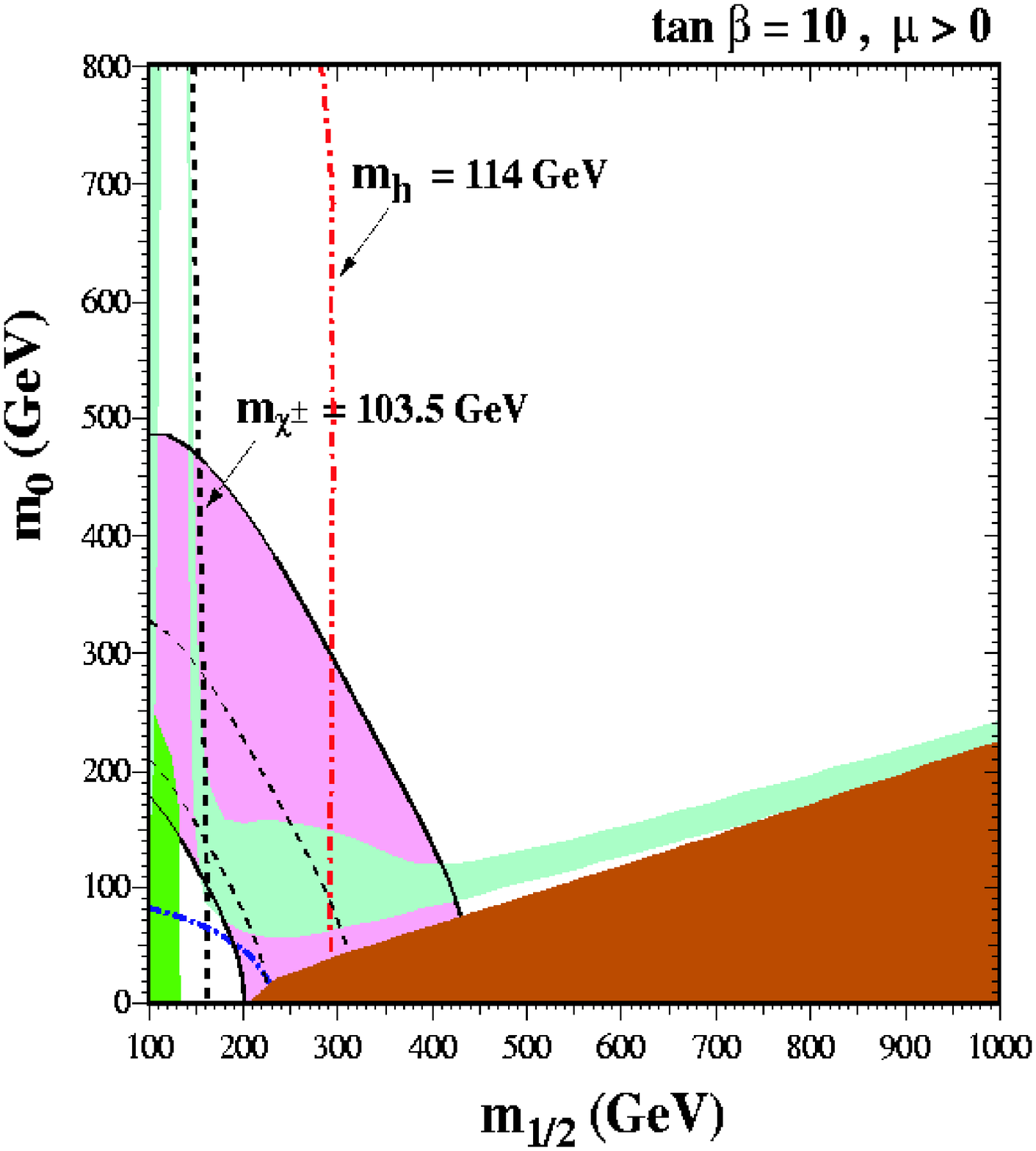,height=3.3in} \hfill
\end{minipage}
\begin{minipage}{8in}
\epsfig{file=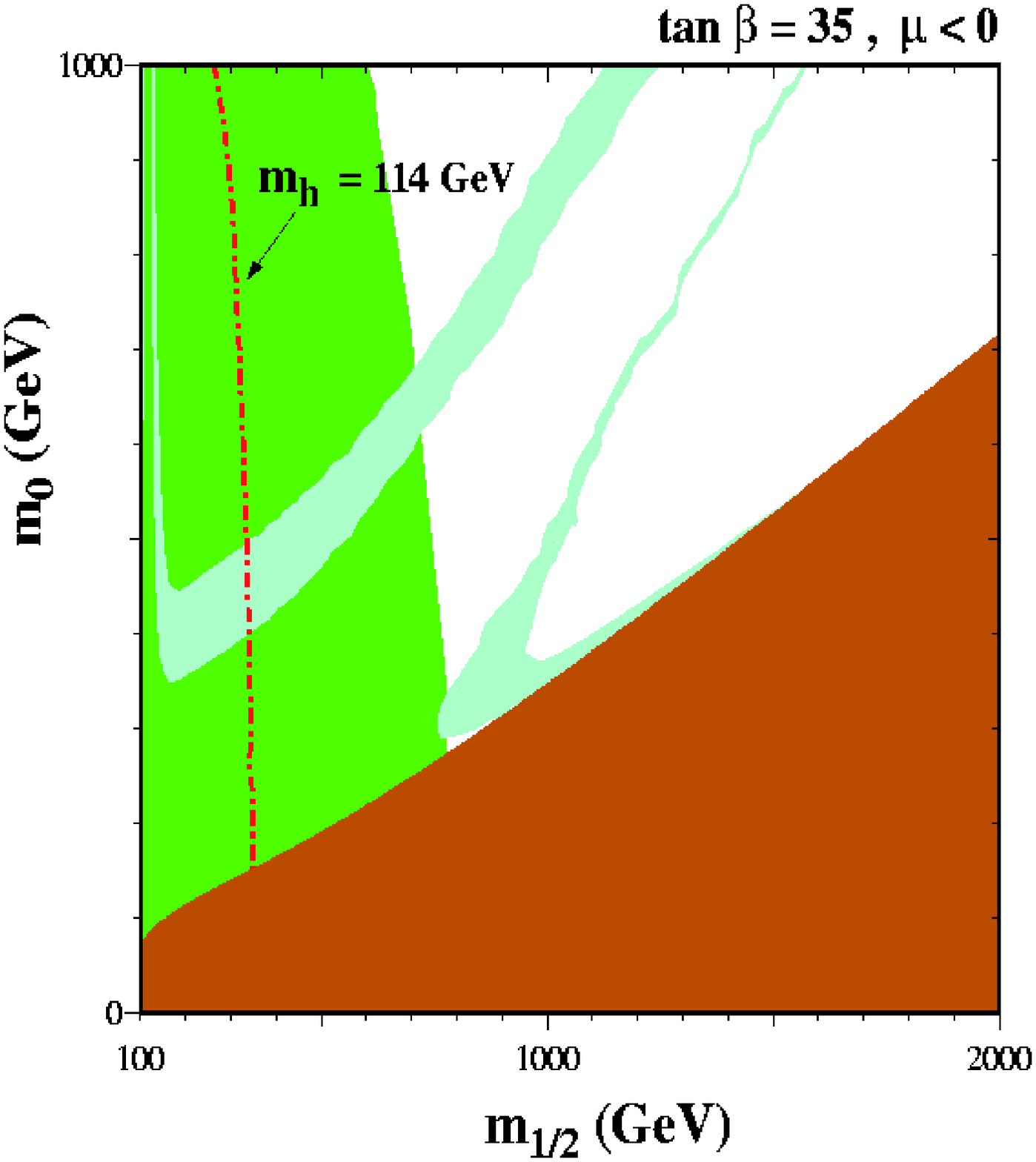,height=3.3in}
\epsfig{file=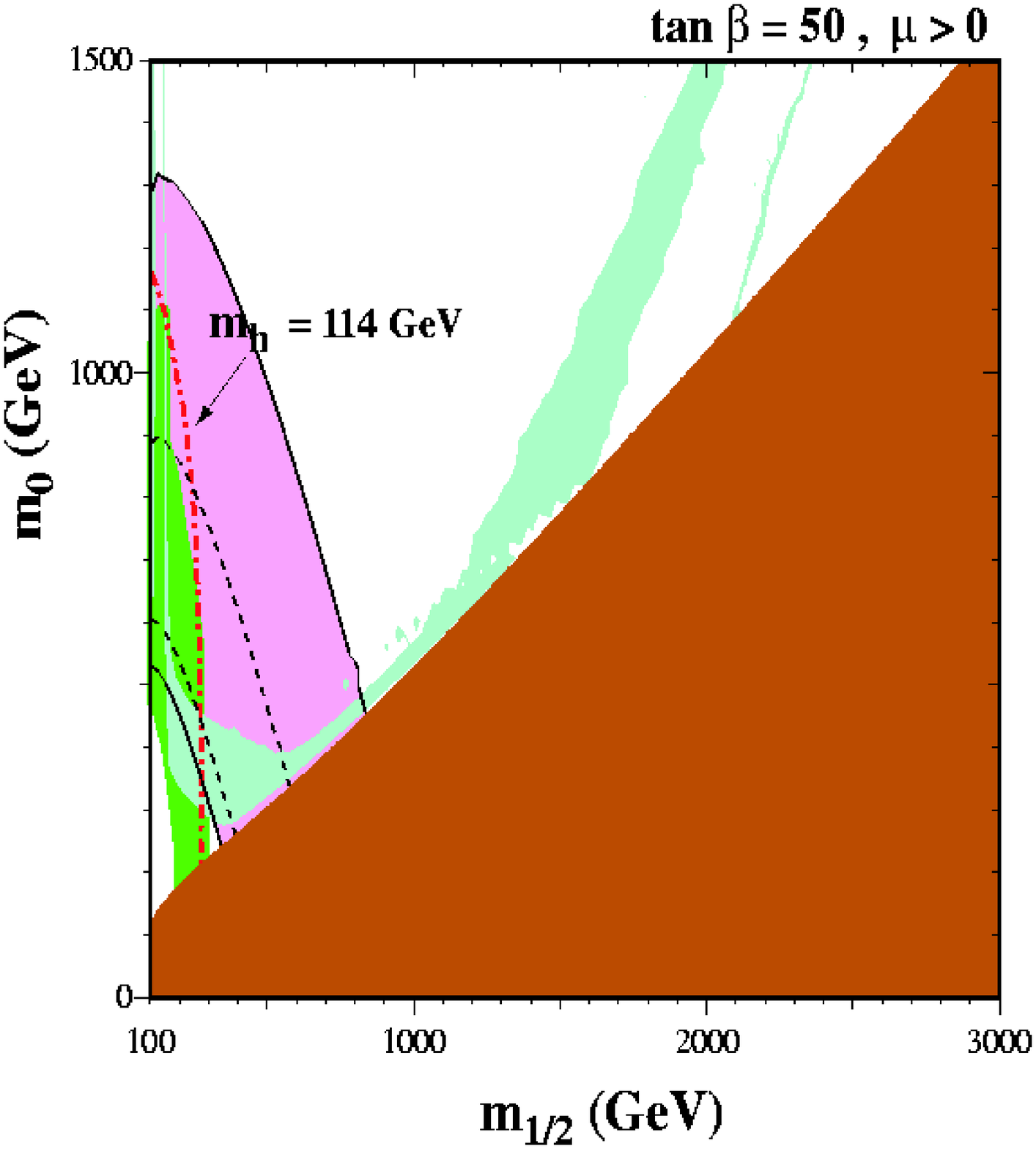,height=3.3in} \hfill
\end{minipage}
\caption{
{\it Compilations of phenomenological constraints on the CMSSM for
(a) $\tan \beta = 10, \mu < 0$,  (b) $\tan \beta = 10, \mu > 0$, (c)
$\tan \beta = 35, \mu < 0$ and (d)  $\tan \beta = 50, \mu > 0$, assuming
$A_0 = 0, m_t = 175$~GeV and $m_b(m_b)^{\overline {MS}}_{SM} = 4.25$~GeV
\cite{EFGOSi}.  The near-vertical lines are the LEP limits
$m_{\chi^\pm} = 103.5$~GeV (dashed and black)~\cite{LEPsusy}, shown in
(b) only, and
$m_h = 114$~GeV (dotted and red)~\cite{LEPHWG}. 
Also, in the lower left corner of (b), we show the $m_{\tilde e} = 99$
GeV contour \protect\cite{LEPSUSYWG_0101}.  In the dark (brick red)
shaded regions, the LSP is the charged
${\tilde
\tau}_1$, so this region is excluded. The light (turquoise) shaded areas
are the cosmologically preferred regions with
\protect\mbox{$0.1\leq\ohsq\leq 0.3$}~\cite{EFGOSi}. The medium (dark
green) shaded regions that are most prominent in panels (a) and (c) are
excluded by $b \to s \gamma$~\cite{bsg}. The shaded (pink) regions in the 
upper right regions show the $\pm 2 \, \sigma$ ranges of $g_\mu -
2$. For $\mu > 0$, the $\pm 2 (1) \, \sigma$ contours are also shown as 
solid (dashed) black lines~\cite{EOSnew}.
}}
\label{fig:CMSSM}
\end{figure}

Important experimental constraints on the MSSM parameter space are
provided by direct searches at LEP and the Tevatron collider, as compiled
in the $(m_{1/2}, m_0)$ planes for different values of $\tan \beta$ and
the sign of $\mu$ in Fig.~\ref{fig:CMSSM}. One of these is the limit
$m_{\chi^\pm} \gappeq$ 103.5 GeV provided by chargino searches at
LEP~\cite{LEPsusy}, where the fourth significant figure depends on other
CMSSM parameters. LEP has also provided lower limits on slepton masses, of
which the strongest is $m_{\tilde e}\gappeq$ 99 GeV \cite{LEPSUSYWG_0101},
again depending only sightly on the other CMSSM parameters, as long as
$m_{\tilde e} - m_\chi \gappeq$ 10 GeV. The most important constraints on
the $u, d, s, c, b$ squarks and gluinos are provided by the FNAL Tevatron
collider: for equal masses $m_{\tilde q} = m_{\tilde g} \gappeq$ 300 GeV.
In the case of the $\tilde t$, LEP provides the most stringent limit when
$m_{\tilde t} - m_\chi$ is small, and the Tevatron for larger $m_{\tilde
t} - m_\chi$~\cite{LEPsusy}.

Another important constraint is provided by the LEP lower limit on the
Higgs mass: $m_H > $ 114.4 GeV \cite{LEPHWG}. This holds in the
Standard Model, for the lightest Higgs boson $h$ in the general MSSM for
$\tan\beta
\lappeq 8$, and almost always in the CMSSM for all $\tan\beta$, at least
as long as CP is conserved~\footnote{The lower bound on the lightest MSSM
Higgs boson may be relaxed significantly if CP violation feeds into the
MSSM Higgs sector~\cite{CEPW}.}. Since $m_h$ is sensitive to sparticle
masses, particularly $m_{\tilde t}$, via loop corrections:
\begin{equation}
\delta m^2_h \propto {m^4_t\over m^2_W}~\ln\left({m^2_{\tilde t}\over
m^2_t}\right)~ + \ldots
\label{nine}
\end{equation}
the Higgs limit also imposes important constraints on the soft 
supersymmetry-breaking CMSSM parameters,
principally $m_{1/2}$~\cite{EGNO} as seen in Fig.~\ref{fig:CMSSM}. The 
constraints are here
evaluated using {\tt FeynHiggs}~\cite{FeynHiggs}, which is estimated to 
have a residual uncertainty of a couple of GeV in $m_h$.

Also shown in Fig.~\ref{fig:CMSSM} is the constraint imposed by
measurements of $b\rightarrow s\gamma$~\cite{bsg}. These agree with the
Standard Model, and therefore provide bounds on MSSM particles, such as
the chargino and charged Higgs masses, in particular. Typically, the
$b\rightarrow s\gamma$ constraint is more important for $\mu < 0$, as seen
in Fig.~\ref{fig:CMSSM}a and c, but it is also relevant for $\mu > 0$,
particularly when $\tan\beta$ is large as seen in Fig.~\ref{fig:CMSSM}d.

The final experimental constraint we consider is that due to the
measurement of the anomolous magnetic moment of the muon. Following its
first result last year~\cite{BNL1}, the BNL E821 experiment has recently
reported a new measurement~\cite{BNL2} of $a_\mu\equiv {1\over 2} (g_\mu
-2)$, which deviates by 3.0 standard deviations from the best available
Standard Model predictions based on low-energy $e^+ e^- \to $ hadrons
data~\cite{Davier}. On the other hand, the discrepancy is more like 1.6 
standard
deviations if one uses $\tau \to $ hadrons data to calculate the Standard
Model prediction. Faced with this confusion, and remembering the chequered
history of previous theoretical calculations~\cite{lightbylight}, it is
reasonable to defer judgement whether there is a significant discrepancy
with the Standard Model. However, either way, the measurement of $a_\mu$
is a significant constraint on the CMSSM, favouring $\mu > 0$ in general,
and a specific region of the $(m_{1/2}, m_0)$ plane if one accepts the
theoretical prediction based on $e^+ e^- \to $ hadrons 
data~\cite{susygmu}. The regions
preferred by the current $g-2$ experimental data and the $e^+ e^- \to $
hadrons data are shown in Fig.~\ref{fig:CMSSM}.

Fig.~\ref{fig:CMSSM} also displays the regions where the supersymmetric 
relic density $\rho_\chi = \Omega_\chi \rho_{critical}$ falls within the 
preferred range
\begin{equation}
0.1 < \Omega_\chi h^2 < 0.3
\label{ten}
\end{equation}
The upper limit on the relic density is rigorous, since astrophysics and 
cosmology tell us that
the total matter density $\Omega_m \lappeq 0.4$~\cite{density}, 
and the Hubble expansion
rate $h \sim 1/\sqrt{2}$ to within about 10 \% (in units of 100 km/s/Mpc). On
the other hand, the lower limit in (\ref{ten}) is optional, since there
could be other important contributions to the overall matter 
density. Smaller values of $\Omega_\chi h^2$ correspond to 
smaller values of $(m_{1/2}, m_0)$, in general.

As is seen in Fig.~\ref{fig:CMSSM}, there are generic regions of the CMSSM
parameter space where the relic density falls within the preferred range
(\ref{ten}). What goes into the calculation of the relic density? It is
controlled by the annihilation cross section~\cite{EHNOS}:
\begin{equation}
\rho_\chi = m_\chi n_\chi \, , \quad n_\chi \sim {1\over
\sigma_{ann}(\chi\chi\rightarrow\ldots)}\, ,
\label{eleven}
\end{equation}
where the typical annihilation cross section $\sigma_{ann} \sim 1/m_\chi^2$.
For this reason, the relic density typically increases with the relic
mass, and this combined with the upper bound in (\ref{ten}) then leads to
the common expectation that $m_\chi \lappeq {\cal O}(1)$~GeV. 

\begin{figure}
\vskip 0.5in
\vspace*{-0.75in}
\begin{minipage}{8in}
\epsfig{file=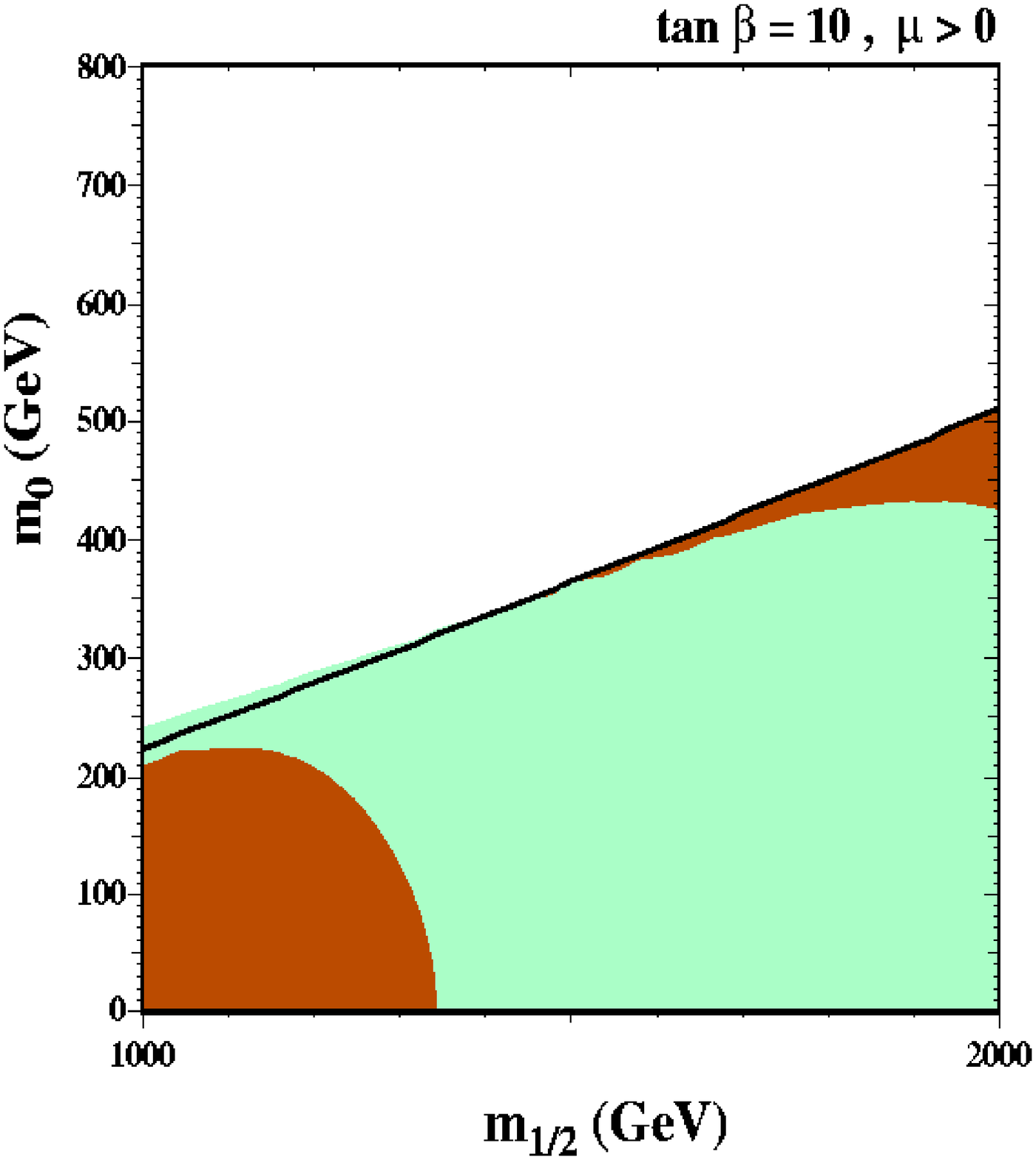,height=3.3in}
\hspace*{-0.17in}
\epsfig{file=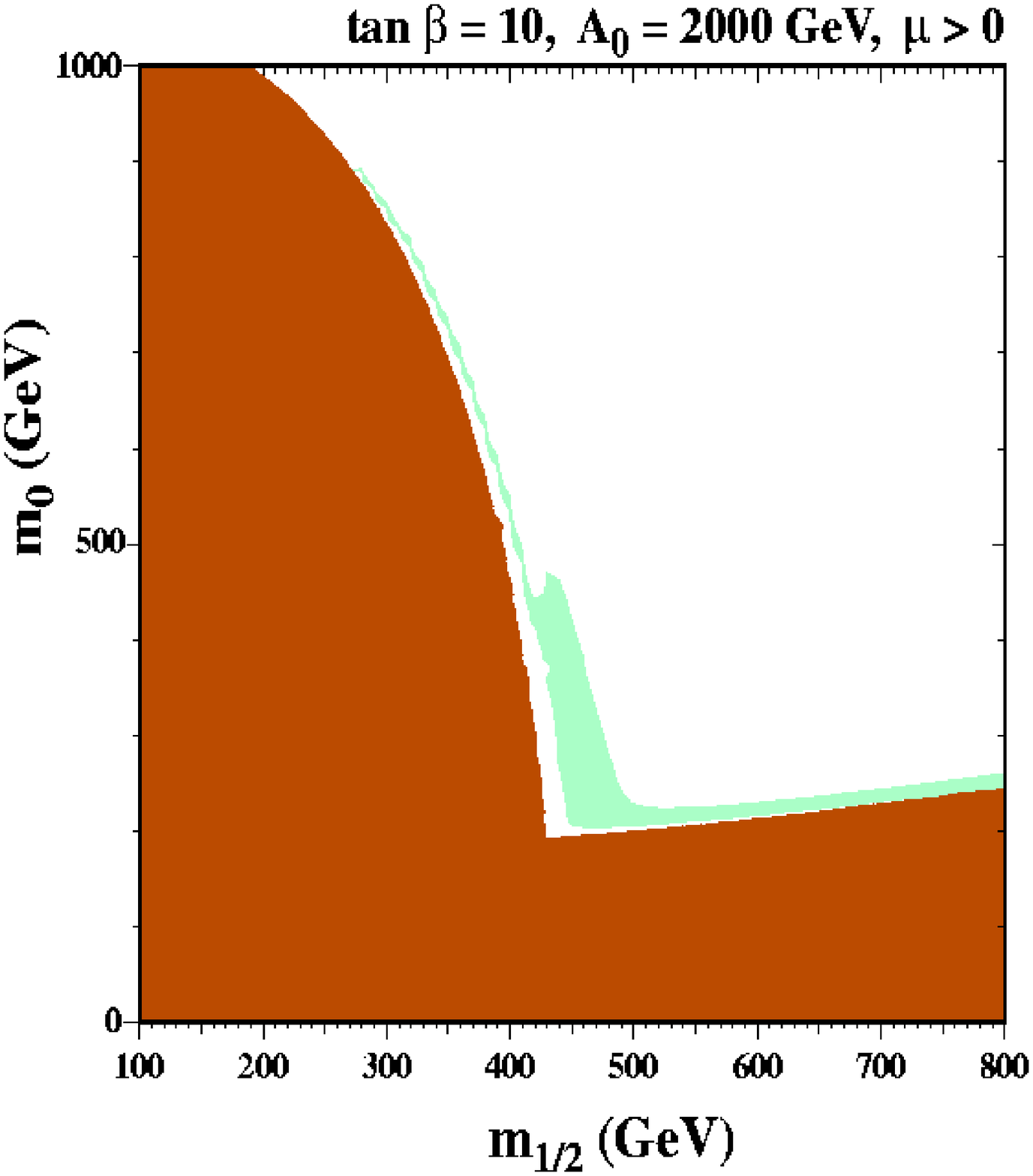,height=3.3in} \hfill
\end{minipage}
\caption[]{\it (a) The large-$m_{1/2}$ `tail' of the $\chi - {\tilde 
\tau_1}$ 
coannihilation region 
for $\tan \beta = 10$, $A = 0$ and $\mu < 0$~\cite{ourcoann}, superimposed 
on the disallowed dark (brick red) shaded region where $m_{\tilde
\tau_1} < m_\chi$, and (b) the $\chi - {\tilde t_1}$ coannihilation region
for $\tan \beta = 10$, $A = 2000$~GeV and $\mu > 0$~\cite{EOS}, exhibiting 
a large-$m_0$ `tail', again with a dark (brick red) shaded region 
excluded because the LSP is charged.}
\label{fig:coann}
\end{figure}

However, there are various ways in which the generic upper bound on
$m_\chi$ can be increased along filaments in the $(m_{1/2},m_0)$ plane.
For example, if the next-to-lightest sparticle (NLSP) is not much heavier
than $\chi$: $\Delta m/m_\chi \lappeq 0.1$, the relic density may be
suppressed by coannihilation: $\sigma (\chi + $NLSP$ \rightarrow \ldots
)$~\cite{coann}.  In this way, the allowed CMSSM region may acquire a
`tail' extending to larger sparticle masses. An example of this
possibility is the case where the NLSP is the lighter stau: $\tilde\tau_1$
and $m_{\tilde\tau_1} \sim m_\chi$, as seen in Figs.~\ref{fig:CMSSM}(a)
and (b) and extended to larger $m_{1/2}$ in
Fig.~\ref{fig:coann}(a)~\cite{ourcoann}. Another example is coannihilation
when the NLSP is the lighter stop~\cite{stopco}, $\tilde t_1$, and
$m_{\tilde t_1} \sim m_\chi$, which may be important in the general MSSM
or in the CMSSM when $A$ is large, as seen in
Fig.~\ref{fig:coann}(b)~\cite{EOS}. In the cases studied, the upper limit
on $m_\chi$ is not affected by stop coannihilation.

\begin{figure}
\hspace*{-.40in}
\begin{minipage}{8in}
\epsfig{file=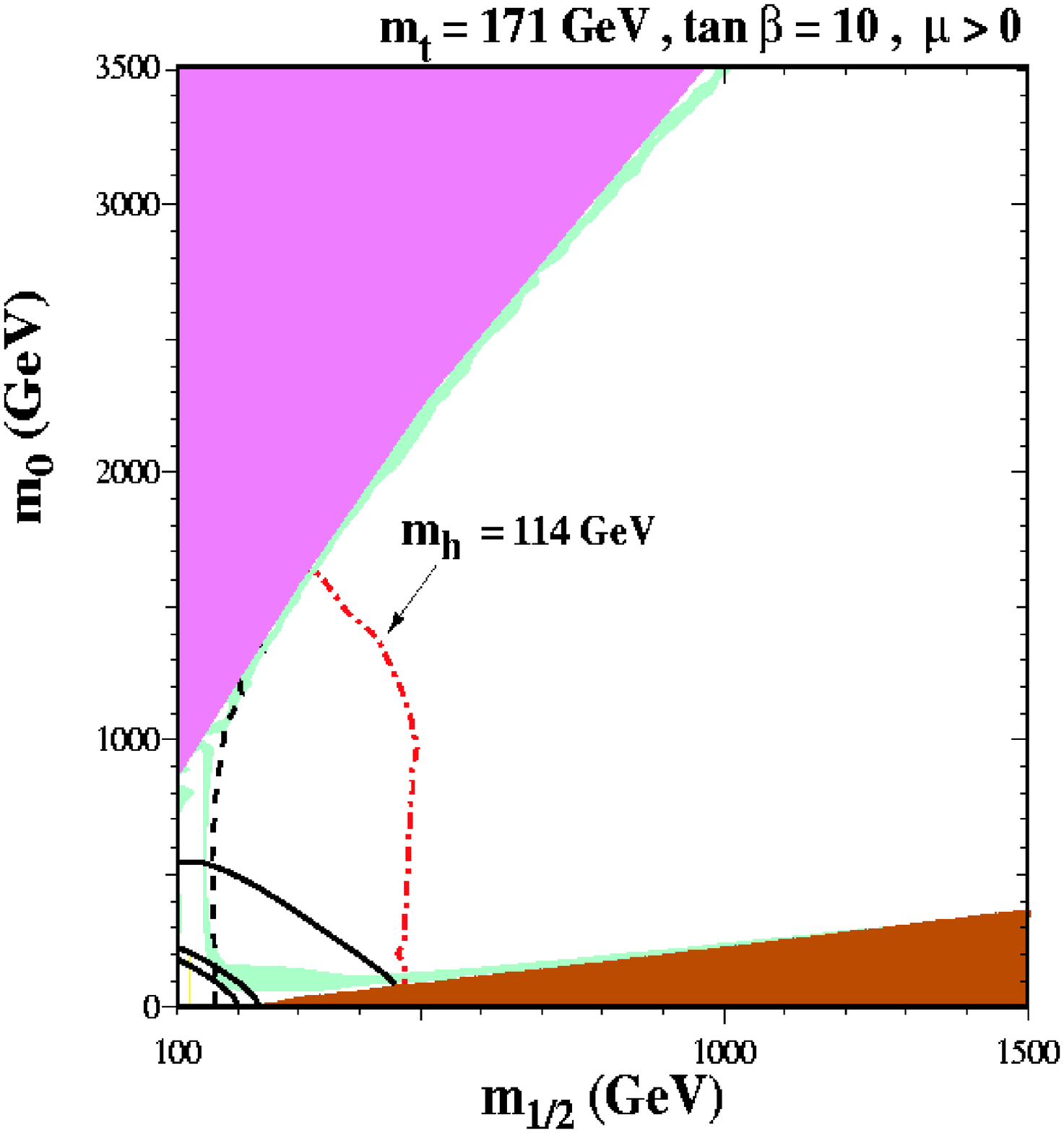,height=3.3in}
\hspace*{-0.17in}
\epsfig{file=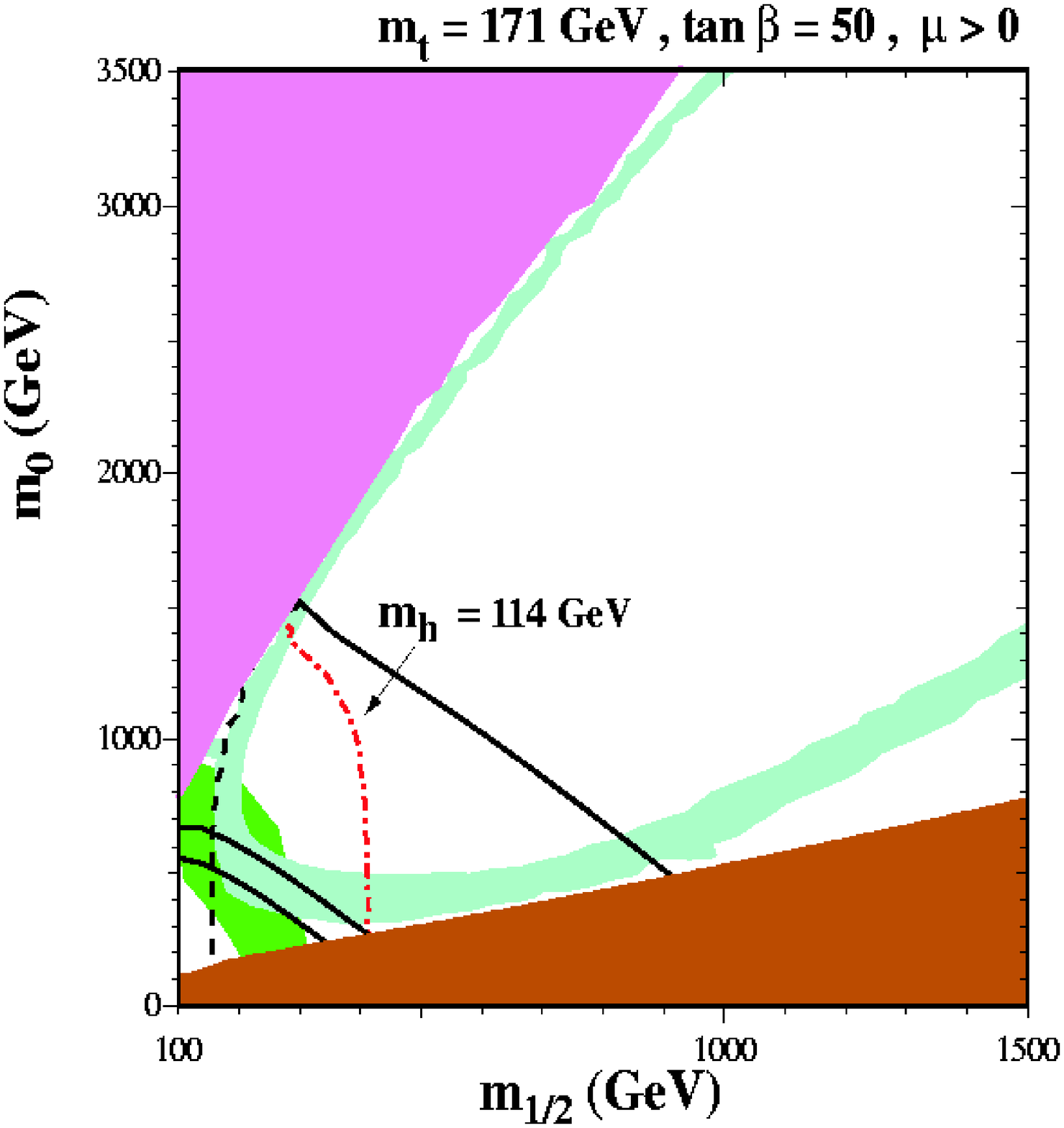,height=3.3in} \hfill
\end{minipage}
\caption[]{\it An expanded view of the $m_{1/2} - m_0$ parameter plane
showing the focus-point regions \protect\cite{focus} at large $m_0$ for 
(a) $tan \beta
= 10$, and (b) $\tan \beta = 50$~\cite{EOSnew}. In the shaded (mauve) 
region in the 
upper
left corner, there are no solutions with proper electroweak symmetry 
breaking, so these are
excluded in the CMSSM.  Note that we have chosen $m_t = 171$ GeV, in 
which case the focus-point region is at lower $m_0$ than when $m_t = 175$ 
GeV, as assumed in the other figures. The position of this region
is very sensitive to $m_t$. The black contours (both dashed and solid)
are as in Fig.~\protect\ref{fig:CMSSM}, we do not shade the preferred
$g-2$ region. }
\label{fig:focus}
\end{figure}

Another mechanism for extending the allowed CMSSM region to large $m_\chi$
is rapid annihilation via a direct-channel pole when $m_\chi \sim {1\over
2} m_{Higgs, Z}$~\cite{funnel,EFGOSi}. This may yield a `funnel' extending
to large $m_{1/2}$ and $m_0$ at large $\tan\beta$, as seen in panels (c)
and (d) of Fig.~\ref{fig:CMSSM}~\cite{EFGOSi}. Yet another allowed region
at large $m_{1/2}$ and $m_0$ is the `focus-point' region~\cite{focus},
which is adjacent to the boundary of the region where electroweak symmetry
breaking is possible, as seen in Fig.~\ref{fig:focus}. The lightest
supersymmetric particle is relatively light in this region.

\subsection{Fine Tuning}

\begin{figure}
\vskip 0.5in
\vspace*{-0.75in}
\hspace*{-.20in}
\begin{minipage}{8in}
\epsfig{file=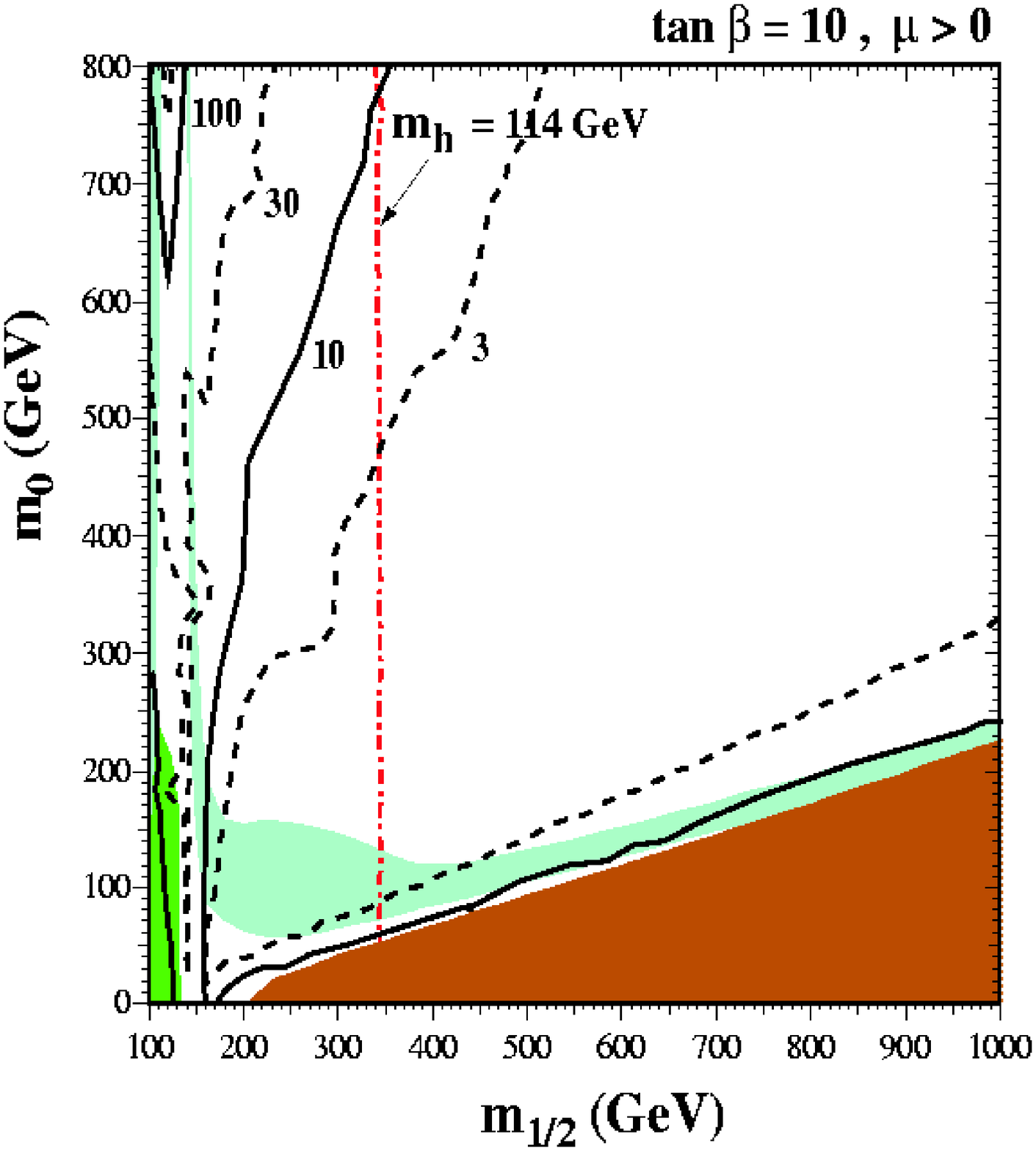,height=3.3in}
\hspace*{+0.10in}
\epsfig{file=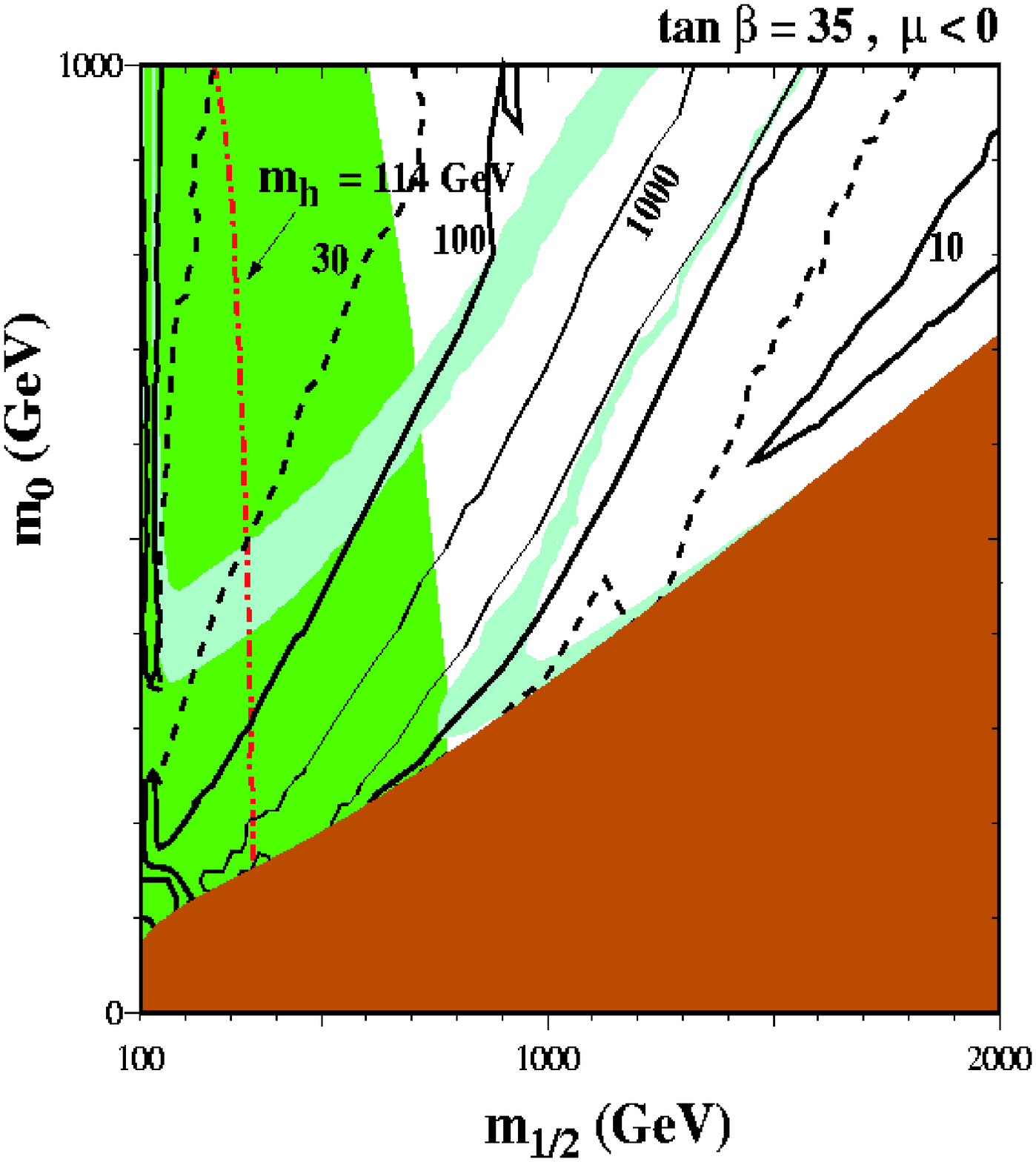,height=3.3in} \hfill
\end{minipage}
\hspace*{-.20in}
\begin{minipage}{8in}
\epsfig{file=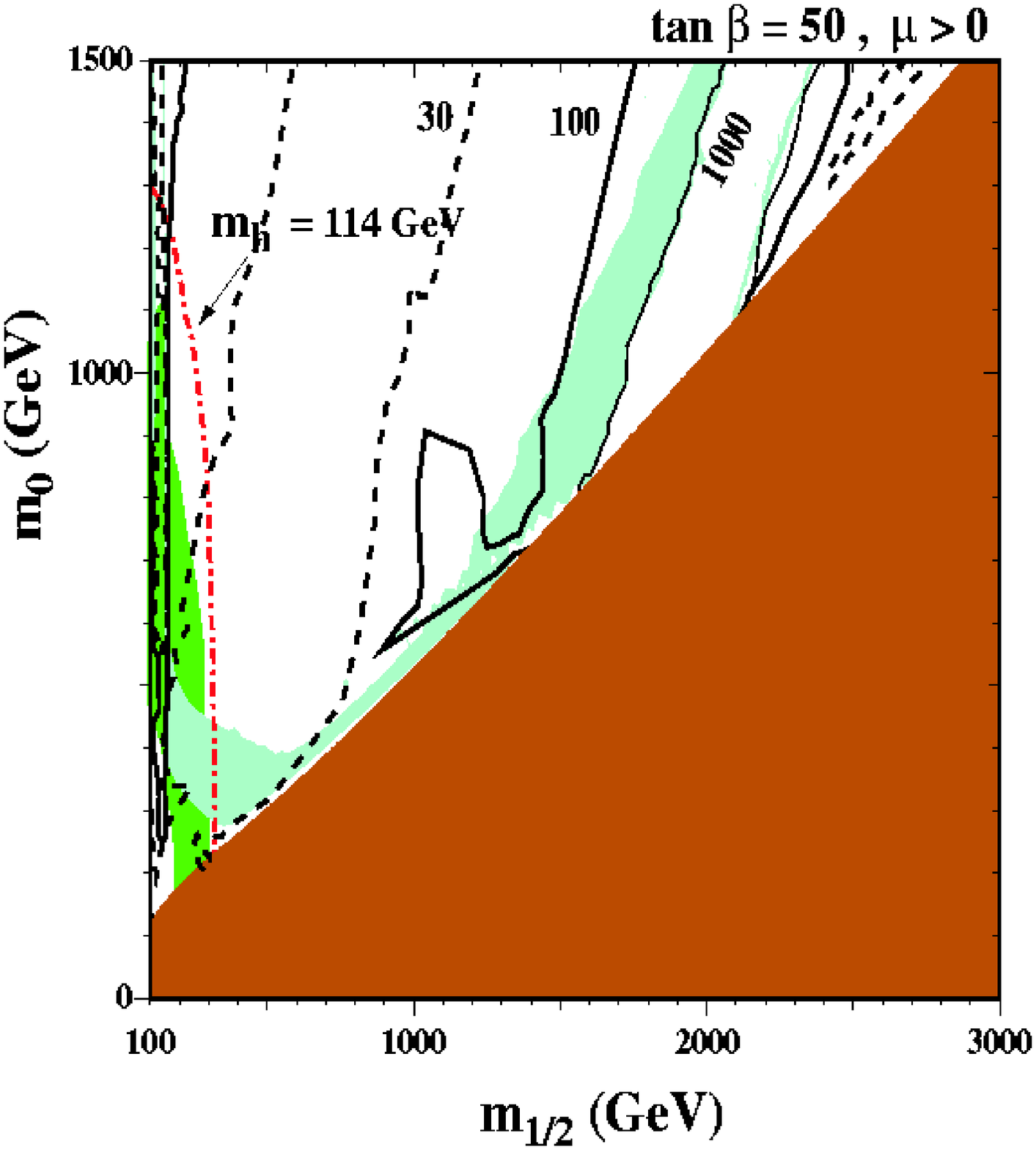,height=3.3in}
\epsfig{file=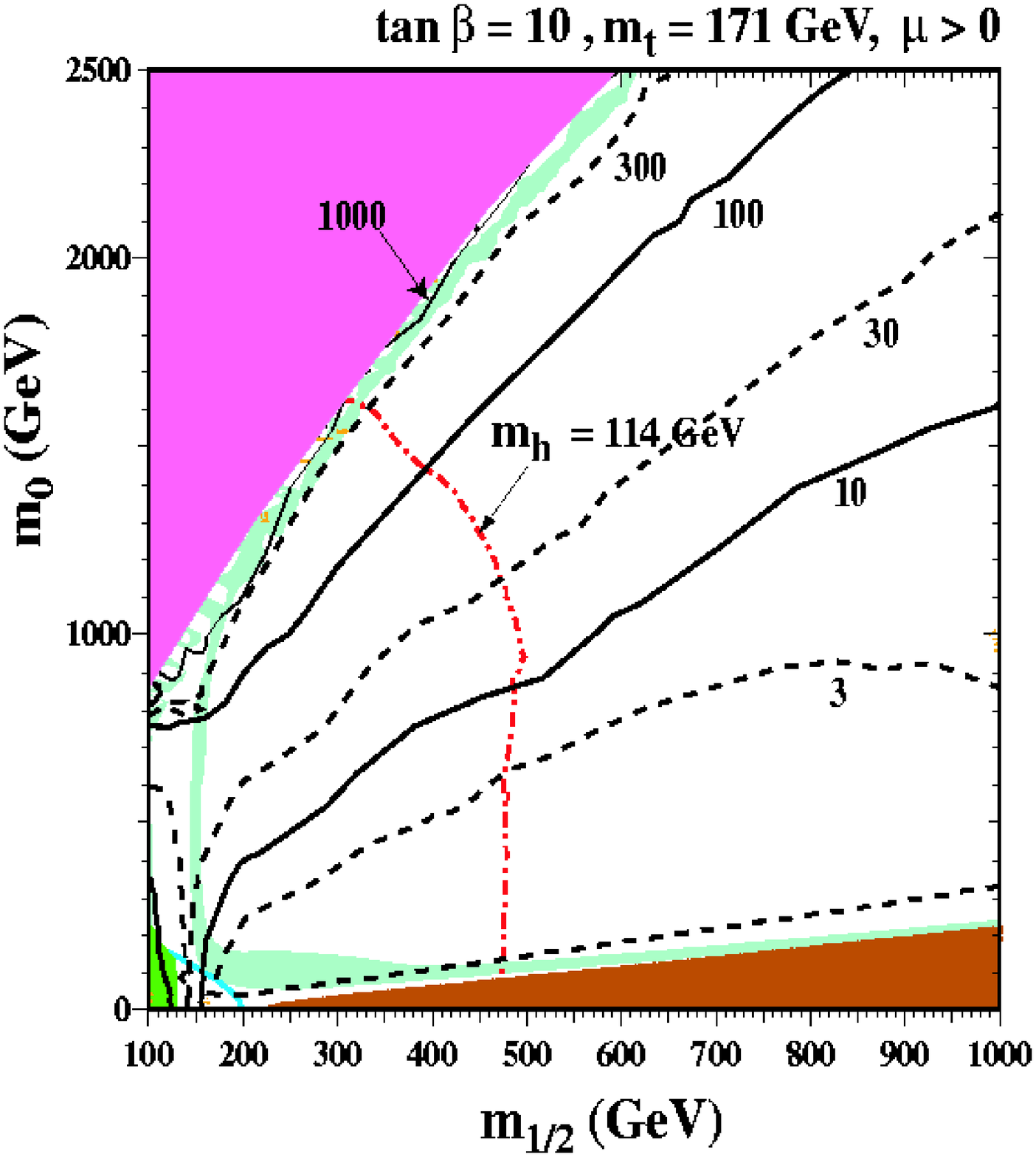,height=3.3in} \hfill
\end{minipage}
\caption{\label{fig:overall}
{\it Contours of the total sensitivity $\Delta^\Omega$ (\ref{twelve}) of 
the relic density in the
$(m_{1/2}, m_0)$ planes for (a) $\tan \beta = 10, \mu > 0, m_t =
175$~GeV, (b) $\tan \beta = 35, \mu < 0, m_t = 175$~GeV, (c)
$\tan \beta = 50, \mu > 0, m_t = 175$~GeV, and (d) $\tan \beta =
10, \mu > 0, m_t = 171$~GeV, all for $A_0 = 0$~\cite{EO}. The light 
(turquoise)
shaded areas are the cosmologically preferred regions with
\protect\mbox{$0.1\leq\ohsq\leq 0.3$}. In the dark (brick red) shaded
regions, the LSP is the charged ${\tilde \tau}_1$, so these regions are
excluded. In panel (d), the medium shaded (mauve) region is excluded by
the electroweak vacuum conditions. }}
\end{figure}

The above-mentioned filaments extending the preferred CMSSM parameter 
space are clearly
exceptional in some sense, so it is important to understand the sensitivity
of the relic density to input parameters, unknown higher-order effects, 
etc. One proposal is the relic-density fine-tuning measure~\cite{EO}
\beq
\Delta^\Omega \equiv \sqrt{\sum_i ~~\left({\partial\ln (\Omega_\chi h^2)\over
\partial
\ln a_i}\right)^2 }
\label{twelve}
\eeq
where the sum runs over the input parameters, which might include
(relatively) poorly-known Standard Model quantities such as $m_t$ and
$m_b$, as well as the CMSSM parameters $m_0, m_{1/2}$, etc. As seen in
Fig.~\ref{fig:overall}, the sensitivity $\Delta^\Omega$ (\ref{twelve}) is 
relatively small
in the `bulk' region at low $m_{1/2}$, $m_0$, and $\tan\beta$. However, it
is somewhat higher in the $\chi - \tilde\tau_1$ coannihilation `tail', and
at large $\tan\beta$ in general. The sensitivity measure $\Delta^\Omega$
(\ref{twelve}) is particularly high in the rapid-annihilation `funnel' and
in the `focus-point' region. This explains why published relic-density
calculations may differ in these regions~\cite{otherOmega}, whereas they 
agree well when
$\Delta^\Omega$ is small: differences may arise because of small
differences in the values and treatments of the inputs.

\begin{figure}
\vskip 0.5in
\vspace*{-0.75in}
\hspace*{-.20in}
\begin{minipage}{8in}
\epsfig{file=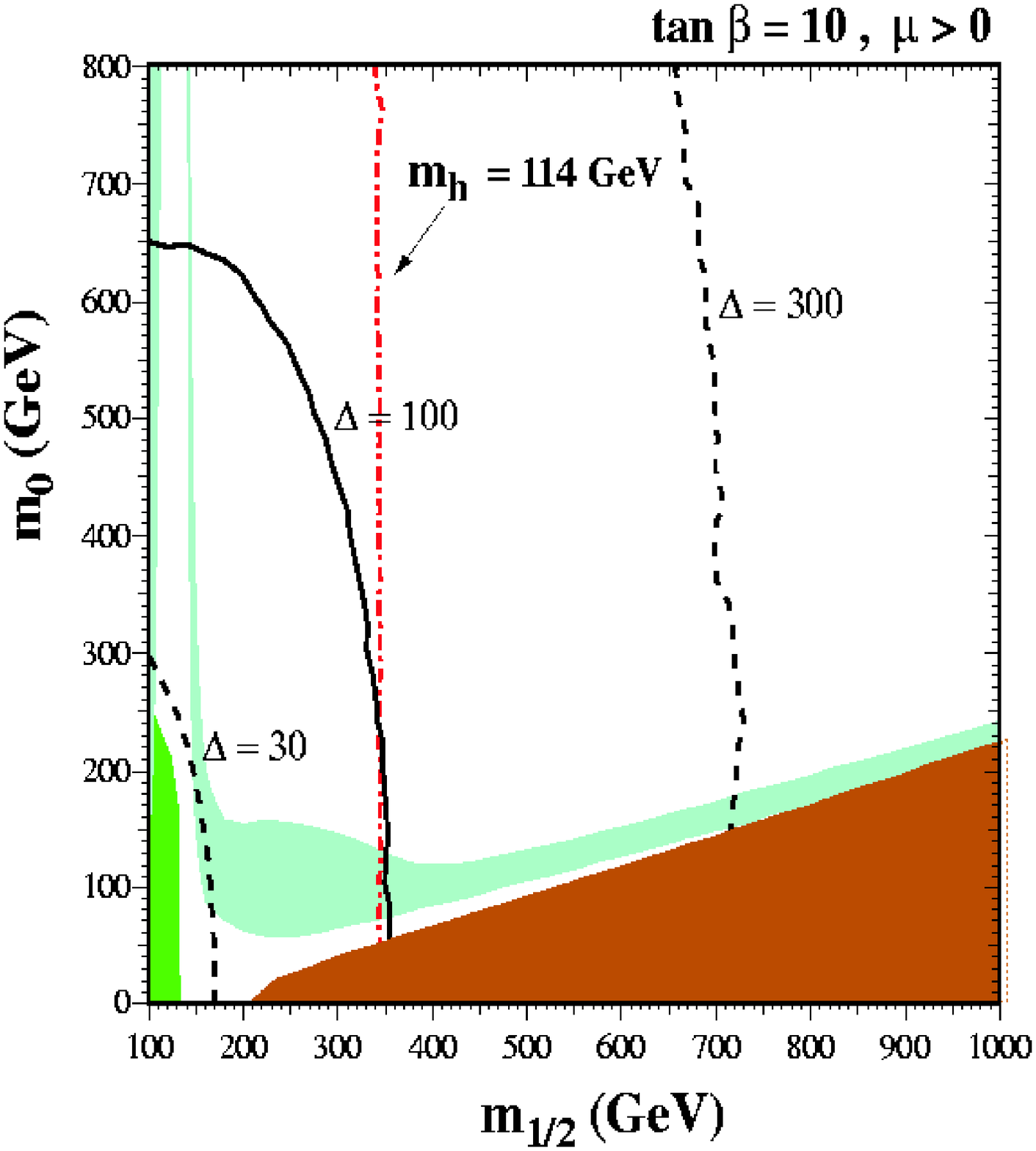,height=3.3in}
\hspace*{+0.10in}
\epsfig{file=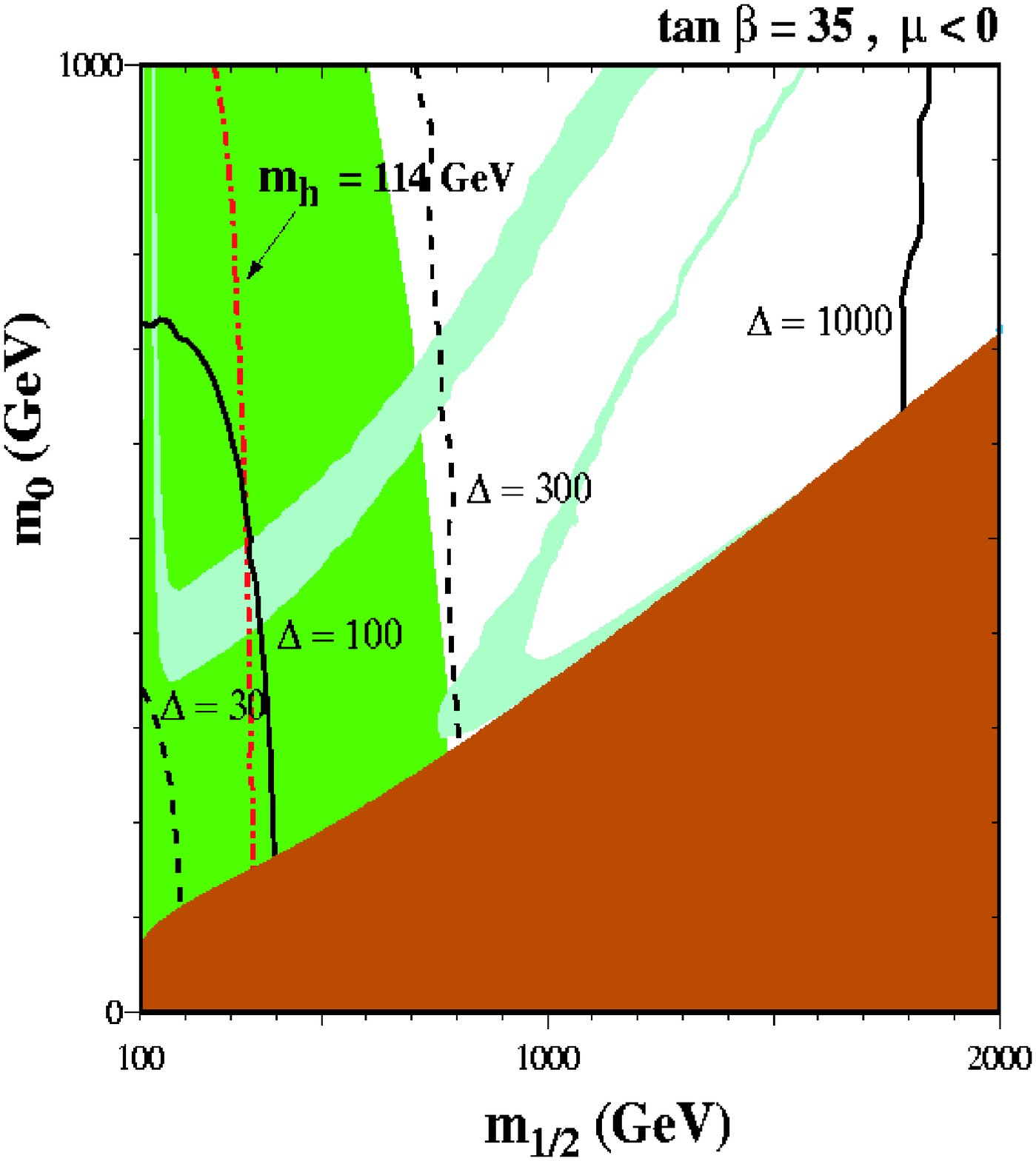,height=3.3in} \hfill
\end{minipage}
\hspace*{-.20in}
\begin{minipage}{8in}
\epsfig{file=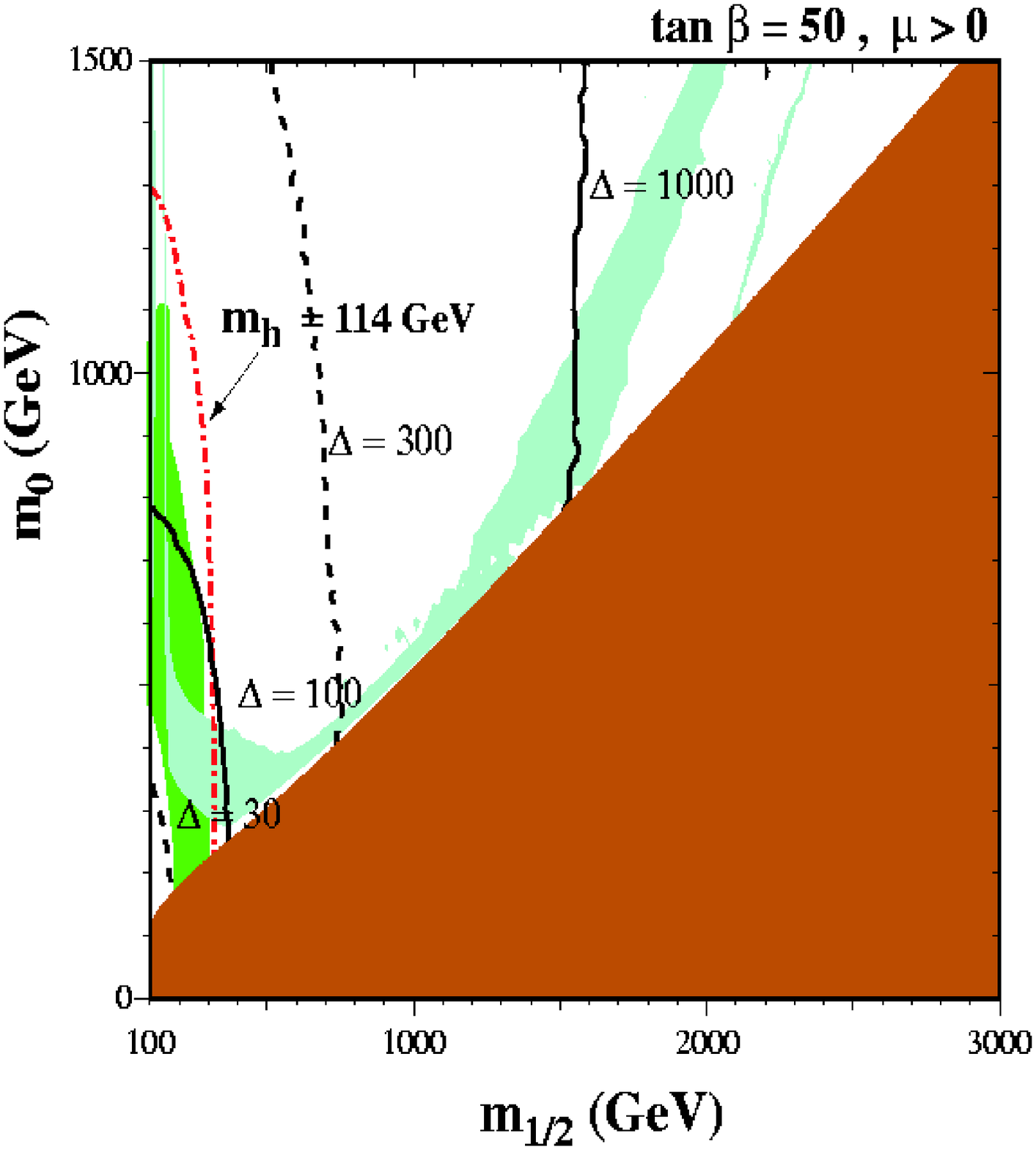,height=3.3in}
\hspace*{-0.10in}
\epsfig{file=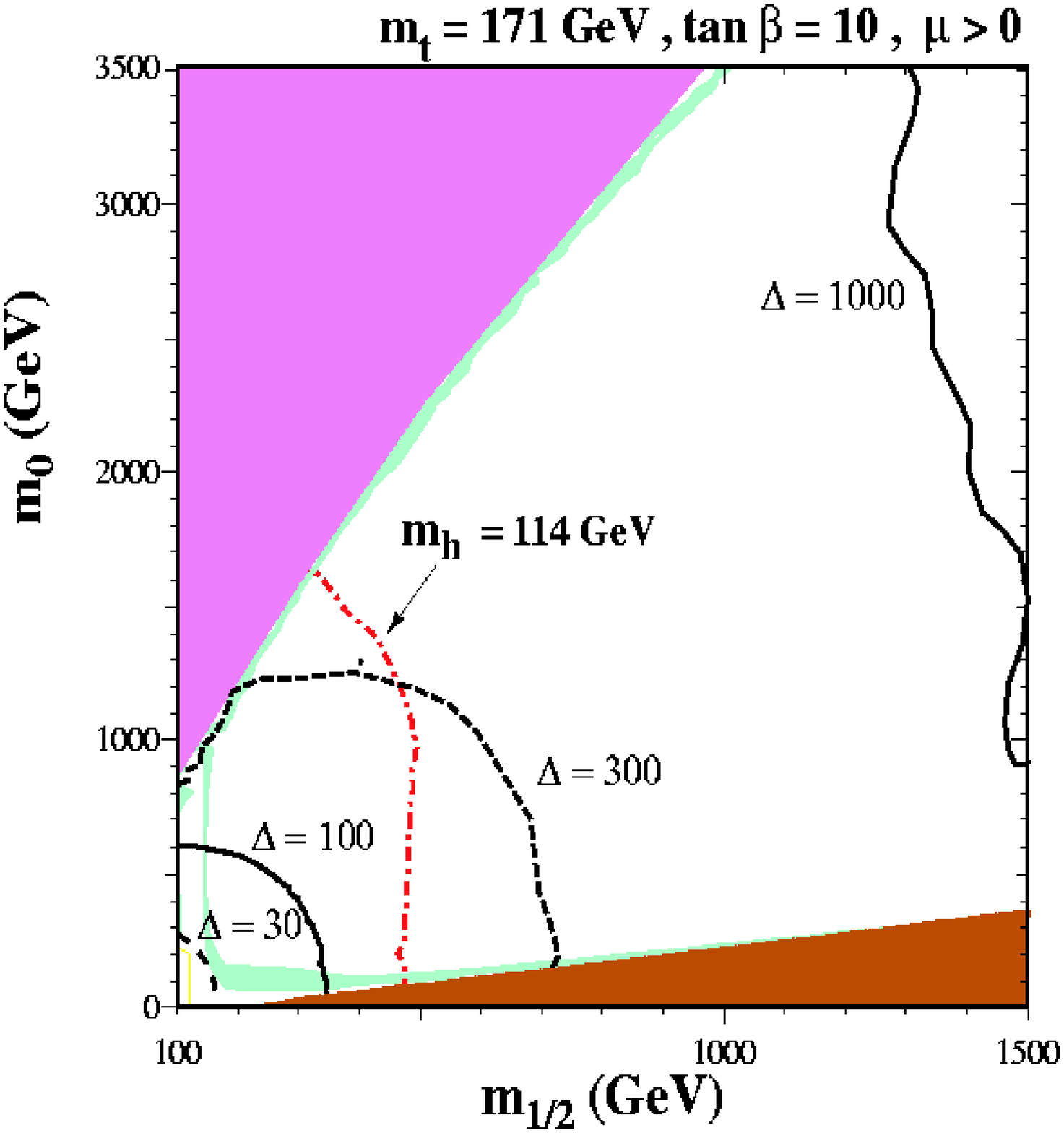,height=3.3in} \hfill
\end{minipage}
\caption{\label{fig:EWFT}
{\it Contours of the electroweak fine-tuning measure $\Delta$ 
(\ref{thirteen}) in the
$(m_{1/2}, m_0)$ planes for (a) $\tan \beta = 10, \mu > 0, m_t =
175$~GeV, (b) $\tan \beta = 35, \mu < 0, m_t = 175$~GeV, (c)
$\tan \beta = 50, \mu > 0, m_t = 175$~GeV, and (d) $\tan \beta =
10, \mu > 0, m_t = 171$~GeV, all for $A_0 = 0$~\cite{EOSnew}. The light 
(turquoise)
shaded areas are the cosmologically preferred regions with
\protect\mbox{$0.1\leq\ohsq\leq 0.3$}. In the dark (brick red) shaded
regions, the LSP is the charged ${\tilde \tau}_1$, so this region is
excluded. In panel (d), the medium shaded (mauve) region is excluded by
the electroweak vacuum conditions. }}
\end{figure}  

It is important to note that the relic-density fine-tuning measure
(\ref{twelve}) is distinct from the traditional measure of the fine-tuning
of the electroweak scale~\cite{EENZ}:
\beq
\Delta = \sqrt{\sum_i ~~\Delta_i^{\hspace{0.05in} 2}}\, , \quad \Delta_i \equiv
{\partial \ln
m_W\over \partial \ln a_i}
\label{thirteen}
\eeq
Sample contours of the electroweak fine-tuning measure are shown 
(\ref{thirteen}) are shown in Figs.~\ref{fig:EWFT}~\cite{EOS}.
This electroweak fine tuning is logically different from
the cosmological fine tuning, and values
of $\Delta$ are not necessarily related to values of
$\Delta^\Omega$, as is apparent when comparing the contours in 
Figs.~\ref{fig:overall} and \ref{fig:EWFT}.  Electroweak fine-tuning is 
sometimes used as a 
criterion
for restricting the CMSSM parameters. However, the interpretation of 
$\Delta$ (\ref{thirteen}) is unclear. How large a value of $\Delta$ is
tolerable? Different people may well have different pain thresholds.
Moreover, correlations between input parameters may reduce its value in
specific models, and the regions allowed by the different constraints can 
become very different when we relax some of the CMSSM assumptions,
e.g., the universality between the input Higgs masses and those of the 
squarks and sleptons, a subject beyond the scope of these Lectures.

\subsection{Benchmark Supersymmetric Scenarios}

\begin{figure}
\begin{centering}
\hspace{2cm}
\epsfig{figure=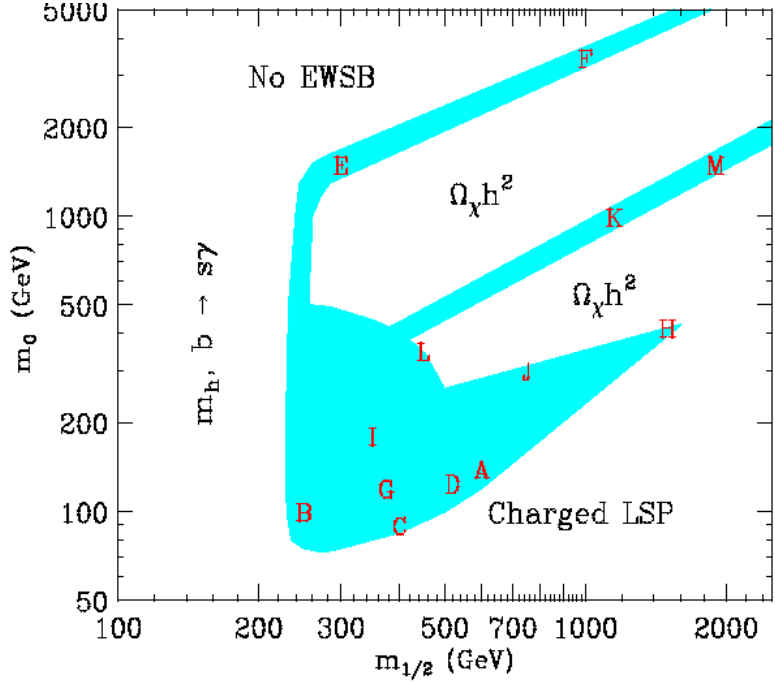,width=4in}
\end{centering}
\hglue3.5cm   
\caption[]{\it The locations of the benchmark points proposed 
in~\cite{Bench} in the region of the
$(m_{1/2}, m_0)$ plane where $\ohsq$ falls within the range preferred by 
cosmology (shaded blue). Note that the filaments of the allowed parameter 
space extending to large $m_{1/2}$ and/or $m_0$ are sampled.}
\label{fig:Bench}
\end{figure}

As seen in Fig.~\ref{fig:CMSSM}, all the experimental, cosmological and
theoretical constraints on the MSSM are mutually compatible. As an aid to
understanding better the physics capabilities of the LHC, various $e^+e^-$
linear collider designs and non-accelerator experiments, a set of
benchmark supersymmetric scenarios have been proposed~\cite{Bench}. 
Their distribution in the $(m_{1/2}, m_0)$ plane is sketched in 
Fig.~\ref{fig:Bench}. These benchmark scenarios
are compatible with all the accelerator constraints mentioned above,
including the LEP searches and $b \to s \gamma$, and yield relic densities
of LSPs in the range suggested by cosmology and astrophysics. The
benchmarks are not intended to sample `fairly' the allowed parameter
space, but rather to illustrate the range of possibilities currently
allowed.

In addition to a number of benchmark points falling in the `bulk' region
of parameter space at relatively low values of the supersymmetric particle
masses, as see in Fig.~\ref{fig:Bench}, we also proposed~\cite{Bench} some
points out along the `tails' of parameter space extending out to larger
masses. These clearly require some degree of fine-tuning to obtain the
required relic density and/or the correct $W^\pm$ mass, and some are also
disfavoured by the supersymmetric interpretation of the $g_\mu - 2$
anomaly, but all are logically consistent possibilities.

\subsection{Prospects for Discovering Supersymmetry}

\begin{figure}
\begin{centering}
\hspace{1.5cm}
\epsfig{figure=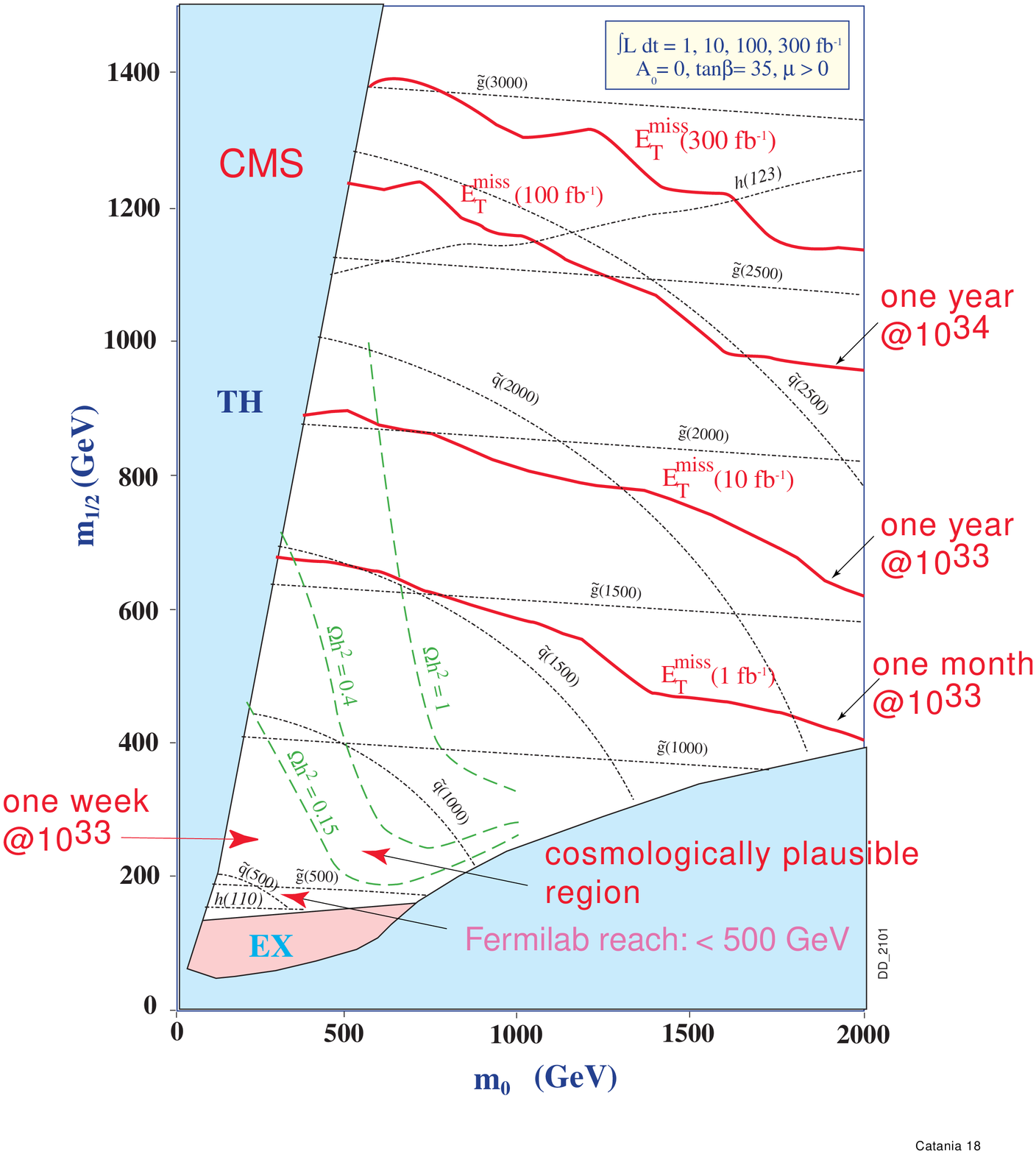,width=5in}
\end{centering}
\caption[]{\it The regions of the $(m_0, m_{1/2})$ plane that can be 
explored by the LHC with various integrated luminosities~\cite{CMS}, using 
the 
missing energy + jets signature~\cite{Paige}.}
\label{fig:Paige1}
\end{figure}

\begin{figure}
\begin{centering}
\hspace{2cm}
\epsfig{figure=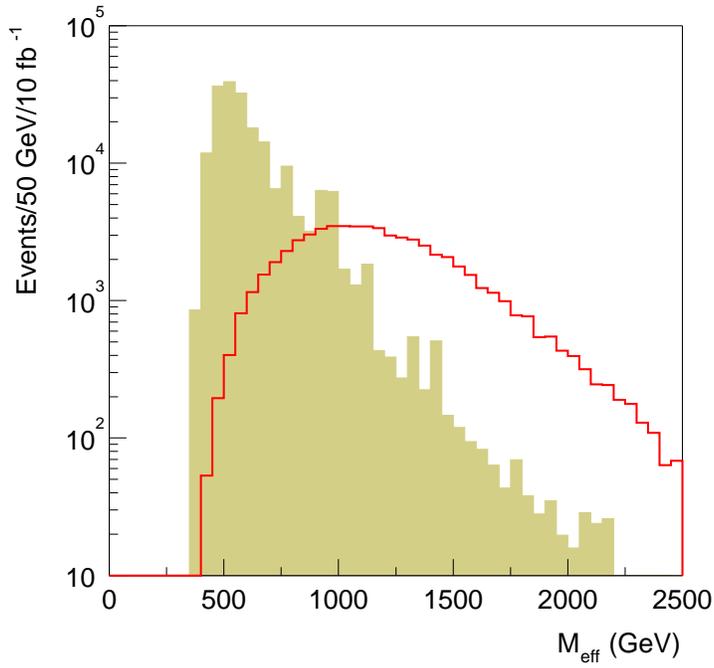,width=4in}
\end{centering}
\hglue3.5cm   
\caption[]{\it The distribution expected at the LHC in the variable 
$M_{\rm eff}$ that combines the jet energies with the missing 
energy~\cite{HP,Tovey,Paige}.}
\label{fig:Paige3}
\end{figure}

\begin{figure}
\begin{centering}
\hspace{2cm}
\epsfig{figure=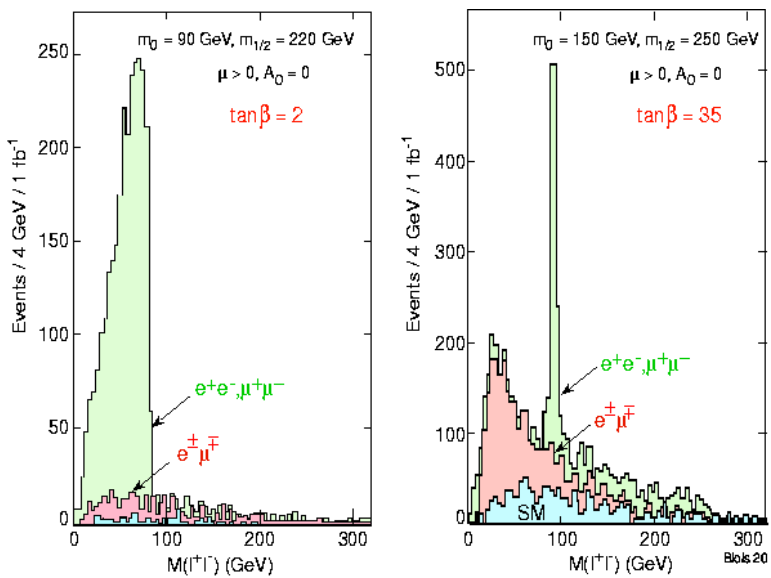,width=4in}
\end{centering}
\hglue3.5cm   
\caption[]{\it The dilepton mass distributions expected at the LHC due to 
sparticle decays in two different supersymmetric 
scenarios~\cite{HP,CMS,Paige}.}
\label{fig:Paige4}
\end{figure}

\begin{figure}
\begin{centering}
\hspace{2cm}
\epsfig{figure=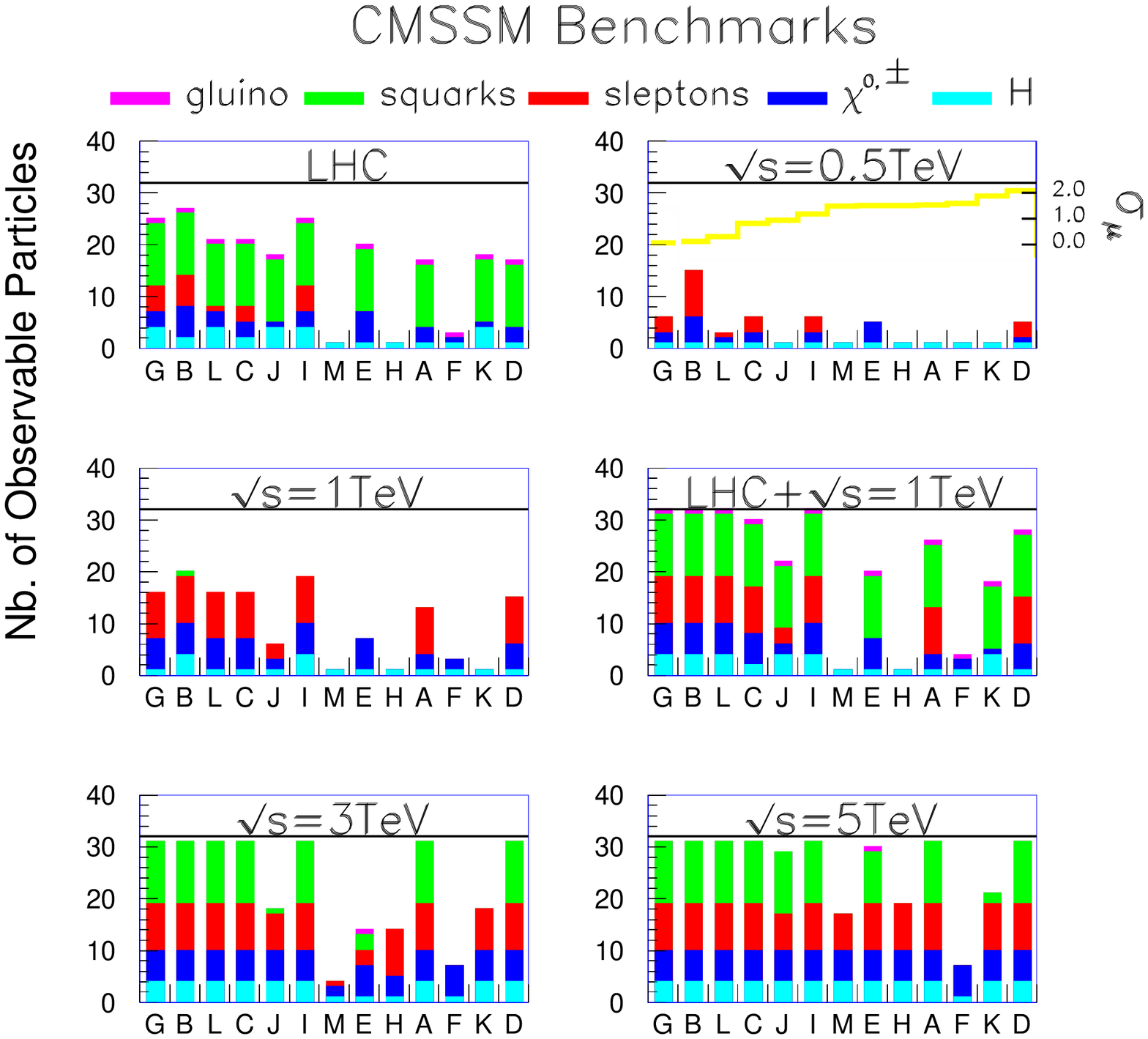,width=4in}
\end{centering}
\hglue3.5cm   
\caption[]{\it The numbers of different sparticles expected to be 
observable at the LHC and/or linear $e^+ e^-$ colliders with various 
energies, in each of the proposed benchmark 
scenarios~\cite{Bench}, 
ordered by their 
difference from the present central experimental value of $g_\mu - 2$.} 
\label{fig:Manhattan} 
\end{figure}

In the CMSSM discussed here, there are just a few prospects for
discovering supersymmetry at the FNAL {\it Tevatron
collider}~\cite{Bench}, but these could be increased in other
supersymmetric models~\cite{BenchKane}. Fig.~\ref{fig:Paige1} shows the
physics reach for observing pairs of supersymmetric particles at the {\it
LHC}. The signature for supersymmetry - multiple jets (and/or leptons)  
with a large amount of missing energy - is quite distinctive, as seen in
Fig.~\ref{fig:Paige3}~\cite{Tovey,Paige}. Therefore, the detection of the
supersymmetric partners of quarks and gluons at the LHC is expected to be
quite easy if they weigh less than about 2.5~TeV~\cite{CMS}. Moreover, in 
many
scenarios one should be able to observe their cascade decays into lighter
supersymmetric particles, as seen in Fig.~\ref{fig:Paige4}~\cite{Rurua}.  
As seen in Fig.~\ref{fig:Manhattan}, large fractions of the supersymmetric
spectrum should be seen in most of the benchmark scenarios, although there
are a couple where only the lightest supersymmetric Higgs boson would be
seen~\cite{Bench}, as seen in Fig.~\ref{fig:Manhattan}.

{\it Electron-positron colliders} provide very clean experimental
environments, with egalitarian production of all the new particles that
are kinematically accessible, including those that have only weak
interactions. Moreover, polarized beams provide a useful analysis tool,
and $e \gamma$, $\gamma \gamma$ and $e^- e^-$ colliders are readily
available at relatively low marginal costs.

The $e^+ e^- \to {\bar t} t$ threshold is known to be at $E_{\rm CM} \sim
350$~GeV. Moreover, if the Higgs boson indeed weighs less than 200~GeV, as
suggested by the precision electroweak data, its production and study
would also be easy at an $e^+ e^-$ collider with $E_{\rm CM} \sim
500$~GeV. With a luminosity of $10^{34}$~cm$^{-2}$s$^{-1}$ or more, many
decay modes of the Higgs boson could be measured very accurately, and one
might be able to find a hint whether its properties were modified by
supersymmetry~\cite{TESLA,EHOW2}.

However, the direct production of supersymmetric particles at such a
collider cannot be guaranteed~\cite{EGO}.
We do not yet know what the supersymmetric threshold energy may be (or
even if there is one!). We may well not know before the operation of the
LHC, although $g_\mu - 2$ might provide an indication~\cite{susygmu}, if
the uncertainties in the Standard Model calculation can be reduced.

\begin{figure}
\hspace{2.5cm}
\epsfig{figure=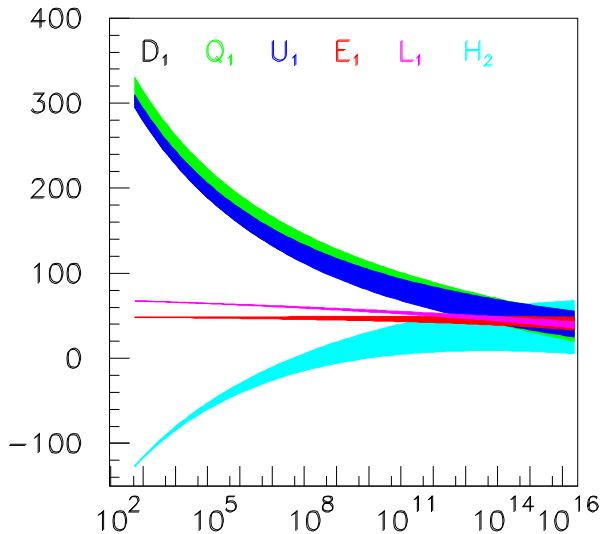,width=7in}
\hglue3.5cm
\vspace{-10cm}   
\caption[]{\it Measurements of sparticle masses at the LHC and a 
linear $e^+ e^-$ linear collider will enable one to check their 
universality at some input GUT scale, and check possible models of 
supersymmetry breaking~\cite{BPZ}. Both axes are labelled in GeV units.} 
\label{fig:BPZ} 
\end{figure}

If an $e^+ e^-$ collider is above the supersymmetric threshold, it will be
able to measure very accurately the sparticle masses. By comparing their
masses with those of different sparticles produced at the LHC, one would
be able to make interesting tests of string and GUT models of
supersymmetry breaking, as seen in Fig.~\ref{fig:BPZ}~\cite{BPZ}. However,
independently from the particular benchmark scenarios proposed, a linear
$e^+ e^-$ collider with $E_{\rm CM} < 1$~TeV would not cover all the
supersymmetric parameter space allowed by cosmology~\cite{EGO,Bench}.

Nevertheless, there are compelling physics arguments for such a linear
$e^+ e^-$ collider, which would be very complementary to the LHC in terms
of its exploratory power and precision~\cite{TESLA}. It is to be hoped
that the world community will converge on a single project with the widest
possible energy range.

CERN and collaborating institutes are studying the possible following step
in linear $e^+ e^-$ colliders, a multi-TeV machine called
CLIC~\cite{CLIC,CLICphys}. This would use a double-beam technique to 
attain
accelerating gradients as high as 150~MV/m, and the viability of
accelerating structures capable of achieving this field has been
demonstrated in the CLIC test facility~\cite{CTF3}. Parameter sets have
been calculated for CLIC designs with $E_{\rm CM} = 3, 5$~TeV and
luminosities of $10^{35}$~cm$^{-2}$s$^{-1}$ or more~\cite{CLIC}.

\begin{figure}
\begin{centering}
\hspace{2.5cm}
\epsfig{figure=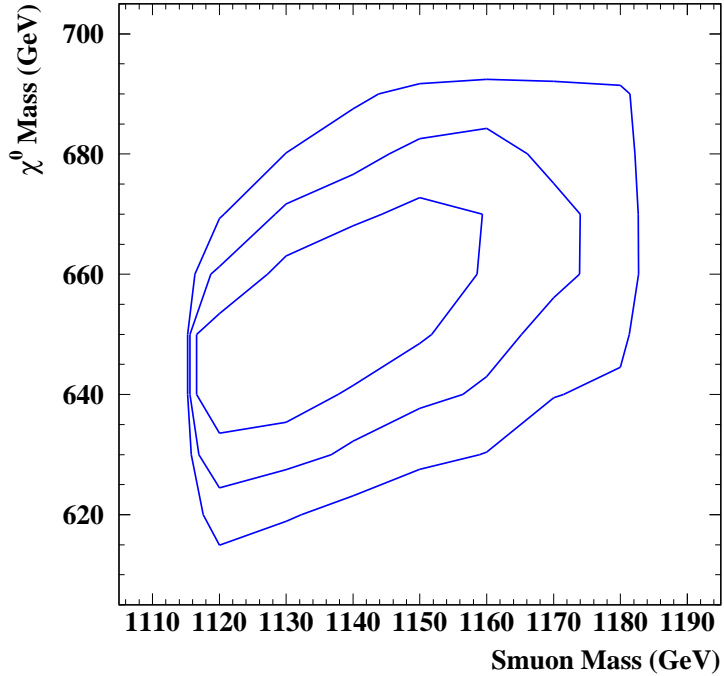,width=10cm}
\end{centering}
\hglue3.5cm   
\caption[]{\it Like lower-energy $e^+ e^-$ colliders, CLIC enables 
very accurate measurements of sparticle masses to be made, in this case the
supersymmetric partner of the muon and the lightest
neutralino $\chi^0$~\cite{Battaglia}.}
\label{fig:Battaglia}
\end{figure}

In many of the proposed benchmark supersymmetric scenarios, CLIC would be
able to complete the supersymmetric spectrum and/or measure in much more
detail heavy sparticles found previously at the LHC, as seen in
Fig.~\ref{fig:Manhattan}~\cite{Bench}. CLIC produces more beamstrahlung
than lower-energy linear $e^+ e^-$ colliders, but the supersymmetric
missing-energy signature would still be easy to distinguish, and accurate
measurements of masses and decay modes could still be made, as seen in
Fig.~\ref{fig:Battaglia}~\cite{Battaglia}.

\subsection{Searches for Dark Matter Particles}

\begin{figure}
\vskip 0.75in
\vspace*{-0.75in}
\hspace*{-.40in}
\begin{minipage}{8in}
\epsfig{file=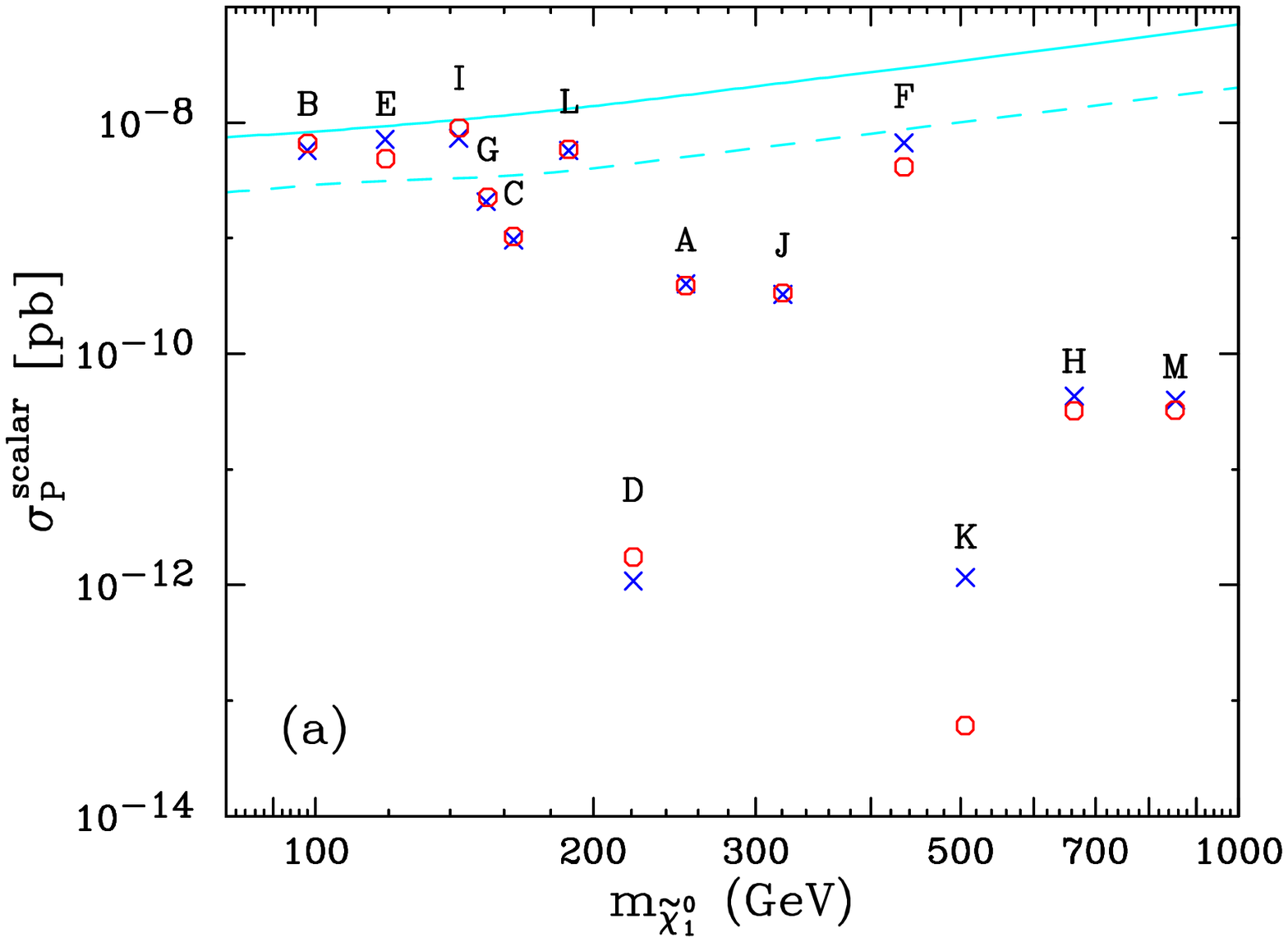,height=2.5in}
\hspace*{-0.17in}
\epsfig{file=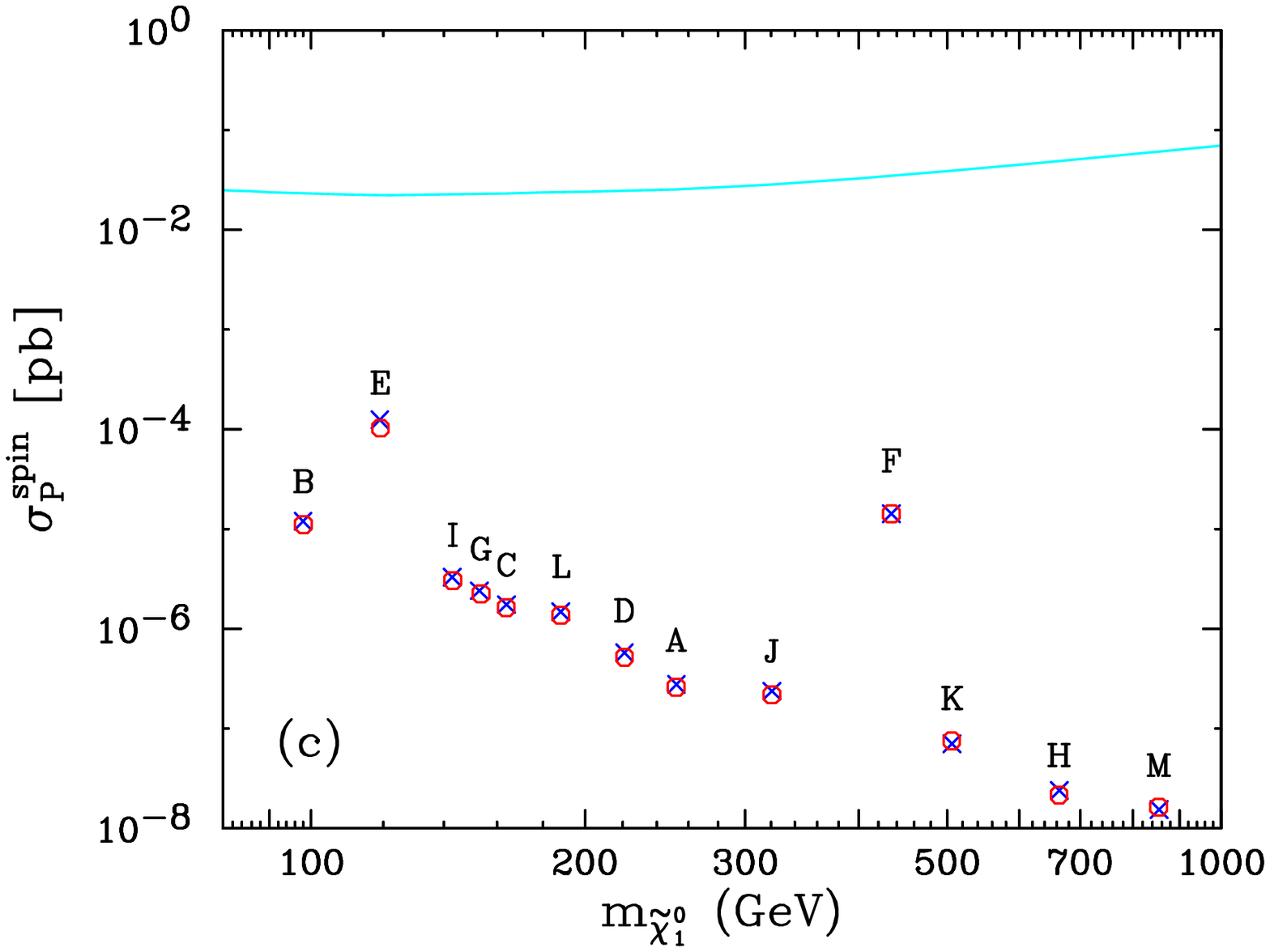,height=2.5in} \hfill
\end{minipage}
\caption[]{\it Left panel: elastic spin-independent scattering  
of supersymmetric relics on protons calculated in 
benchmark scenarios~\cite{EFFMO}, compared with the 
projected sensitivities for CDMS
II~\cite{Schnee:1998gf} and CRESST~\cite{Bravin:1999fc} (solid) and
GENIUS~\cite{GENIUS} (dashed).
The predictions of the {\tt SSARD} code (blue
crosses) and {\tt Neutdriver}\cite{neut} (red circles) for
neutralino-nucleon scattering are compared.
The labels A, B, ...,L correspond to the benchmark points as shown in 
Fig.~\protect\ref{fig:Bench}. Right panel: prospects for detecting 
elastic spin-independent scattering in the benchmark scenarios, which are 
less bright.}
\label{fig:DM}
\end{figure}

\begin{figure}
\vskip 0.75in
\vspace*{-0.75in}
\hspace*{-.40in}
\begin{minipage}{8in}
\epsfig{file=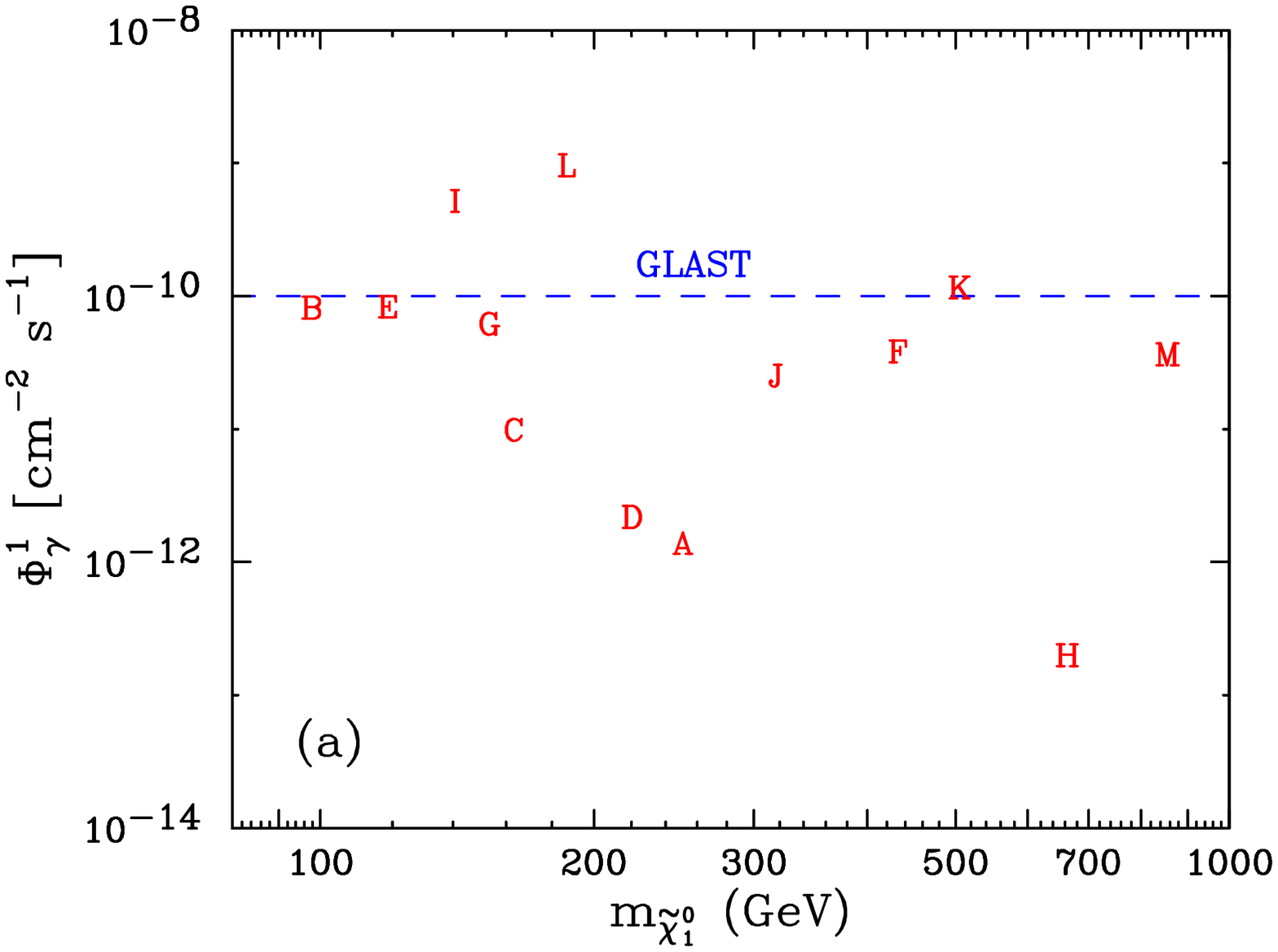,height=2.5in}
\hspace*{-0.17in}
\epsfig{file=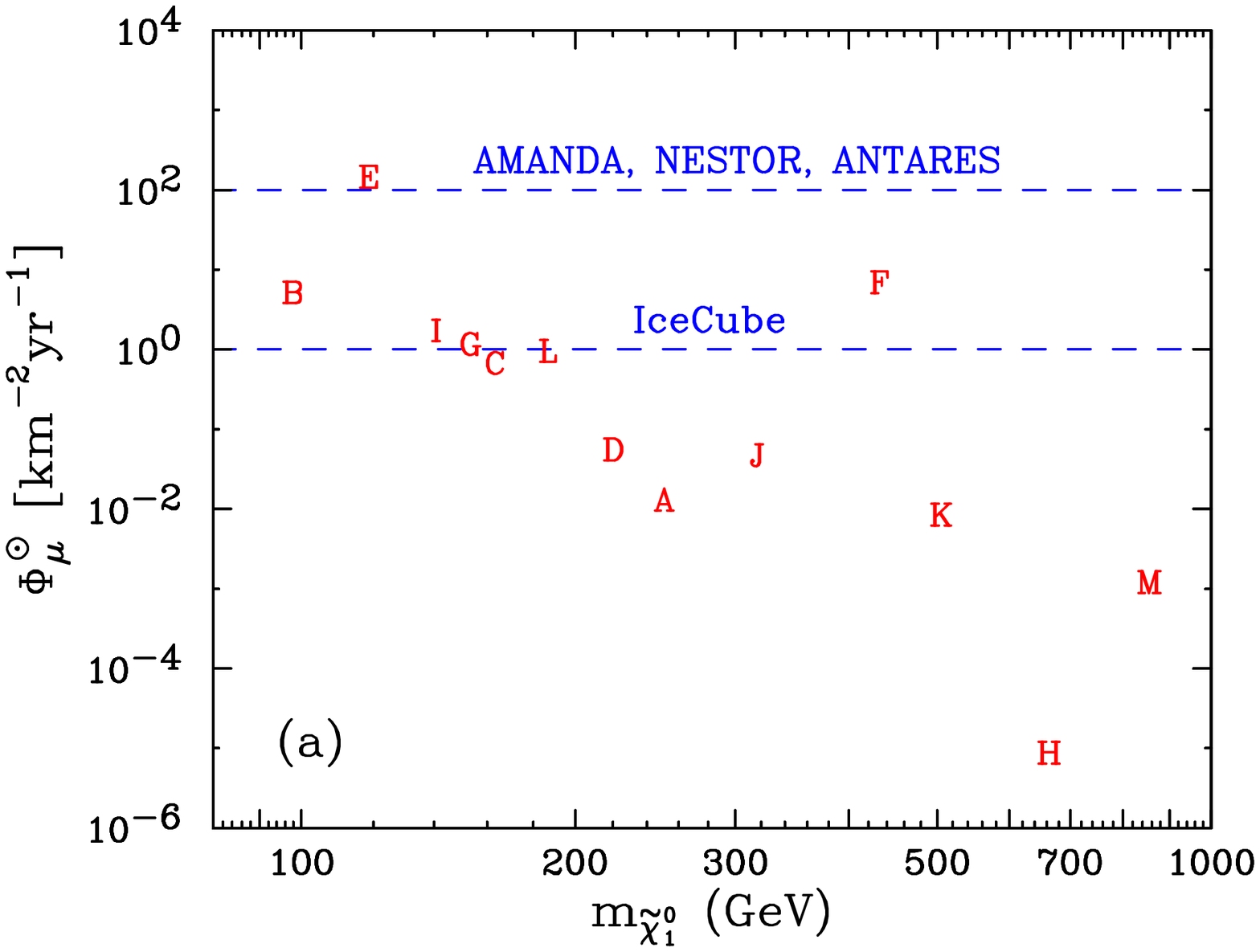,height=2.5in} \hfill
\end{minipage}
\caption[]{\it Left panel: prospects for detecting photons with energies 
above 
1~GeV 
from annihilations in the centre of the galaxy, assuming a moderate 
enhancement there of the overall halo density, and right panel: prospects 
for detecting 
muons from energetic solar neutrinos produced by relic annihilations in 
the Sun, as calculated~\cite{EFFMO} in the benchmark scenarios using {\tt 
Neutdriver}\cite{neut}.}
\label{fig:indirectDM}
\end{figure}

In the above discussion, we have paid particular attention to the region
of parameter space where the lightest supersymmetric particle could
constitute the cold dark matter in the Universe~\cite{EHNOS}. How easy
would this be to detect? Fig.~\ref{fig:DM} shows rates for the elastic
spin-independent scattering of supersymmetric relics~\cite{EFFMO},
including the projected sensitivities for CDMS II~\cite{Schnee:1998gf} and
CRESST~\cite{Bravin:1999fc} (solid) and GENIUS~\cite{GENIUS} (dashed).
Also shown are the cross sections calculated in the proposed benchmark
scenarios discussed in the previous section, which are considerably below
the DAMA \cite{DAMA} range ($10^{-5} - 10^{-6}$~pb), but may be within
reach of future projects. The prospects for detecting elastic
spin-independent scattering are less bright, as also shown in
Fig.~\ref{fig:DM}. Indirect searches for supersymmetric dark matter via
the products of annihilations in the galactic halo or inside the Sun also
have prospects in some of the benchmark scenarios~\cite{EFFMO}, as seen in
Fig.~\ref{fig:indirectDM}.

\section{Lepton Flavour Violation}

\subsection{Why not?}

There is no good reason why either the total lepton number $L$ or the
individual lepton flavours $L_{e, \mu, \tau}$ should be
conserved~\cite{Marciano}. We have learnt that the only conserved quantum
numbers are those associated with exact gauge symmetries, just as the
conservation of electromagnetic charge is associated with $U(1)$ gauge
invariance. On the other hand, there is no exact gauge symmetry associated
with any of the lepton numbers.

Moreover, neutrinos have been seen to oscillate between their different
flavours~\cite{SK,SNO}, showing that the separate lepton flavours $L_{e,
\mu, \tau}$ are indeed not conserved, though the conservation of total
lepton number $L$ is still an open question. The observation of such
oscillations strongly suggests that the neutrinos have different masses.
Again, massless particles are generally associated with exact gauge
symmetries, e.g., the photon with the $U(1)$ symmetry of the Standard
Model, and the gluons with its $SU(3)$ symmetry. In the absence of any
leptonic gauge symmetry, non-zero lepton masses are to be expected, in
general.

The conservation of lepton number is an accidental symmetry of the 
renormalizable terms in the Standard Model lagrangian. However, one could 
easily add to the Standard Model non-renormalizable terms that would 
generate neutrino masses, even without introducing a `right-handed' 
neutrino field. For example, a non-renormalizable term of the 
form~\cite{BEG}
\begin{equation}
{1 \over M} \nu H \cdot \nu H,
\label{nonren}
\end{equation}
where $M$ is some large mass beyond the scale of the Standard Model, would 
generate a neutrino mass term:
\begin{equation}
m_\nu \nu \cdot \nu:
\; m_\nu \; = \; {\langle 0 \vert H \vert 0 \rangle^2 \over M}.
\label{smallm}
\end{equation}
Of course, a non-renormalizable interaction such as (\ref{nonren}) seems
unlikely to be fundamental, and one should like to understand the origin 
of the large mass scale $M$.

The minimal renormalizable model of neutrino masses requires the
introduction of weak-singlet `right-handed' neutrinos $N$. These will in
general couple to the conventional weak-doublet left-handed neutrinos via
Yukawa couplings $Y_\nu$ that yield Dirac masses $m_D \sim m_W$. In
addition, these `right-handed' neutrinos $N$ can couple to themselves via
Majorana masses $M$ that may be $\gg m_W$, since they do not require 
electroweak summetry breaking. Combining the two types of
mass term, one obtains the seesaw mass matrix~\cite{seesaw}:
\begin{eqnarray}
\left( \nu_L, N\right) \left(
\begin{array}{cc}
0 & M_{D}\\ 
M_{D}^{T} & M
\end{array}
\right)
\left(
\begin{array}{c}
\nu_L \\
N
\end{array}
\right),
\label{seesaw}
\end{eqnarray}
where each of the entries should be understood as a matrix in generation 
space.

In order to provide the two measured differences in neutrino
masses-squared, there must be at least two non-zero masses, and hence at
least two heavy singlet neutrinos $N_i$~\cite{Frampton,Morozumi}.  
Presumably, all three light neutrino masses are non-zero, in which case
there must be at least three $N_i$. This is indeed what happens in simple
GUT models such as SO(10), but some models~\cite{fSU5} have more singlet
neutrinos~\cite{EGLLN}. In this Lecture, for simplicity we consider just
three $N_i$.

As we discuss in the next Section, this seesaw model can accommodate the
neutrino mixing seen experimentally, and naturally explains the small
differences in the masses-squared of the light neutrinos. By itself, it
would lead to unobservably small transitions between the different
charged-lepton flavours. However, supersymmetry may enhance greatly the
rates for processes violating the different charged-lepton flavours,
rendering them potentially observable, as we discuss in subsequent
Sections.

\subsection{Neutrino Masses and Mixing in the Seesaw Model}

The effective mass matrix for light neutrinos in the seesaw model may be 
written as:
\begin{equation}
{\cal M}_\nu \; = \; Y_\nu^T {1 \over M} Y_\nu v^2 \left[ \sin^2 \beta 
\right]
\label{seesawmass}
\end{equation}
where we have used the relation $m_D = Y_\nu v \left[ \sin \beta \right]$
with $v \equiv \langle 0 \vert H \vert 0 \rangle$, and the factors of 
$\sin \beta$ appear in the supersymmetric version of the seesaw model. It 
is convenient to work in the field basis where the charged-lepton masses 
$m_{\ell^\pm}$ and the heavy singlet-neutrino mases $M$ are real and 
diagonal. The seesaw neutrino mass matrix ${\cal M}_\nu$ 
(\ref{seesawmass}) may then be diagonalized by a unitary transformation 
$U$:
\begin{equation}
U^T {\cal M}_\nu U \; = \; {\cal M}_\nu^d.
\label{diag}
\end{equation}
This diagonalization is reminiscent of that required for the quark mass 
matrices in the Standard Model. In that case, it is well known that one 
can redefine the phases of the quark fields~\cite{EGN} so that the mixing 
matrix 
$U_{CKM}$ has just one CP-violating phase~\cite{KM}. However, in the 
neutrino case, 
there are fewer independent field phases, and one is left with three 
physical CP-violating parameters:
\begin{equation}
U \; = \; {\tilde P}_2 V P_0: \; P_0 \equiv {\rm Diag} \left( 
e^{i\phi_1},  
e^{i\phi_2}, 1 \right).
\label{MNSP}
\end{equation}
Here ${\tilde P}_2 = {\rm Diag} \left( e^{i\alpha_1}, e^{i\alpha_2}, 
e^{i\alpha_3} \right)$ contains three phases that can be removed by 
phase rotations and are unobservable in 
light-neutrino physics, $V$ is the light-neutrino mixing matrix 
first considered by Maki, Nakagawa and Sakata (MNS)~\cite{MNS}, and $P_0$ 
contains 2 observable CP-violating 
phases $\phi_{1,2}$. The MNS matrix describes neutrino oscillations
\begin{eqnarray}
V \; = \; \left(
\begin{array}{ccc}
c_{12} & s_{12} & 0 \\
- s_{12} & c_{12} & 0 \\
0 & 0 & 1
\end{array}
\right)
\left(
\begin{array}{ccc}
1 & 0 & 0 \\
0 & c_{23} & s_{23} \\
0 & - s_{23} & c_{23}
\end{array}
\right)
\left(
\begin{array}{ccc}
c_{13} & 0 & s_{13} \\
0 & 1 & 0 \\
- s_{13} e^{- i \delta} & 0 & c_{13} e^{- i \delta}
\end{array}
\right).
\label{MNSmatrix}
\end{eqnarray}
The Majorana phases $\phi_{1,2}$ are in principle observable in 
neutrinoless double-$\beta$ decay, whose matrix element is proportional to
\begin{equation}
\langle m_\nu \rangle_{ee} \; \equiv \; \Sigma_i U_{ei}^* m_{\nu_i} 
U_{ie}^\dagger.
\label{doublebeta}
\end{equation}
Later we discuss how other observable quantities might be sensitive 
indirectly to the Majorana phases.

\begin{figure}
\hspace{2.5cm}
\epsfig{figure=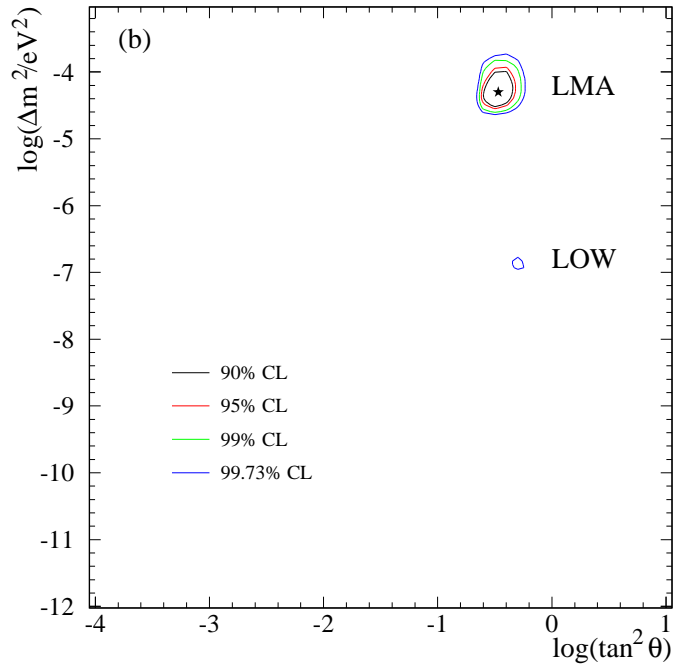,width=10cm}
\hglue3.5cm   
\caption[]{\it A global fit to solar neutrino data, following the SNO 
measurements of
the total neutral-current reaction rate, the energy spectrum and the
day-night asymmetry, favours large mixing and $\Delta m^2 \sim 6 \times
10^{-5}$~eV$^2$~\cite{SNO}.}
\label{fig:SNO}
\end{figure}

\begin{figure}
\hspace{2cm}
\epsfig{figure=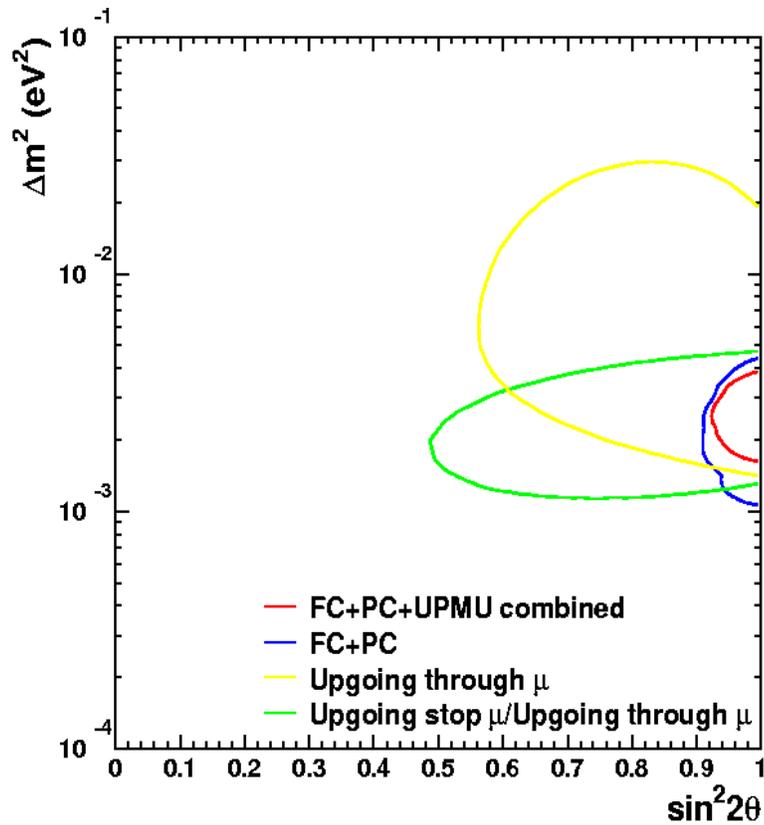,width=10cm}
\hglue3.5cm   
\caption[]{\it A fit to the Super-Kamiokande data on atmospheric
neutrinos~\cite{SK} indicates near-maximal $\nu_\mu - \nu_\tau$ mixing
with $\Delta m^2 \sim 2.5 \times 10^{-3}$~eV$^2$.}
\label{fig:SK}
\end{figure}

The first matrix factor in (\ref{MNSmatrix}) is measurable in solar 
neutrino
experiments. As seen in Fig.~\ref{fig:SNO}, the recent data from SNO 
\cite{SNO} and
Super-Kamiokande~\cite{SKsolar} prefer quite strongly the
large-mixing-angle (LMA) solution to the solar neutrino problem with
$\Delta m_{12}^2 \sim 6 \times 10^{-5}$~eV$^2$, though the LOW solution
with lower $\delta m^2$ cannot yet be ruled out. The data favour large but
non-maximal mixing: $\theta_{12} \sim 30^o$. The second matrix factor in
(\ref{MNSmatrix}) is measurable in atmospheric neutrino experiments. As 
seen 
in Fig.~\ref{fig:SK}, the data
from Super-Kamiokande in particular~\cite{SK} favour maximal mixing of
atmospheric neutrinos:  $\theta_{23} \sim 45^o$ and $\Delta m_{23}^2 \sim
2.5 \times 10^{-3}$~eV$^2$. The third matrix factor in (\ref{MNSmatrix}) 
is basically unknown, with experiments such as Chooz~\cite{Chooz} and
Super-Kamiokande only establishing upper limits on $\theta_{13}$, and {\it
a fortiori} no information on the CP-violating phase $\delta$.

\begin{figure}
\hspace{2cm}
\epsfig{figure=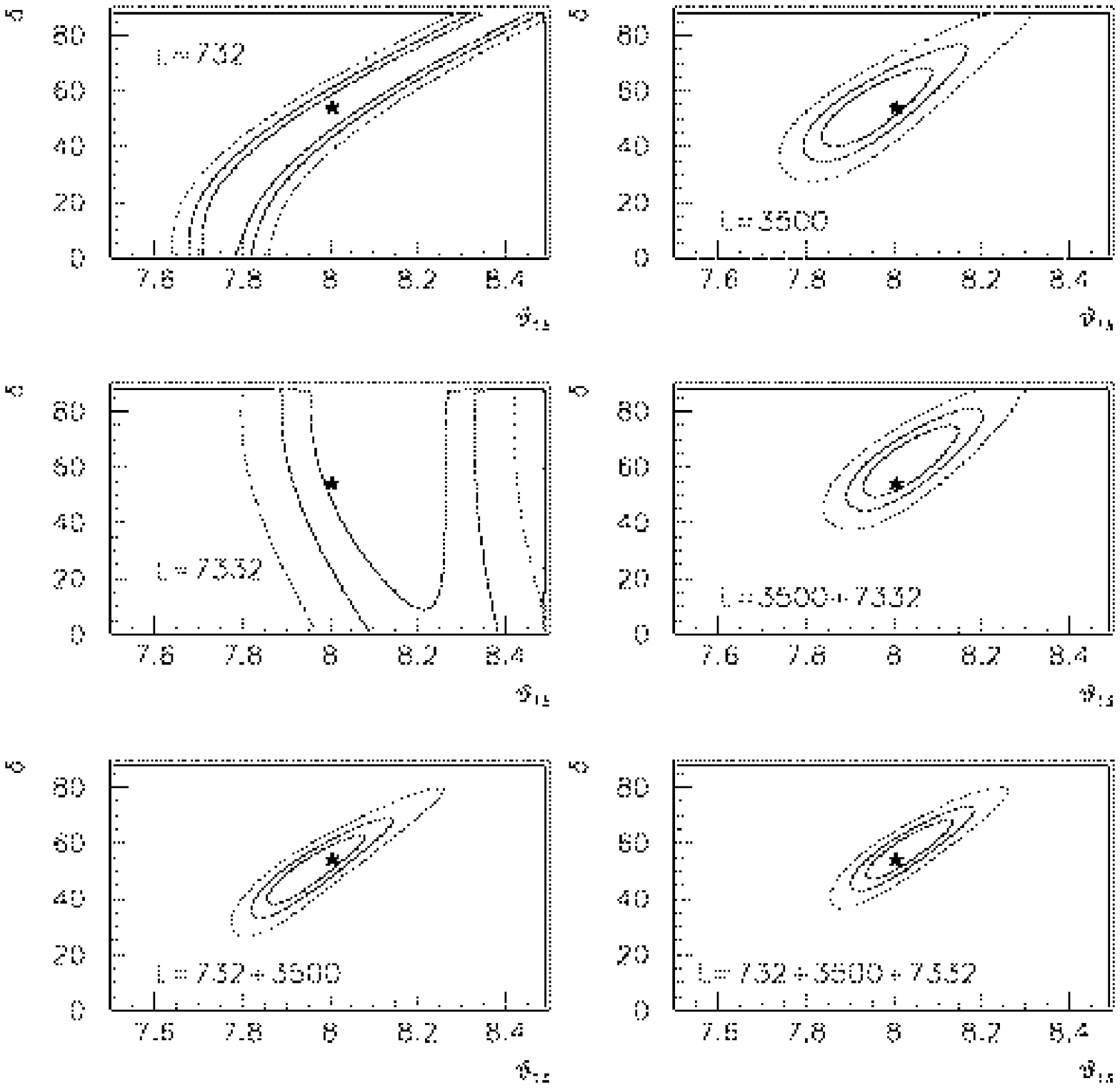,width=10cm}
\hglue3.5cm   
\caption[]{\it Correlations in a simultaneous fit of $\theta_{13}$ and
$\delta$,
using a neutrino energy threshold of about 10 GeV.
Using a single baseline correlations are very strong, but can be 
largely reduced by combining information from different baselines and 
detector techniques~\cite{golden}, enabling the CP-violating phase 
$\delta$ to be extracted.}
\label{fig:cpnu}
\end{figure}

The phase $\delta$ could in principle be measured by comparing the 
oscillation 
probabilities for neutrinos and antineutrinos and computing the 
CP-violating asymmetry~\cite{DGH}:
\begin{eqnarray}
P \left( \nu_e \to \nu_\mu \right) - P \left( {\bar \nu}_e \to 
{\bar \nu}_\mu \right) \; & & = \; 
16 s_{12} c_{12} s_{13} c^2_{13} s_{23} c_{23} \sin \delta \\ \nonumber
& & \sin \left( {\Delta m_{12}^2 \over 4 E} L \right)
\sin \left( {\Delta m_{13}^2 \over 4 E} L \right)
\sin \left( {\Delta m_{23}^2 \over 4 E} L \right),
\label{CPV}
\end{eqnarray}
as seen in Fig.~\ref{fig:cpnu}~\cite{golden,Frejus}.
This is possible only if $\Delta m_{12}^2$ and $s_{12}$ are large enough -
as now suggested by the success of the LMA solution to the solar neutrino 
problem, and if $s_{13}$ is large enough - which remains an open question.

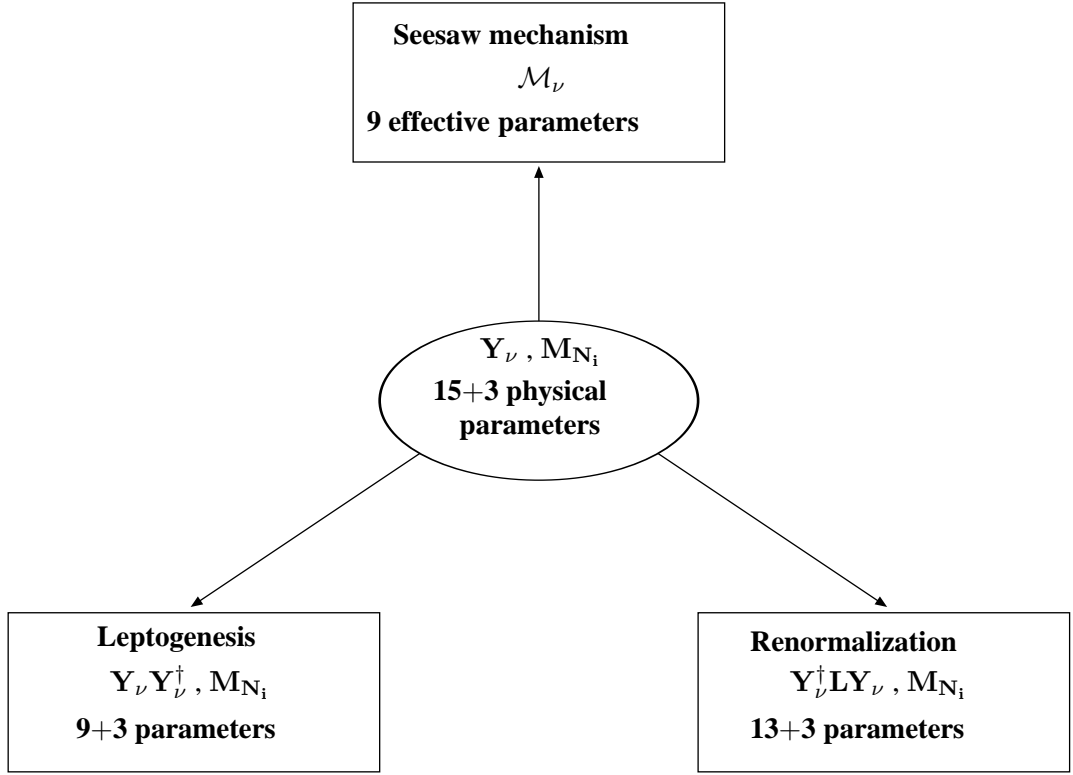
\begin{figure}[t]
\begin{center}
\begin{picture}(400,300)(-200,-150)
\Oval(0,0)(30,60)(0)
\Text(-25,13)[lb]{ ${\bf Y_\nu}$  ,  ${\bf M_{N_i}}$}
\Text(-40,-2)[lb]{{\bf 15$+$3 physical}}
\Text(-30,-15)[lb]{{\bf parameters}}
\EBox(-70,90)(70,150)
\Text(-55,135)[lb]{{\bf Seesaw mechanism}}  
\Text(-8,117)[lb]{${\bf {\cal M}_\nu}$}
\Text(-65,100)[lb]{{\bf 9 effective parameters}}
\EBox(-200,-140)(-60,-80)
\Text(-167,-95)[lb]{{\bf Leptogenesis}}
\Text(-165,-113)[lb]{ ${\bf Y_\nu Y_\nu^\dagger}$ , ${\bf M_{N_i}}$}
\Text(-175,-130)[lb]{{\bf 9$+$3 parameters}}
\EBox(60,-140)(200,-80)
\Text(80,-95)[lb]{{\bf Renormalization}}
\Text(95,-113)[lb]{${\bf Y_\nu^\dagger L Y_\nu}$ , ${\bf M_{N_i}}$}
\Text(80,-130)[lb]{{\bf 13$+$3 parameters}}
\LongArrow(0,30)(0,87)
\LongArrow(-45,-20)(-130,-77)
\LongArrow(45,-20)(130,-77)
\end{picture}
\end{center}
\caption{\it
Roadmap for the physical observables derived from $Y_\nu$ and 
$N_i$~\cite{ER}.
}
\label{fig:map}
\end{figure}

We have seen above that the effective low-energy mass matrix for the light
neutrinos contains 9 parameters, 3 mass eigenvalues, 3 real mixing angles
and 3 CP-violating phases. However, these are not all the parameters in
the minimal seesaw model. As shown in Fig.~\ref{fig:map}, this model has a
total of 18 parameters~\cite{Casas,EHLR}. Most of the rest of this Lecture
is devoted to understanding better the origins and possible manifestations
of the remaining parameters, many of which may have controlled the
generation of matter in the Universe via leptogenesis~\cite{FY} and may be
observable via renormalization in supersymmetric
models~\cite{DI,EHLR,EHRS,EHRS2}.

To see how the extra 9 parameters appear~\cite{EHLR}, we reconsider the 
full lepton sector, assuming that we have diagonalized the charged-lepton mass 
matrix:
\begin{equation}
\left( Y_\ell \right)_{ij} \; = \; Y^d_{\ell_i} \delta_{ij},
\end{equation}
as well as that of the heavy singlet neutrinos:
\begin{equation}
M_{ij} \; = M^d_i \delta_{ij}.
\label{diagM}
\end{equation}
We can then parametrize the neutrino Dirac coupling matrix $Y_\nu$ in 
terms of its real and diagonal eigenvalues and unitary rotation matrices:
\begin{equation}
Y_\nu \; = \; Z^* Y^d_{\nu_k} X^\dagger,
\label{diagYnu}
\end{equation}
where $X$ has 3 mixing angles and one CP-violating phase, just like the 
CKM matrix, and we can write $Z$ in the form
\begin{equation}
Z \; = \; P_1 {\bar Z} P_2,
\label{PZP}
\end{equation}
where ${\bar Z}$ also resembles the
CKM matrix, with  3 mixing angles and one CP-violating phase, and the 
diagonal matrices $P_{1,2}$ each have two CP-violating phases:
\begin{equation}
P_{1,2} \; = \; {\rm Diag} \left( e^{i\theta_{1,3}}, e^{i\theta_{2,4}}, 1 
\right). 
\label{PP}
\end{equation}
In this parametrization, we see explicitly that the neutrino 
sector has 18 parameters: the 3 heavy-neutrino mass eigenvalues 
$M^d_i$, the 3 real eigenvalues of $Y^D_{\nu_i}$, the $6 = 3 + 3$ real 
mixing angles in $X$ and ${\bar Z}$, and the $6 = 1 + 5$ CP-violating 
phases in $X$ and ${\bar Z}$~\cite{EHLR}.

As we discuss later in more detail, leptogenesis~\cite{FY} is proportional 
to the product
\begin{equation}
Y_\nu Y^\dagger_\nu \; = \; P_1^* {\bar Z}^* \left( Y^d_\nu \right)^2 
{\bar Z}^T P_1,
\label{leptog}
\end{equation}
which depends on 13 of the real parameters and 3 CP-violating phases, 
whilst the leading renormalization of soft supersymmetry-breaking masses 
depends on the combination
\begin{equation}
Y_\nu^\dagger Y_\nu \; = \; X \left( Y^d_\nu \right)^2 X^\dagger,
\label{renn}
\end{equation}
which depends on just 1 CP-violating phase, with two 
more phases appearing in higher orders, when one allows the heavy singlet 
neutrinos to be non-degenerate~\cite{EHRS}.

In order to see how the low-energy sector is embedded in this full 
parametrization, we first recall that the 3 phases in ${\tilde P}_2$ 
(\ref{MNSP}) become observable when one also considers high-energy 
quantities. Next, we introduce a complex orthogonal matrix
\begin{equation}
R \; \equiv \; \sqrt{M^d}^{-1} Y_\nu U \sqrt{M^d}^{-1} \left[ v \sin \beta 
\right],
\label{defR}
\end{equation}
which has 3 real mixing angles and 3 phases: $R^T R = 1$. These 6 
additional parameters may be used to characterize $Y_\nu$, by inverting 
(\ref{defR}):
\begin{equation}
Y_\nu \; = \; {\sqrt{M^d} R \sqrt{M^d} U^\dagger \over \left[ v \sin \beta 
\right]},
\label{invertY}
\end{equation}
giving us the same grand total of $18 = 9 + 3 +6$ parameters~\cite{EHLR}.
The leptogenesis observable (\ref{leptog}) may now be written in the form
\begin{equation}
Y_\nu Y^\dagger_\nu \; = \; { \sqrt{M^d} R {\cal M}^d_\nu R^\dagger 
\sqrt{M^d} \over \left[ v^2 \sin^2 \beta \right]},
\label{newleptog}
\end{equation}
which depends on the 3 phases in $R$, but {\it not} the 3 low-energy 
phases $\delta, \phi_{1,2}$, {\it nor} the 3 real MNS mixing 
angles~\cite{EHLR}!
Conversely, the leading renormalization observable (\ref{renn}) may be 
written in the form
\begin{equation}
Y^\dagger_\nu Y_\nu \; = \; U {\sqrt{{\cal M}^d_\nu} R^\dagger M^d R 
\sqrt{{\cal M}^d_\nu} \over \left[ v^2 \sin^2 \beta \right]} U^\dagger,
\label{newrenn}
\end{equation}
which depends explicitly on the MNS matrix, including the 
CP-violating phases $\delta$ and $\phi_{1,2}$, but only one of the three 
phases in ${\tilde P}_2$~\cite{EHLR}.

\subsection{Renormalization of Soft Supersymmetry-Breaking Parameters}

Let us now discuss the renormalization of soft supersymmetry-breaking 
parameters $m_0^2$ and $A$ in more detail, assuming that the input values 
at the GUT scale are flavour-independent. If they are not, there will be 
additional sources of flavour-changing processes, beyond those discussed 
in this and subsequent sections~\cite{FCNI,Masiero}. In the 
leading-logarithmic 
approximation, and assuming degenerate heavy singlet neutrinos, one finds
the following radiative corrections to the soft 
supersymmetry-breaking terms for sleptons:
\begin{eqnarray}
\left( \delta m_{\tilde L}^2 \right)_{ij} \; &=& \;
- { 1 \over 8 \pi^2} \left( 3 m_0^2 + A_0^2 \right) \left( Y_\nu^\dagger 
Y_\nu \right)_{ij} {\rm Ln} \left( {M_{GUT} \over M} \right), \nonumber \\
\left( \delta A_\ell \right)_{ij} \; &=& \;
- { 1 \over 8 \pi^2} A_0 Y_{\ell_i} \left( Y_\nu^\dagger 
Y_\nu \right)_{ij} {\rm Ln} \left( {M_{GUT} \over M} \right),
\label{leading}
\end{eqnarray}
where we have intially assumed that the heavy singlet neutrinos are 
approximately degenerate with $M \ll M_{GUT}$. In this case, there is a 
single analogue of the Jarlskog invariant of the Standard 
Model~\cite{Jarlskog}:
\begin{equation}
J_{\tilde L} \; \equiv \; {\rm Im} \left[ \left( m_{\tilde L}^2 
\right)_{12} \left( m_{\tilde L}^2
\right)_{23} \left( m_{\tilde L}^2
\right)_{31} \right],
\label{J}
\end{equation}
which depends on the single phase that is observable in this 
approximation. There are other Jarlskog invariants defined analogously in 
terms of various combinations with the $A_\ell$, but these are all 
proportional~\cite{EHLR}.

There are additional contributions if the heavy singlet neutrinos are not 
degenerate:
\begin{equation}
\left( {\tilde \delta} m_{\tilde L}^2 \right)_{ij} \; = \;
- { 1 \over 8 \pi^2} \left( 3 m_0^2 + A_0^2 \right) \left( Y_\nu^\dagger
L Y_\nu \right)_{ij}: \; L \equiv {\rm Ln} \left( {{\bar M} \over M_i} 
\right) \delta_{ij},
\label{morerenn}
\end{equation}
where ${\bar M} \equiv {^3}\sqrt{M_1 M_2 M_3}$, with $\left( {\tilde 
\delta} A_\ell \right)_{ij}$ being defined analogously. These new 
contributions contain the matrix factor
\begin{equation}
Y^\dagger L Y \; = \; X Y^d P_2 {\bar Z}^T L {\bar Z}^* P_2^* y^d 
X^\dagger,
\label{YLY}
\end{equation}
which introduces dependences on the phases in ${\bar Z} P_2$, though not 
$P_1$. In this way, the renormalization of the soft supersymmetry-breaking 
parameters becomes sensitive to a total of 3 CP-violating 
phases~\cite{EHRS}.

\subsection{Exploration of Parameter Space}

Now that we have seen how the 18 parameters in the minimal supersymmetric
seesaw model might in principle be observable, we would like to explore
the range of possibilities in this parameter space. This requires
confronting two issues: the unwieldy large dimensionality of the parameter
space, and the inclusion of the experimental information already obtained
(or obtainable) from low-energy studies of neutrinos. Of the 9 parameters 
accessible to these experiments: $m_{\nu_1}, m_{\nu_2}, m_{\nu_3}, 
\theta_{12}, \theta_{23}, \theta_{31}, \delta, \phi_1$ and $\phi_2$, we 
have measurements of 4 combinations: $\Delta m_{12}^2, \Delta 
m_{23}^2, \theta_{12}$ and $\theta_{23}$, and upper limits on 
the overall light-neutrino mass scale, $\theta_{13}$ and the 
double-$\beta$ decay observable (\ref{doublebeta}).

The remaining 9 parameters not measurable in low-energy neutrino physics 
may be characterized by an auxiliary Hermitean matrix of the following 
form~\cite{DI,EHRS2}:
\begin{equation}
H \; \equiv \; Y^\dagger_\nu D Y_\nu,
\label{defH}
\end{equation}
where $D$ is an arbitrary real and diagonal matrix. Possible choices for 
$D$ include ${\rm Diag}( \pm 1, \pm 1, \pm 1)$ and the logarithmic matrix 
$L$ defined in (\ref{morerenn}). Once one specifies the 9 parameters in 
$H$, either in a statistical survey or in some definite model, one can 
calculate
\begin{equation}
H^\prime \; \equiv \; \sqrt{{\cal M}^d_\nu} U^\dagger H U \sqrt{{\cal 
M}^d_\nu},
\label{defHprime}
\end{equation}
which can then be diagonalized by a complex orthogonal matrix $R^\prime$:
\begin{equation}
H^\prime \; = \; {R^\prime}^\dagger {\cal M}^{\prime d} R^\prime: \; 
{R^\prime}^T R^\prime = 1.
\label{defRprime}
\end{equation}
In this way, we can calculate all the remaining physical parameters:
\begin{equation}
( {\cal M}_\nu, H) \to ( {\cal M}_\nu, {\cal M}^{\prime d}, R^\prime) \to 
(Y_\nu, M_i)
\label{procedure}
\end{equation}
and then go on to calculate leptogenesis, charged-lepton violation, 
etc~\cite{DI,EHRS2}.

A freely chosen model will in general violate the experimental upper limit 
on $\mu \to e \gamma$~\cite{PDG}. It is easy to avoid this problem using 
the parametrization (\ref{defH})~\cite{EHRS2}. If one chooses $D = L$ and 
requires the 
entry $H_{12} = 0$, the leading contribution to $\mu \to e \gamma$ from 
renormalization of the soft supersymmetry-breaking masses will be 
suppressed. To suppress $\mu \to e \gamma$ still further, one may impose 
the constraint $H_{13} H_{23} = 0$. This condition evidently has two 
solutions: either $H_{13} = 0$, in which case $\tau \to e \gamma$ is 
suppressed but not $\tau \to \mu \gamma$, or alternatively $H_{23} = 0$, 
which favours $\tau \to e \gamma$ over $\tau \to \mu \gamma$. Thus we may 
define two generic textures $H^1$ and $H^2$:
\begin{eqnarray}
H^1 \; \equiv \; \left(
\begin{array}{ccc}
a & 0 & 0 \\
0 & b & d \\
0 & d^\dagger & c
\end{array}
\right), \; H^2 \; \equiv \; \left(
\begin{array}{ccc}
a & 0 & d \\
0 & b & 0 \\
d^\dagger & 0 & c
\end{array}
\right).
\label{H1H2}
\end{eqnarray}
We use these as guides in the following, whilst recalling that they 
represent extremes, and the truth may not favour one $\tau \to \ell 
\gamma$ decay mode so strongly over the other.

\subsection{Leptogenesis}

In addition to the low-energy neutrino constraints, we frequently employ
the constraint that the model parameters be compatible with the
leptogenesis scenario for creating the baryon asymmetry of the
Universe~\cite{FY}.  We recall that the baryon-to-entropy ratio $Y_B$ in
the Universe today is found to be in the range $10^{-11} < Y_B < 3 \times
10^{-10}$. This is believed to have evolved from a similar asymmetry in
the relative abundances of quarks and antiquarks before they became
confined inside hadrons when the temperature of the Universe was about
$100$~MeV. In the leptogenesis scenario~\cite{FY}, non-perturbative
electroweak interactions caused this small asymmetry to evolve out of a
similar small asymmetry in the relative abundances of leptons and
antileptons that had been generated by CP violation in the decays of heavy
singlet neutrinos.

The total decay rate of such a heavy neutrino $N_i$ may be written in the 
form
\begin{equation}
\Gamma_i \; = \; {1 \over 8 \pi} \left( Y_\nu Y^\dagger_\nu \right)_{ii} 
M_i.
\label{gammai}
\end{equation}
One-loop CP-violating diagrams involving the exchange of heavy 
neutrino $N_j$ would generate an asymmetry in $N_i$ decay of the form:
\begin{equation}
\epsilon_{ij} \; = \; {1 \over 8 \pi} {1 \over \left( Y_\nu Y^\dagger_\nu 
\right)_{ii}} {\rm Im} \left( \left( Y_\nu Y^\dagger_\nu \right)_{ij} 
\right)^2 f \left( {M_j \over M_i} \right),
\label{epsilon}
\end{equation}
where $f ( M_j / M_i )$ is a known kinematic function.

As already remarked, the relevant combination
\begin{equation}
\left( Y_\nu Y^\dagger_\nu \right) \; = \; \sqrt{M^d} R {\cal M}^d 
R^\dagger \sqrt{M^d}
\label{YY}
\end{equation}
{\it is independent of $U$ and hence of the light neutrino mixing angles 
and 
CP-violating phases}. The basic reason for this is that one makes a 
unitary 
sum over all the light lepton species in evaluating the 
asymmetry $\epsilon_{ij}$. It is easy to derive a compact expression for 
$\epsilon_{ij}$ in terms of the heavy neutrino masses and the complex 
orthogonal matrix $R$:
\begin{equation}
\epsilon_{ij} \; = \; {1 \over 8 \pi} M_j f \left( {M_j \over M_i} \right)
{ {\rm Im} \left( \left( R {\cal M}_\nu^d R^\dagger \right)_{ij} \right)^2 
\over \left( R {\cal M}_\nu^d R^\dagger \right)_{ii} }.
\label{epsilonR}
\end{equation}
This depends explicitly on the extra phases in $R$: how can we measure 
them?

The basic principle of a strategy to do this is the
following~\cite{EHLR,EHRS,EHRS2}. The renormalization of soft
supersymmetry-breaking parameters, and hence flavour-changing interactions
and CP violation in the lepton sector, depend on the leptogenesis
parameters as well as the low-energy neutrino parameters $\delta,
\phi_{1,2}$. If one measures the latter in neutrino experiments, and the
discrepancy in the soft supersymmetry-breaking determines the leptogenesis
parameters.

An example how this could work is provided by the two-generation version 
of the supersymmetric seesaw model~\cite{EHLR}. In this case, we have 
${\cal M}^d_\nu 
= {\rm Diag}(m_{\nu_1}, m_{\nu_1})$ and $M^d = {\rm Diag}(M_1, M_2)$, and 
we may parameterize
\begin{eqnarray}
R \; = \; \left(
\begin{array}{cc}
\cos (\theta_r + i \theta_i ) & \sin (\theta_r + i \theta_i ) \\
- \sin (\theta_r + i \theta_i ) & \cos (\theta_r + i \theta_i )
\end{array}
\right).
\label{twodR}
\end{eqnarray}
In this case, the leptogenesis decay asymmetry is proportional to
\begin{equation}
{\rm Im} \left( \left( Y_\nu Y^\dagger_\nu \right)^{21} \right)^2 \; = \;
{ \left( m^2_{\nu_1} - m^2_{\nu_2} \right) M_1 M_2 \over 2 v^4 \sin^4 
\beta} {\rm sinh} 2 \theta_i sin 2 \theta_r.
\label{leptog2}
\end{equation}
We see that this is related explicitly to the CP-violating phase and 
mixing angle in $R$ (\ref{twodR}), and is independent of the low-energy 
neutrino parameters. Turning now to the renormalization of the soft 
supersymmetry-breaking parameters, assuming for simplicity maximal mixing 
in the MNS matrix $V$ and setting the diagonal Majorana phase matrix $P_0 
= {\rm Diag}(e^{-i \phi}, 1)$, we find that
\begin{eqnarray}
{\rm Re} \left[ \left( Y^\dagger_\nu Y_\nu \right)^{12} \right] \; &=& \;
- { (m_{\nu_2} - m_{\nu_1} ) \over 4 v^2 \sin^2 \beta} (M_1 + M_2) {\rm 
cosh} 2 \theta_i \; + \; \cdots, \nonumber \\
{\rm Im} \left[ \left( Y^\dagger_\nu Y_\nu \right)^{12} \right] \; &=& \;
{\sqrt{m_{\nu_2} m_{\nu_1}} \over 2 v^2 \sin^2 \beta} (M_1 + M_2) {\rm 
sinh} 2 \theta_i \cos \phi \; + \; \cdots.
\label{twodYY}
\end{eqnarray}
In this case, the strategy for relating leptogenesis to low-energy 
observables would be: (i) use double-$\beta$ decay to determine $\phi$,
(ii) use low-energy observables sensitive to ${\rm Re, Im} \left[ \left( 
Y^\dagger_\nu Y_\nu \right)^{12} \right]$ to determine $\theta_r$ and 
$\theta_i$ (\ref{twodYY}), which then (iii) determine the leptogenesis 
asymmetry (\ref{leptog2}) in this two-generation model.

\begin{figure}
\begin{centering}
\hspace{2.5cm}
\epsfig{figure=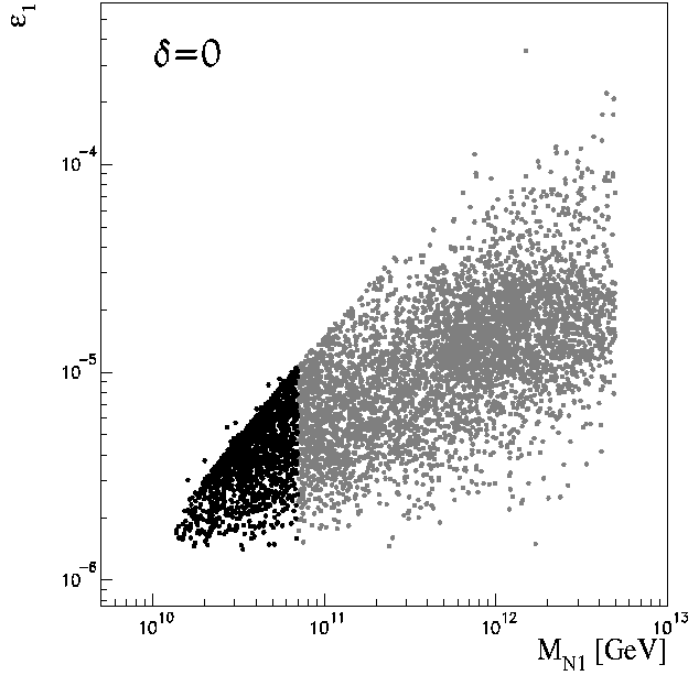,width=10cm}
\end{centering}
\hglue3.5cm   
\caption[]{\it Heavy singlet neutrino decay may exhibit a CP-violating 
asymmetry, leading to leptogenesis and hence baryogenesis, even if the 
neutrino oscillation phase $\delta$ vanishes~\cite{ER}.} 
\label{fig:nodelta}
\end{figure}

\begin{figure}
\begin{centering}
\hspace{2.5cm}
\epsfig{figure=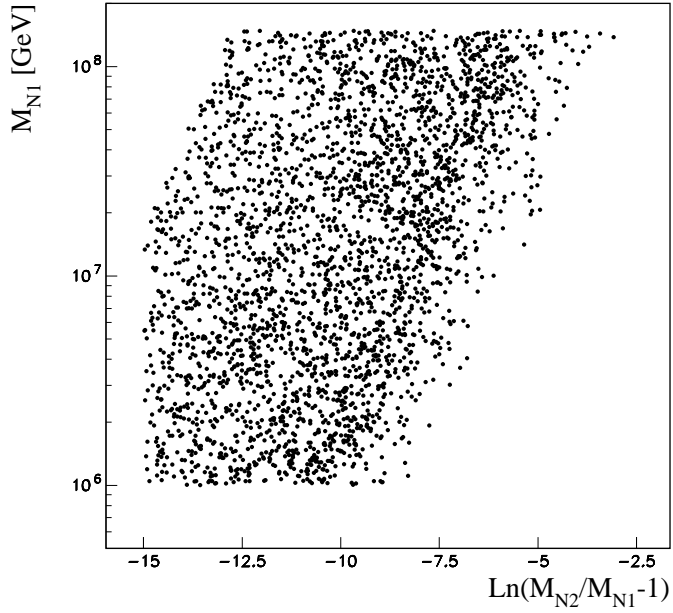,width=10cm}
\end{centering}
\hglue3.5cm   
\caption[]{\it The lower limit on the mass of the lightest heavy singlet 
neutrino may be significantly reduced if the two lightest singlet 
neutrinos are almost degenerate~\cite{ERY}.} 
\label{fig:ERY}
\end{figure}

In general, one may formulate the following strategy for calculating 
leptogenesis in terms of laboratory observables:
\begin{itemize}
\item{Measure the neutrino oscillation phase $\delta$ and the Majorana 
phases $\phi_{1,2}$,}
\item{Measure observables related to the renormalization of soft 
supersymmetry-breaking parameters, that are functions of $\delta, 
\phi_{1,2}$ and the leptogenesis phases,}
\item{Extract the effects of the known values of $\delta$ and 
$\phi_{1,2}$, and isolate the leptogenesis parameters.}
\end{itemize}
In the absence of complete information on the first two steps above, we 
are currently at the stage of preliminary explorations of the 
multi-dimensional parameter space. As seen in Fig.~\ref{fig:nodelta}, the 
amount of the leptogenesis asymmetry is explicitly independent of 
$\delta$~\cite{ER}. An important observation is that there is a 
non-trivial lower 
bound on the mass of the lightest heavy singlet neutrino $N$:
\begin{equation}
M_{N_1} \; \gappeq \; 10^{10}~{\rm GeV}
\label{normal}
\end{equation}
if the light neutrinos have the conventional hierarchy of masses, and
\begin{equation}   
M_{N_1} \; \gappeq \; 10^{11}~{\rm GeV}
\label{inverted}
\end{equation}
if they have an inverted hierarchy of masses~\cite{ER}. This observation 
is 
potentially important for the cosmological abundance of gravitinos, which 
would be problematic if the cosmological temperature was once high enough 
for leptogenesis by thermally-produced singlet neutrinos weighing as much 
as (\ref{normal}, \ref{inverted})~\cite{gravitino}. However, these bounds 
could be relaxed 
if the two lightest $N_i$ were near-degenerate, as seen in 
Fig.~\ref{fig:ERY}~\cite{ERY}. Striking aspects of this scenario include 
the 
suppression of $\mu \to e \gamma$, the relatively large value of $\tau \to 
\mu \gamma$, and a preferred value for the neutrinoless double-$\beta$ 
decay observable:
\begin{equation}
\langle m \rangle_{ee} \; \sim \; \sqrt{\Delta m^2_{solar}} \sin^2 
\theta_{12}.
\end{equation}

\subsection{Flavour-Violating Decays of Charged Leptons}

Several such decays can be studied within this framework, including
$\mu \to e \gamma, \tau \to e \gamma, \tau \to \mu \gamma, \mu \to 3 e$, 
and $\tau \to 3 \mu/e$~\cite{others}.

The effective Lagrangian for $\mu \to e \gamma$ and $\mu \to 3 e$ can be 
written in the form~\cite{effL,EHLR}:
\begin{eqnarray}
{\cal L} &=& -\frac{4G_F}{\sqrt{2}}\{
        {m_{\mu }}{A_R}\overline{\mu_{R}}
        {{\sigma }^{\mu \nu}{e_L}{F_{\mu \nu}}}
       + {m_{\mu }}{A_L}\overline{\mu_{L}}
        {{\sigma }^{\mu \nu}{e_R}{F_{\mu \nu}}} \nonumber \\
    &&   +{g_1}(\overline{{{\mu }_R}}{e_L})
              (\overline{{e_R}}{e_L}) 
       + {g_2}(\overline{{{\mu }_L}}{e_R})
              (\overline{{e_L}}{e_R}) \nonumber \\
    &&   +{g_3}(\overline{{{\mu }_R}}{{\gamma }^{\mu }}{e_R})
              (\overline{{e_R}}{{\gamma }_{\mu }}{e_R})
       + {g_4}(\overline{{{\mu }_L}}{{\gamma }^{\mu }}{e_L})
              (\overline{{e_L}}{{\gamma }_{\mu }}{e_L})  \nonumber \\
    &&   +{g_5}(\overline{{{\mu }_R}}{{\gamma }^{\mu }}{e_R})
              (\overline{{e_L}}{{\gamma }_{\mu }}{e_L})
       + {g_6}(\overline{{{\mu }_L}}{{\gamma }^{\mu }}{e_L})
              (\overline{{e_R}}{{\gamma }_{\mu }}{e_R})
       +  h.c. \}.
\label{effL}
\end{eqnarray}
The decay $\mu \to e \gamma$ is related directly to the coefficients 
$A_{L,R}$:
\begin{equation}
{Br}(\mu^{+} \rightarrow e^{+}\gamma)=384 \pi^2
\left(|A_L|^2+|A_R|^2 \right),
\label{muegamma}
\end{equation}
and the branching ratio for $\mu \to 3 e$ is given by
\begin{equation}
B(\mu \to e \gamma) \; = \; 2(C_1 + C_2) + C_3 + C_4 + 
32\left( {\rm ln} {m_\mu^2 \over m_e^2} - {11 \over 4} \right) (C_5 + C_6) 
\nonumber \\
+ 16(C_7 + C_8) + 8(C_9 + C_{10}),
\label{mu3e}
\end{equation}
where
\begin{eqnarray}
&& C_{1} = \frac{|g_{1}|^{2}}{16} + |g_{3}|^{2},~
C_{2} = \frac{|g_{2}|^{2}}{16} + |g_{4}|^{2},~ \nonumber 
\\
&& C_{3} = |g_{5}|^{2},~ C_{4} = |g_{6}|^{2},C_{5} = 
|eA_{R}|^{2},~
C_{6}   =   |eA_{L}|^{2},~
 C_{7}   =   {\rm Re}(eA_{R}g_{4}^{*}), \nonumber \\
&&C_{8}   =   {\rm Re}(eA_{L}g_{3}^{*}),~
C_{9}   =   {\rm Re}(eA_{R}g_{6}^{*}),~
C_{10}   =   {\rm Re}(eA_{L}g_{5}^{*})\,.
\label{Cintermsofg}.
\end{eqnarray}
These coefficients may easily be calculated using the 
renormalization-group equations for soft supersymmetry-breaking 
parameters~\cite{EHLR,EHRS2}.

\begin{figure}
\hspace{2cm}
\epsfig{figure=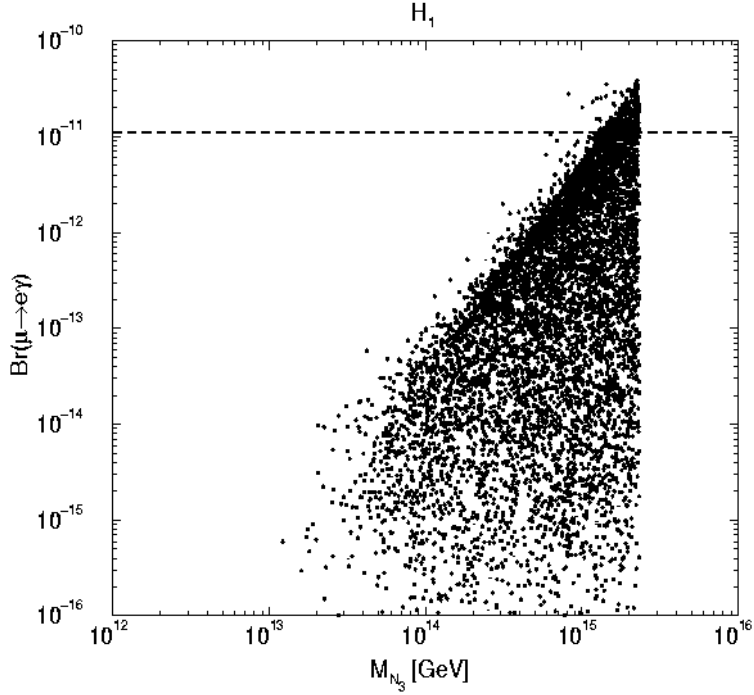,width=10cm}
\hglue3.5cm   
\caption[]{\it Scatter plot of the branching ratio for $\mu \to e \gamma$ 
in the supersymmetric seesaw model for various values of its unknown 
parameters~\cite{EHRS2}.} 
\label{fig:muegamma}
\end{figure}

\begin{figure}
\begin{centering}
\hspace{2cm}
\epsfig{figure=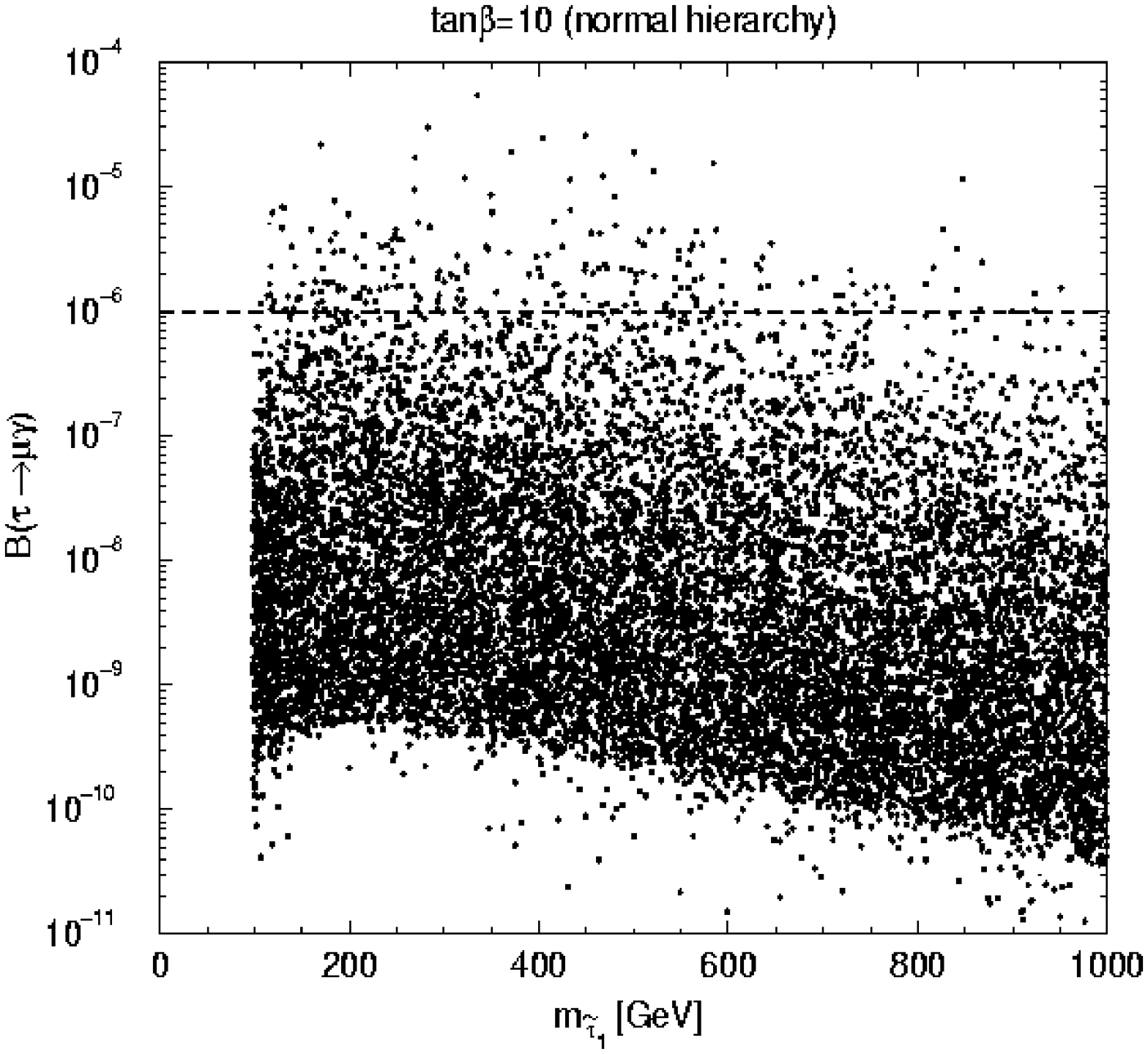,width=10cm}
\end{centering}
\hglue3.5cm   
\caption[]{\it Scatter plot of the branching ratio for $\tau \to \mu 
\gamma$ in one variant of the supersymmetric seesaw model for various 
values of its unknown parameters~\cite{EHRS2}.}
\label{fig:taumugamma}
\end{figure}

\begin{figure}
\begin{centering}
\hspace{2cm}
\epsfig{figure=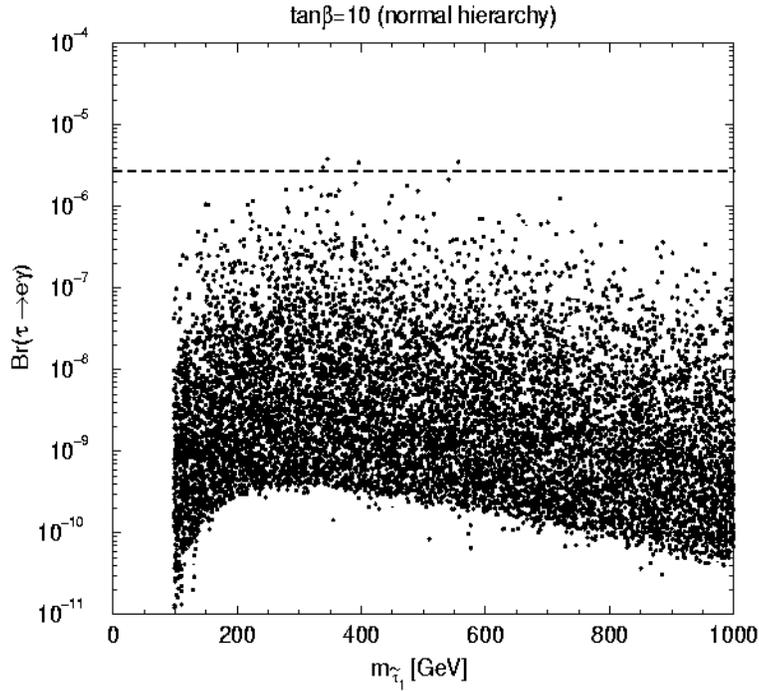,width=10cm}
\end{centering}
\hglue3.5cm   
\caption[]{\it Scatter plot of the branching ratio for $\tau \to e
\gamma$ in a variant the supersymmetric seesaw model for various values of 
its unknown parameters~\cite{EHRS2}.}
\label{fig:tauegamma}
\end{figure}

Fig.~\ref{fig:muegamma} displays a scatter plot of $B(\mu \to e \gamma)$ 
in the texture $H^1$ mentioned earlier, as a function of the 
singlet neutrino mass $M_{N_3}$. We see that $\mu \to e \gamma$ may 
well have a branching ratio close to the present experimental upper limit, 
particularly for larger $M_{N_3}$. Predictions 
for $\tau \to \mu \gamma$ and $\tau \to e \gamma$ decays are shown in 
Figs.~\ref{fig:taumugamma} and \ref{fig:tauegamma} for the textures $H^1$ 
and $H^2$, 
respectively. As advertized earlier, the $H^1$ texture favours $\tau \to 
\mu \gamma$ and the $H^2$ texture favours $\tau \to e \gamma$. We see that 
the branching ratios decrease with increasing sparticle masses, but that 
the range due to variations in the neutrino parameters is considerably 
larger than that due to the sparticle masses. The present 
experimental upper limits on $\tau \to \mu \gamma$, in particular,
already exclude significant numbers of parameter choices.

\begin{figure}
\begin{centering}
\hspace{2cm}
\epsfig{figure=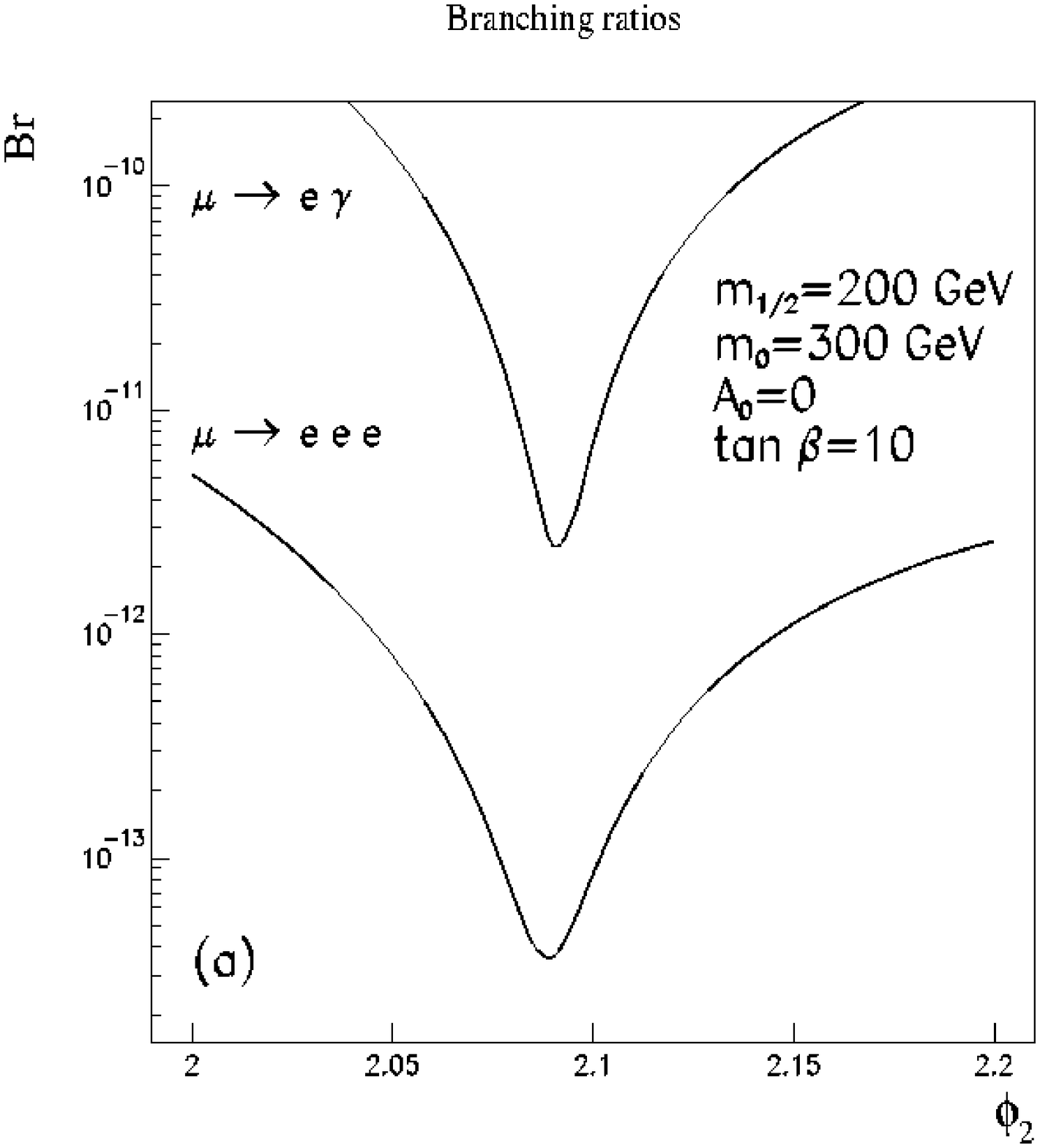,width=10cm}
\end{centering}
\hglue3.5cm   
\caption[]{\it The branching ratio for $\mu \to e \gamma$ may be 
suppressed for some particular values of the model parameters, in 
which case the branching ratio for $\mu \to 3e$ gets significant
contributions form other diagrams besides the photonic penguin 
diagram~\cite{EHLR}.} 
\label{fig:mu3e} 
\end{figure}

The branching ratio for $\mu \to 3 e$ is usually dominated by the photonic
penguin diagram, which contributes the $C_{5,6}$ terms in (\ref{mu3e}),
yielding an essentially constant ratio for $B(\mu \to 3 e) / B(\mu \to e
\gamma)$. However, if $\mu \to e \gamma$ decay is parametrically
suppressed, as it may have to be in order to respect the experimental
upper bound on this decay, then other diagrams may become important in
$\mu \to 3 e$ decay. In this case, the ratio $B(\mu \to 3 e)  / B(\mu \to
e \gamma)$ may be enhanced, as seen in Fig.~\ref{fig:mu3e}. 

\begin{figure}
\begin{centering}
\hspace{2cm}
\epsfig{figure=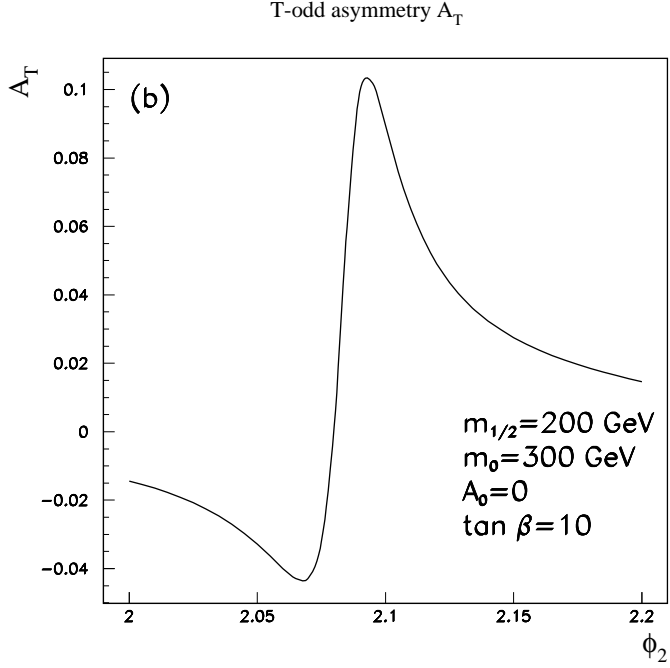,width=10cm}
\end{centering}
\hglue3.5cm   
\caption[]{\it The T-violating asymmetry $A_T$ in $\mu \to 3 e $ 
decay is enhanced in the regions of parameter space shown in 
Fig.~\ref{fig:mu3e} where the branching ratio for $\mu \to e 
\gamma$ is suppressed, and different diagrams may interfere in the 
$\mu \to 3 e $ decay amplitude~\cite{EHLR}.} 
\label{fig:AT} 
\end{figure}

As a result, interference between the photonic penguin diagram and the
other diagrams may in principle generate a measurable T-odd asymmetry in 
$\mu \to 3
e$ decay. This is sensitive to the CP-violating parameters in the 
supersymmetric seesaw model, and is in principle observable in polarized 
$\mu^+ \to e^+ e^- e^+$ decay: 
\begin{equation}
A_T (\mu^+ \to e^+ e^- e^+) = {3 \over 2 {\cal B}} \left( 2.0 C_{11} - 1.6 
C_{12} 
\right),
\label{AT}
\end{equation}
where
\begin{equation}
C_{11} = {\rm Im} \left( e A_R g_4^* + e A_L g_3^* \right),
C_{12} = {\rm Im} \left( e A_R g_6^* + e A_L g_5^* \right),
\label{C1112}
\end{equation}
and ${\cal B}$ is the $\mu \to 3 e$ branching ratio with an optimized 
cutoff for the more energetic positron:
\begin{equation}
{\cal B} = 1.8 (C_1+C_2)+0.96 (C_3+C_4) + 88 (C_5+C_6)
+14 (C_7+C_8) +8 (C_9+C_{10}).
\label{calB}
\end{equation}
As seen in Fig.~\ref{fig:AT}, the T-odd asymmetry is 
enhanced in regions of parameter space where $B(\mu \to e \gamma)$ is 
suppressed~\cite{EHLR}. If/when $\mu \to e \gamma$ and/or $\mu \to 3 e$ 
decays are 
observed, measuring $A_T$ (\ref{AT}) may provide an interesting window on 
CP violation in the seesaw model.

\subsection{Lepton Electric Dipole Moments}

This CP violation may also be visible in electric dipole moments for the
electron and muon $d_e$ and $d_\mu$~\cite{dedmu}. It is usually thought 
that these are
unobservably small in the minimal supersymmetric seesaw model, and that
$|d_e/d_\mu| = m_e/m_\mu$. However, $d_e$ and $d_\mu$ may be strongly
enhanced if the heavy singlet neutrinos are not degenerate~\cite{EHRS}, 
and depend on new phases that contribute to leptogenesis~\footnote{This 
effect makes lepton electric dipole moments possible even in a 
two-generation model.}. The leading 
contributions to $d_e$ and $d_\mu$ in the presence of non-degenerate  
heavy-singlet neutrinos are produced by the following terms in the 
renormalization of soft supersymmetry-breaking parameters:
\begin{eqnarray}
\left( {\tilde \delta} m^2_{\tilde L} \right)_{ij} &=& {18 \over (4 
\pi)^4} \left( m^2_0 + A_e^2 \right) \{ Y^\dagger_\nu L Y_\nu, 
Y^\dagger_\nu Y_\nu \}_{ij} {\rm ln} \left( {M_{GUT} \over {\bar M}} 
\right), \nonumber \\
\left( {\tilde A_e} \right)_{ij} &=& {1 \over (4 \pi)^4} A_0 \left[ 11
\{ Y^\dagger_\nu L Y_\nu, Y^\dagger_\nu Y_\nu \}  + 7 [ 
Y^\dagger_\nu L Y_\nu, Y^\dagger_\nu Y_\nu ] \right]_{ij} {\rm ln} 
\left( {M_{GUT} \over 
{\bar M}} \right),
\label{nondeg}
\end{eqnarray}
where the mean heavy-neutrino mass ${\bar M} \equiv {^3}\sqrt{M_1 M_2 
M_3}$ and the matrix $L \equiv 
{\rm ln}({\bar M} / M_i) \delta_{ij}$ were introduced in 
(\ref{morerenn}).

\begin{figure}
\begin{centering}
\hspace{2.5cm}
\epsfig{figure=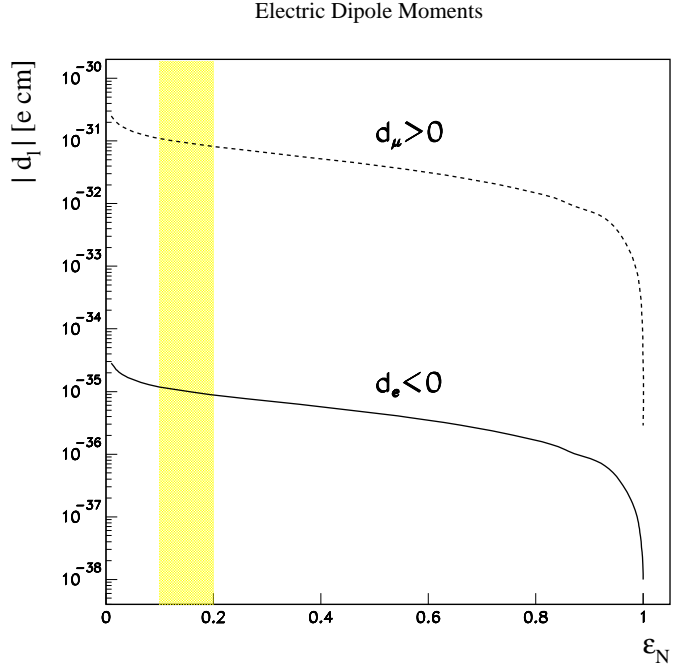,width=10cm}
\end{centering}
\hglue3.5cm   
\caption[]{\it The electric dipole moments of the electron and 
muon, $d_e$ and $d_\mu$, may be enhanced if the heavy singlet 
neutrinos are non-degenerate. The horizontal axis parameterizes 
the breaking of their degeneracy, and the vertical strip indicates 
a range favoured in certain models~\cite{EHRS}.} 
\label{fig:increase} \end{figure}

\begin{figure}
\hspace{1cm}
\epsfig{figure=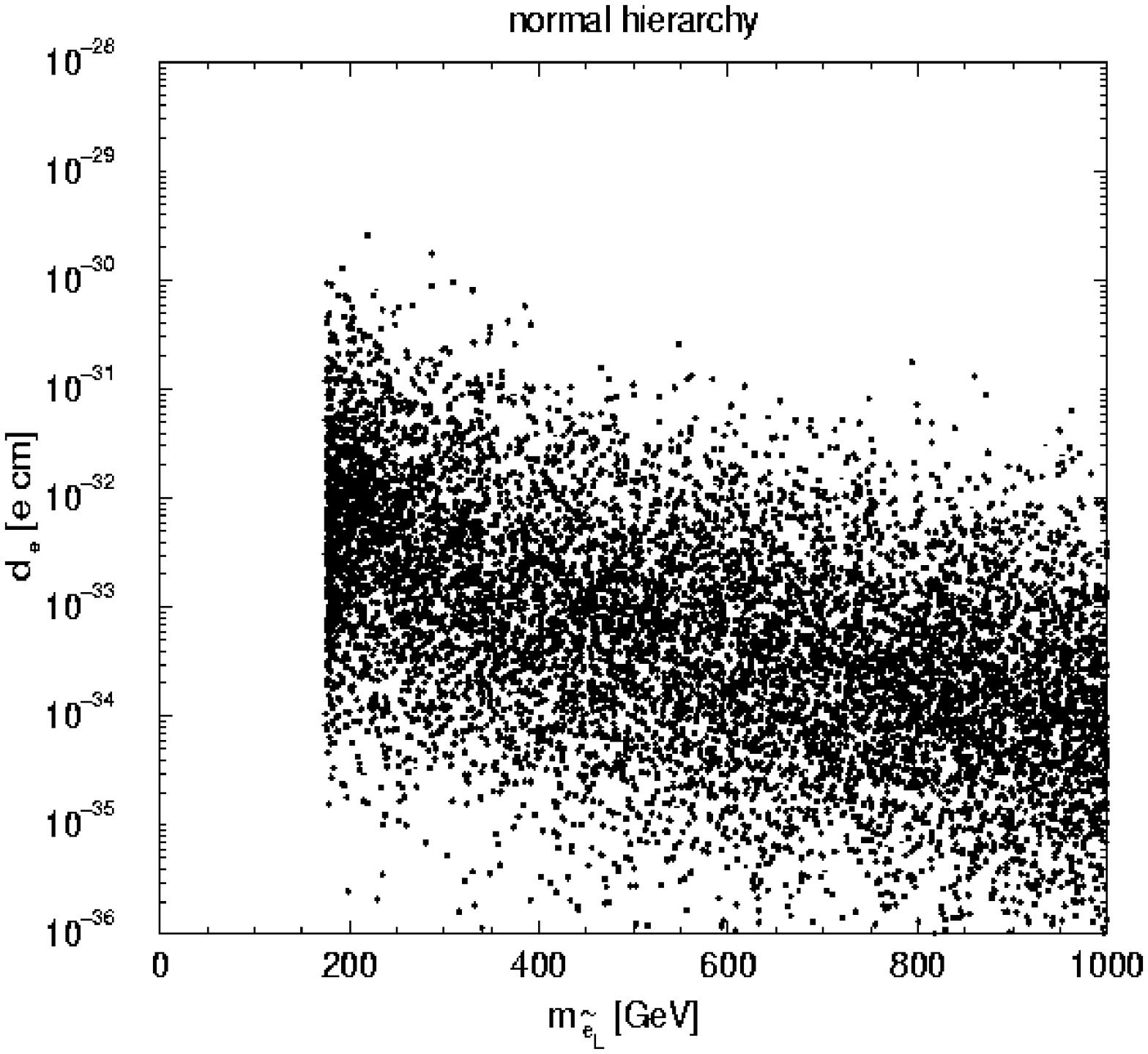,width=7cm}
\epsfig{figure=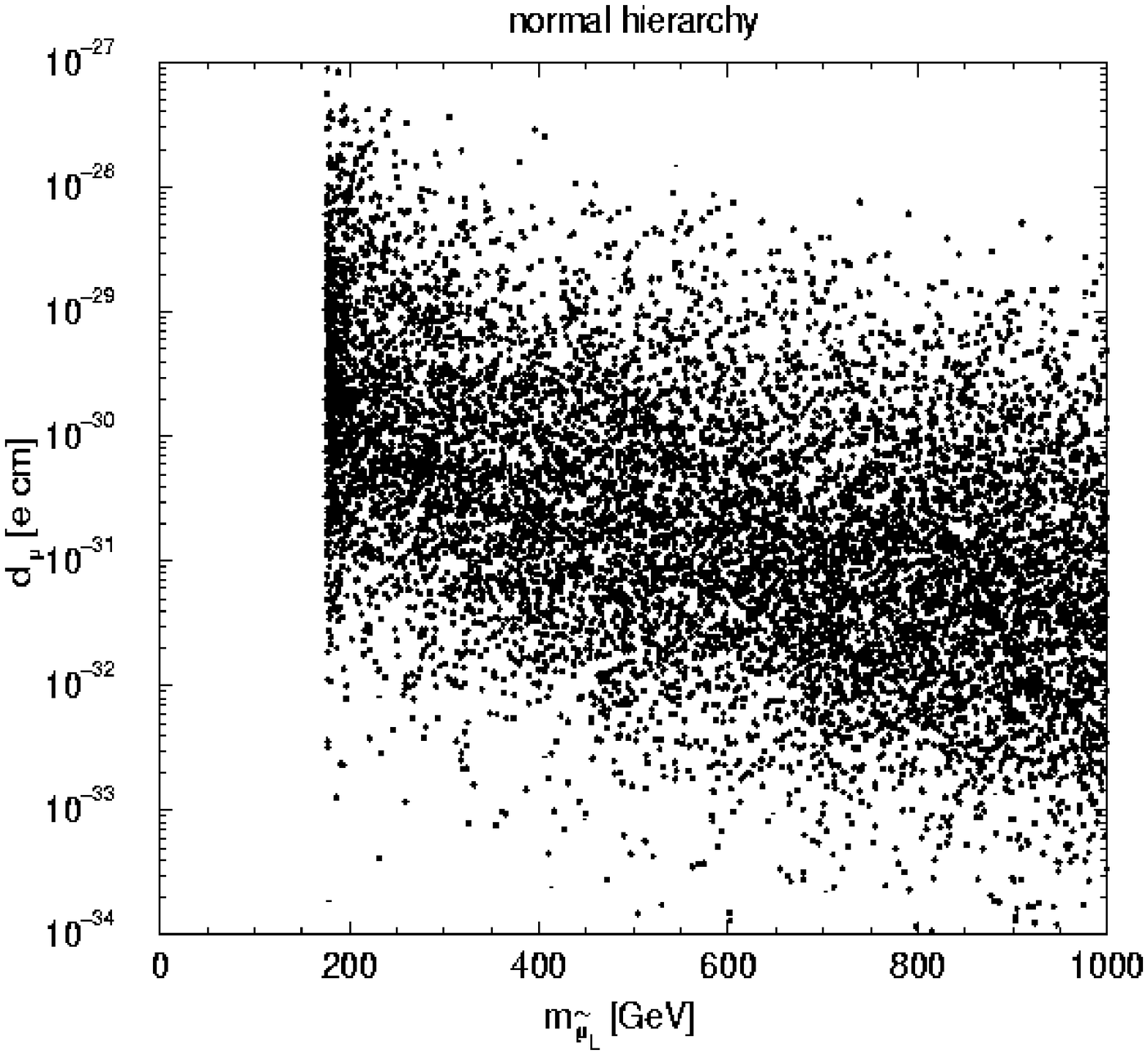,width=7cm}
\hglue3.5cm   
\caption[]{\it Scatter plots of $d_e$ and $d_\mu$ in variants of 
the supersymmetric seesaw model, for different values of the 
unknown parameters~\cite{EHRS2}.} 
\label{fig:dedmu}
\end{figure}

It should be emphasized that non-degenerate heavy-singlet 
neutrinos are actually expected in most models of neutrino masses. 
Typical examples are texture models of the form
\begin{eqnarray}
Y_\nu \sim Y_0 \left(
\begin{array}{ccc}
0 & c \epsilon_\nu^3 & d \epsilon_\nu^3 \nonumber \\
c \epsilon_\nu^3 & a \epsilon_\nu^2 & b \epsilon_\nu^2 \nonumber \\
d \epsilon_\nu^3 & b \epsilon_\nu^2 & e^{i\psi}
\end{array}
\right),
\label{hierarchymodel}
\end{eqnarray}
where $Y_0$ is an overall scale, $\epsilon_\nu$ characterizes the 
hierarchy, $a, b, c$ and $d$ are ${\cal O}(1)$ complex numbers, and $\psi$ 
is an arbitrary phase. For example, there is an SO(10) GUT model of this 
form with $d = 0$ and a flavour SU(3) model with $a=b$ and $c=d$. The 
hierarchy of heavy-neutrino masses in such a model is
\begin{equation}
M_1 : M_2 : M_3 \; = \; \epsilon_N^6 :  \epsilon_N^4 : 1,
\label{ratios}
\end{equation}
and indicative ranges of the hierarchy parameters are
\begin{equation}
\epsilon_\nu \sim \sqrt{{\Delta m^2_{solar} \over \Delta m^2_{atmo}}} \; , 
\; \epsilon_N \sim 0.1 ~ {\rm to} ~ 0.2.
\label{values}
\end{equation}
Fig.~\ref{fig:increase} shows how much $d_e$ and $d_\mu$ may be increased 
as soon as the degeneracy between the heavy neutrinos is broken:
$\epsilon \ne 1$. We also see that $|d_\mu / d_e| \gg m_\mu / m_e$ when 
$\epsilon_N \sim 0.1 ~ {\rm to} ~ 0.2$. Scatter plots of $d_e$ and $d_\mu$ 
are shown in Fig.~\ref{fig:dedmu}, where we see that values as large as 
$d_\mu \sim 10^{-27}$~e.cm and $d_e \sim 3 \times 10^{-30}$~e.cm are 
possible. For comparison, 
the present experimental upper limits are $d_e < 1.6 \times 
10^{-27}$~e.cm~\cite{de} 
and $d_\mu < 10^{-18}$~e.cm~\cite{BNL1}. An ongoing series of experiments 
might be able to reach $d_e < 3 \times 10^{-30}$~e.cm, and a type of 
solid-state experiment 
that might 
be sensitive to $d_e \sim 10^{-33}$~e.cm has been 
proposed~\cite{Lamoreaux}. Also, $d_\mu 
\sim 10^{-24}$~e.cm might be accessible with the PRISM experiment proposed 
for the JHF~\cite{PRISM}, and $d_\mu \sim 5 \times 10^{-26}$~e.cm might be 
attainable 
at the front end of a neutrino factory~\cite{nufact}. It therefore seems 
that $d_e$ 
might be measurable with foreseeable experiments, whilst $d_\mu$ would 
present more of a challenge.

\subsection{(Not so) Rare Sparticle Decays}

The suppression of rare lepton-flavour-violating (LFV) $\mu$ and $\tau$ 
decays in the supersymmetric
seesaw model is due to loop effects and the small masses of the leptons
relative to the sparticle mass scale. The intrinsic slepton mixing may not
be very small, in which case there might be relatively large amounts of
LFV observable in sparticle decays. An example that might be detectable at
the LHC is $\chi_2 \to \chi_1 \ell^\pm \ell^{\prime \mp}$, where $\chi_1
(\chi_2)$ denotes the (next-to-)lightest neutralino~\cite{HPLFV}. The
largest LFV effects might be in $\chi_2 \to \chi_1 \tau^\pm \mu^{\mp}$ and
$\chi_2 \to \chi_1 \tau^\pm e^{\mp}$~\cite{CEGLR}, though $\chi_2 \to
\chi_1 e^\pm \mu^{\mp}$ would be easier to detect.

\begin{figure}
\begin{centering}
\hspace{2.5cm}
\epsfig{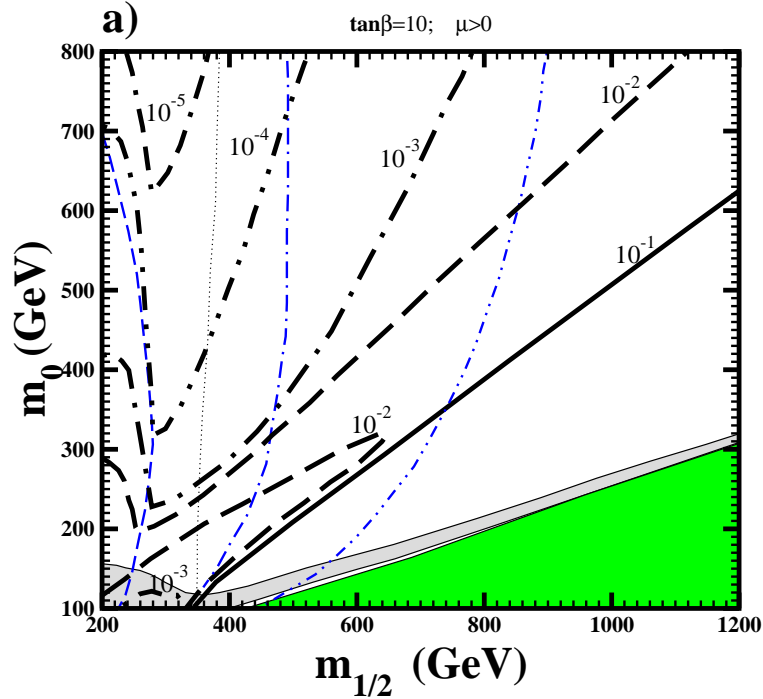}
\end{centering}
\hglue3.5cm   
\caption[]{\it Contours of the possible ratio of the branching ratios for 
$\chi_2 \to \chi_1 \tau^\pm \mu^{\mp}$ and $\chi_2 \to \chi_1 \mu^\pm
\mu^{\mp}$ (black lines) and of the branching ratio for $\tau \to \mu 
\gamma$ (near-vertical grey/blue lines).~\cite{CEGLR}.}
\label{fig:CEGLR}
\end{figure}

As shown in Fig.~\ref{fig:CEGLR}~\cite{CEGLR}, these decays are likely to
be enhanced in a region of CMSSM parameter space complementary to that
where $\tau \to e/\mu \gamma$ decys are most copious. This is because the
interesting $\chi_2 \to \chi_1 \tau^\pm \mu^{\mp}$ and $\chi_2 \to \chi_1
\tau^\pm e^{\mp}$ decays are mediated by slepton exchange, which is
maximized when the slepton mass is close to $m_{\chi_1}$. This happens in
the coannihilation region where the LSP relic density may be in the range
preferred by astrophysics and cosmology, even if $m_{\chi_1}$ is
relatively large. Thus searches for LFV $\chi_2 \to \chi_1 \tau^\pm
\mu^{\mp}$ and $\chi_2 \to \chi_1 \tau^\pm e^{\mp}$ decays are quite
complementary to those for $\tau \to e/\mu \gamma$.

\subsection{Possible CERN Projects beyond the LHC}

What might come after the LHC at CERN? One possibility is the LHC itself,
in the form of an energy or luminosity upgrade~\cite{upgrade}. It seems
that the possibilities for the former are very limited: a substantial
energy upgrade would require a completely new machine in the LHC tunnel,
with even higher-field magnets and new techniques for dealing with
synchrotron radiation. On the other hand, a substantial increase in
luminosity seems quite feasible, though it would require some rebuilding
of (at least the central parts of) the LHC detectors.

The mainstream project for CERN after the LHC is CLIC, the multi-TeV
linear $e^+ e^-$ collider~\cite{CLIC}. CERN is continuing R\&D on this
project, with a view to being able to assess its feasibility when the LHC
starts to produce data, e.g., specifying the energy scale of supersymmetry
or extra dimensions. CLIC would complement the work of the LHC and any
first-generation sub-TeV linear $e^+ e^-$ collider, e.g., by detailed
studies of heavier sparticles such as heavier charginos, neutralinos and
strongly-interacting sparticles~\cite{Battaglia,CLICphys}.

A possible alternative that has attracted considerable enthusiasm in
Europe is to develop neutrino physics beyond the current CNGS
project~\cite{CNGS}. A first step might be an off-axis experiment in the
CNGS beam, which could have interesting sensitivity to
$\theta_{13}$~\cite{offaxis}. A second might be a super-beam produced by
the SPL~\cite{SPL} at CERN and sent to a large detector in the Fr\'jus
tunnel~\cite{Frejus}. A third step could be a storage ring for unstable
ions, whose decays would produce a `$\beta$ beam' of pure $\nu_e$ or
${\bar \nu}_e$ neutrinos that could also be observed in a Fr\'ejus
experiment. These experiments might be able to measure $\delta$ via CP
and/or T violation in neutrino oscillations~\cite{betabeam}. A fourth step
could be a full-fledged neutrino factory based on a muon storage ring,
which would produce pure $\nu_\mu$ and ${\bar \nu}_e$ (or $\nu_e$ and
${\bar \nu}_\mu$ beams and provide a greatly enhanced capability to search
for or measure $\delta$ via CP violation in neutrino
oscillations~\cite{nufact}. Further steps might then include $\mu^+ \mu^-$
colliders with various centre-of-mass energies, from the mass of the
lightest Higgs boson, through those of the heavier MSSM Higgs bosons $H,
A$, to the multi-TeV energy frontier~\cite{mucoll}.

This is an ambitious programme that requires considerable R\&D. CERN
currently does not have the financial resources to support this, but it is
hoped that other European laboratories and the European Union might
support a network of interested physicists. Such an ambitious neutrino
programme would also require wide support in the physics community. In
addition to the neutrino physics itself, many might find enticing the
other experimental possibilities offered by the type of intense proton
driver required. These could include some of the topics discussed in this
Lecture, including rare decays of slow or stopped muons~\cite{nufact},
such as $\mu \to e \gamma$ and anomalous $\mu \to e$ conversion on a
nucleus, measurements of $g_\mu -2$ and $d_\mu$, rare K
decays~\cite{Buchalla}, short-baseline deep-inelastic neutrino experiments
with very intense beams~\cite{Mangano}, muonic atoms, etc., etc..
Physicists interested in such a programme, which nicely complements the
`core business' of the neutrino factory, should get together and see how a
coalition of interested parties could be assembled.  A large investment in
neutrino physics will require a broad range of support.

\end{document}